%% file: thesis.tex
\documentclass[openany,10pt]{book}

\usepackage{mathrsfs}
\usepackage{amsmath}
\usepackage{amssymb}
\usepackage{makeidx}
\usepackage{times}
\usepackage{epsfig}
\usepackage{array}
\usepackage{color}
\usepackage{natbib}
\usepackage{float}
\usepackage{subfigure}
\usepackage{graphicx}

\newcommand{\clearemptydoublepage}{\newpage{\pagestyle{empty}\cleardoublepage}}

\DeclareMathOperator{\Tr}{Tr}

\DeclareMathOperator{\Res}{Res}
\DeclareMathOperator{\Imag}{Im}
\DeclareMathOperator{\Real}{Re}
\DeclareMathOperator{\diag}{diag}

\bibpunct{[}{]}{,}{a}{}{,}

\makeindex

\begin{document}

\title{Mesoscopic Coulomb Drag}
\author{Niels Asger Mortensen}
\date{Ph.D. Thesis\\\emph{Mikroelektronik Centret}\\\emph{Technical University of Denmark}}

\maketitle
\clearemptydoublepage
\pagenumbering{roman}
\setcounter{page}{1}

\include{fwd}

\include{cv}

\include{papers}

\tableofcontents

\listoffigures

\listoftables

\clearemptydoublepage
\pagenumbering{arabic}
\setcounter{page}{1}

\include{introduction}\clearemptydoublepage

\include{kubo}\clearemptydoublepage

\include{matsubara}\clearemptydoublepage

\include{wick}

\clearemptydoublepage

\include{numerical}

\clearemptydoublepage

\include{wires}

\clearemptydoublepage

\include{dots}

\clearemptydoublepage

\include{summary}

\clearemptydoublepage










\bibliographystyle{prsty}
\bibliography{meso}

\printindex

\end{document}

%% file: fwd.tex
\section*{Foreword}

The present thesis is submitted in candidacy for the Ph.D. degree
within the {\it Physics Program} at the Technical University of
Denmark. The thesis describes parts of the work that I have carried
out under supervision of Antti-Pekka Jauho from Mikroelektronik
Centret, Technical University of Denmark and Karsten Flensberg from
the \O rsted Laboratory, Niels Bohr Institute, University of
Copenhagen.

\vspace{10mm} I am grateful to my supervisors for the extremely
stimulating atmosphere and their guiding --- although never directing
--- attitude. I also owe thanks to H. Bruus, K. Flensberg, A.-P.
Jauho, and P. Hedeg\aa rd for introducing me to many-particle methods
and phenomena in mesoscopic physics and condensed matter physics in
general.

During my research I have benefitted from discussions with many
colleagues, but I especially wish to thank my close collaborators G.
Bastian, H. Bruus, J.~C. Egues, J. Erland, T.~S. Jensen, K. Johnsen,
V. Mizeikis, H.~M, R\o nnow, H. Schomerus, and M. Titov.  For useful
and stimulating discussions, I wish to thank first of all C.~W.~J.
Beenakker, but also M. Brandbyge, M.  B\"{u}ttiker, S. Datta, A.
Kristensen, and V.~V. Cheianov.

Apart from the support given by the Technical University of Denmark I
have benefitted from support by the NorFA network, ``Ingeni\o
rvidenskabelig Fond og G.~A. Hagemanns Mindefond'', the Mesoscopic
Physics Group at Instituut--Lorentz in Leiden, the Visitors Program of
the Max-Planck-Institut f\"{u}r Physik komplexer Systeme in Dresden,
and finally the Advanced Research Workshop on Quantum Transport in
Semiconductors (Maratea, Italy, June 2001). I acknowledge the
hospitality of the \O rsted Laboratory, Instituut-Lorentz, and the
Max-Planck-Institut f\"{u}r Physik komplexer Systeme.

I would like to thank M. B\"{u}ttiker, M. Jonson, and K.~W. Jacobsen for serving on my dissertation committee and M.~Bollinger, J. Kutchinsky, and J.
Taylor for their assistance with the manuscript. Finally, this thesis
is dedicated to my son and to my wife who have been extremely patient
and supporting.

\vfill
\begin{center}
  \hspace{6cm}Lyngby, August 7, 2001\\~\\\hspace{6cm}Niels Asger Mortensen
\end{center}
\vfill
\noindent  Document typeset in \LaTeXe

%% file: cv.tex
\section*{Curriculum vit\ae}

After obtaining the {\it baccalaureate} in natural sciences from
Sor\o\ Akademis Skole in 1992 I matriculated at the Technical
University of Denmark within the Applied Physics Program. My
fascination with condensed matter physics grew rapidly during my
studies at Department of Physics. My bachelor's thesis described an
experimental study of charge transport in hybrid
semimetal-superconductor structures [1] with J\o rn
Bindslev Hansen and Rafael Taboryski as supervisors. Later on I was
involved in atomic force microscopy studies of solid--liquid
interfaces [2].

My master's thesis involved the theory of Andreev scattering at
semiconductor--superconductor interfaces [3,\,4] with
J\o rn Bindslev, Antti-Pekka Jauho (Mikroelektronik Centret, Technical
University of Denmark), and Karsten Flensberg (Danish Institute of
Fundamental Metrology) as supervisors. After receiving the Master of
Science degree in September 1998 I held a short-term position as
research assistant in the Theory Group at Mikroelektronik Centret.

Along with the studies for the master degree I worked as a
teaching assistant at the Department of Physics in the Semiconductor
Physics and Mechanics \& Waves courses (both master level).

February 1999 I was given support by the Technical University of
Denmark for a three year position within the Physics Ph.D. Program
with Antti-Pekka Jauho and Karsten Flensberg (\O rsted Laboratory,
Niels Bohr Institute, University of Copenhagen) as supervisors. My
research has been conducted as a member of the Theory Group (by now
the Science \& Education Group) at Mikroelektronik Centret and partly
the Condensed Matter Theory Group at the \O rsted Laboratory.

My focus has mainly been on mesoscopic physics which is a field within
solid-state physics which deals with systems at the borderline between
the microscopic and macroscopic world. The first part of my work was a
continued theoretical study of various semiconductor--superconductor
hybrids [5,\,6,\,7,\,8]. Other mesoscopic
problems include the contact problem of carbon nanotubes [9] and spin-polarized transport [10,\,11].

I have also been involved in work on magnetic modes in SO(5) theory of
high-$T_c$ superconductivity [12], optical properties of
semiconductor microcavity polaritons [13,\,14], and
Coulomb drag in periodically modulated systems [15].
During the fall of 2000 I was a guest in the Mesoscopic Physics Group
of Carlo Beenakker (Instituut--Lorentz, Universiteit Leiden) where I
worked on the physics of Andreev levels [16]. Finally,
during the last two years I have been working on extending the theory
of Coulomb drag to the mesoscopic regime [17,\,18,\,19,\,20].

During my Ph.D. studies I have been a teaching assistant in the
course Advanced Semiconductor Physics (master and Ph.D. level) and two
three-weeks courses: Applications of Electron Structure Theory and
Applications of Electron Transport Theory (both master level). I was
also a project-supervisor on the BEST Copenhagen summer-school 2000
(master and Ph.D. level). Finally, I have besides been refereeing for a
year for Physical Review Letters and Physical Review B.

\newpage

\begin{itemize}
\item[{[1]}] M.~V. Bollinger, K.~R. Bukh, N.~A. Mortensen, and
  M.~P. Sager, {\it Low temperature studies of hybrid
    superconductor--semimetal components}, Bachelor's thesis
  (Technical University of Denmark, 1996).
  
\item[{[2]}] N.~A. Mortensen, A. K\"{u}hle, and K.~A. M\o rch, {\it
    Interfacial tension in water at solid surfaces}, in {\it Proceedings of the Third International Symposium on Cavitation}, edited by J.~M. Michel and H.
  Kato (Grenoble, 1998), pp. 87-91. [phy\-sics/9901014]
  
\item[{[3]}] N.~A. Mortensen, {\it Theoretical models of transport
    in macroscopic and mesoscopic NS structures}, Master's thesis
  (Technical University of Denmark, 1998).
  
\item[{[4]}] N.~A. Mortensen, K. Flensberg, and A.-P. Jauho,
  {\it Angle dependence of Andreev scattering at
    semiconductor-superconductor interfaces}, Phys. Rev. B {\bf 59},
  10176 (1999).
  
\item[{[5]}] N.~A. Mortensen, A.-P. Jauho, K. Flensberg, and H.
  Schomerus, {\it Conductance enhancement in quantum point
    contact-semiconductor-supercon\-ductor devices}, Phys. Rev. B {\bf
    60}, 13762 (1999).
  
\item[{[6]}] N.~A. Mortensen, A.-P. Jauho, and K. Flensberg,
  {\it Andreev scattering and conductance enhancement in mesoscopic
    semiconductor-superconductor junctions}, in {\it Extended
    abstracts of Electron Transport in Mesoscopic Systems}, edited by
  P. Delsing, T. Henning, E. H\"{u}rfeld, and T. Nord (G\"{o}teborg, 1999), pp. 120-121. [cond-mat/9911372]

\item[{[7]}] N.~A. Mortensen, A.-P. Jauho, and K. Flensberg,
  {\it Dephasing in semiconductor-superconductor structures by
    coupling to a voltage probe}, Superlattice Microstr. {\bf 28}, 67
  (2000).
  
\item[{[8]}] N.~A. Mortensen and G. Bastian, {\it Side-gate
    modulation of critical current in mesoscopic Josephson junction},
  Superlattice Microstr. {\bf 28}, 231 (2000).
  
\item[{[9]}] N.~A. Mortensen, K. Johnsen, A.-P. Jauho, and K.
  Flensberg, {\it Contact resistance of quantum tubes}, Superlattice
  Microst. {\bf 26}, 351 (1999).

\item[{[10]}] K. Flensberg, T.~S. Jensen, and N.~A. Mortensen, {\it Diffusion equation and spin drag in spin-polarized transport}, Phys. Rev. B, in press [cond-mat/0107149]
  
\item[{[11]}] N.~A. Mortensen and J.~C. Egues (unpublished).

\item[{[12]}] N.~A. Mortensen, H.~M. R\o nnow, H. Bruus, and P.
  Hedeg\aa rd, {\it The magnetic neutron scattering resonance of
    high-$T_c$ superconductors in external magnetic fields: an SO(5)
    study}, Phys. Rev. B {\bf 62}, 8703 (2000).
  
\item[{[13]}] J. Erland, V. Mizeikis, W. Langbein, J.~R.
  Jensen, N.~A. Mortensen, and J.~M. Hvam, {\it Seeding of Polariton
    Stimulation in a Homogeneously Broadened Microcavity}, Phys. Stat.
  Sol. (b) {\bf 221}, 115 (2000).
  
\item[{[14]}] V. Mizeikis, J. Erland, J.~R. Jensen, N.~A.
  Mortensen, and J.~M. Hvam, {\it Stimulation of polariton emission in
    a homogeneously broadened semiconductor microcavity}, Springer
  Proc. in Phys. {\bf 87}, 687 (2001).
  
\item[{[15]}] B.~Y.-K. Hu, K. Flensberg, A.-P. Jauho, H. Smith,
  and N.~A. Mortensen (unpublished).
  
\item[{[16]}] M. Titov, N.~A. Mortensen, H. Schomerus, and
  C.~W.~J. Beenakker, {\it Andreev levels in a single-channel
    conductor}, Phys. Rev. B {\bf 64}, 134206 (2001).
  
\item[{[17]}] N.~A. Mortensen, K. Flensberg, and A.-P. Jauho,
  {\it Coulomb Drag in Coherent Mesoscopic Systems}, Phys. Rev. Lett.
  {\bf 86}, 1841 (2001).
  
\item[{[18]}] N.~A. Mortensen, K. Flensberg, and A.-P. Jauho,
  {\it Coulomb drag in phase-coherent mesoscopic structures}, Springer
  Proc. in Phys. {\bf 87}, 1347 (2001).
        
\item[{[19]}] N.~A. Mortensen, K. Flensberg, and A.~-P. Jauho,
  {\it Coulomb drag in the mesoscopic regime}, Phys. Scripta, in press. [cond-mat/0108203] 
  
\item[{[20]}] N.~A. Mortensen, K. Flensberg, and A.-P. Jauho,
  {\it Mesoscopic fluctuations of Coulomb drag between quasi-ballistic
    1D--wires}, Phys. Rev. B, in press. [cond-mat/0108263] 

\end{itemize}

%% file: papers.tex
\section*{Included papers}

Representative papers  [A,\,B,\,C,\,D] from my ``early period'' in mesoscopic physics
are included in this thesis. They are all based on the scattering approach to mesoscopic charge-transport.

In Ref.~[A] a simple mode-matching model for the contact resistance of quantum tubes is given. It is motivated by the, at least at that point, small attention paid to the conditions for a good transmission between carbon nanotubes and metallic contacts. 

Refs.~[B,\,C,\,D] all focus on the modification of the charge-transport in semiconductors due to coupling to superconducting reservoirs where electrons are Andreev reflected into holes. Ref.~[B] studies deviations from the two-probe conductance quantization in phase-coherent quantum-point-contact semiconductor--superconductor devices. Ref.~[C] describes the two-probe conductance in the presence of dephasing caused by coupling to an additional voltage probe. In Ref.~[D] it is studied how an additional side-gate can be used in modifying the critical current in a Josephson junction.

The mesoscopic
Coulomb drag work, which forms the core of this thesis, will be described
in detail in some introductory chapters which also contain unpublished
work as well as results from [E,\,F,\,G,\,H].

\vspace{1cm}
\begin{itemize}
\item[{[A]}] Superlattice Microst. {\bf 26}, 351 (1999).

\item[{[B]}] Phys. Rev. B {\bf 60}, 13762 (1999).

\item[{[C]}] Superlattice Microst. {\bf 28}, 67 (2000).

\item[{[D]}] Superlattice Microst. {\bf 28}, 231 (2000).

\item[{[E]}] Phys. Rev. Lett. {\bf 86}, 1841 (2001).

\item[{[F]}] Phys. Scripta, in press. [cond-mat/0108203]  

\item[{[G]}] Phys. Rev. B, in press. [cond-mat/0108263] 

\item[{[H]}] Springer Proc. in Phys. {\bf 87}, 1347 (2001).

\end{itemize}

%% file: introduction.tex
\chapter{Introduction}

This thesis describes how the merging of the two fields of Coulomb
drag and mesoscopic physics gives rise to interesting physics.  The
present studies are motivated by the fundamental interest in Coulomb
coupled systems, but the phenomena that we find might as well in the
future become important in the context of the increasing
packing-density in electronic devices on the same chip. Studies of
Coulomb drag are also motivated in the context of small MOSFETS
\cite{fischetti2001} and drag effects utilized for nano-scale liquid
flow detectors have been suggested recently \cite{kral2001}.

Below we give a brief introduction to the two fields and summarize
some of the important properties of the electron gas with emphasize on
situations where Coulomb drag has been studied. Finally, we give an
outline for the rest of this thesis.

\section{Mesoscopic physics}

Mesoscopic physics has from the field of solid-state physics
\cite{kohn1999} grown into a rich and surprising field itself, see
{\it e.g.}  \cite{sohn1997,beenakker1997,imry,ferry,datta}. It deals
with systems of often reduced dimensionality at the borderline between
the microscopic regime of atoms and molecules and the macroscopic
world.  When the dimensions become shorter than the phase-breaking
length\index{length scales!, phase-breaking length} $\ell_\phi$ (typically on the
sub-micron scale at liquid Helium temperature for semiconductor
heterostructures) the wave nature of the electrons reveals itself
fully in terms of significant corrections to the classical behavior of
the electrons as well as completely new phenomena. The laws of
classical mechanics do not fully succeed in describing the observed
phenomena, but on the other hand a full quantum mechanical
description based on a scaling up of the microscopic description is
often impractical due to too many degrees of freedom. A successful
description is in many cases based on concepts from both classical
electro-dynamics, quantum mechanics, and statistical physics.

The list of mesoscopic phenomena is long and we shall only mention a
few important examples: In ballistic wires and quantum point contacts
the conductance is quantized in units of $2e^2/h$
\cite{vanwees1988,wharam1988} and in disordered wires the
sample-to-sample fluctuations of the conductance have a universal
magnitude of $e^2/h$ \cite{altshuler1985,lee1985}. Interestingly,
these effects have their counterparts in short ($L\ll \xi_0=\hbar
v_F/\pi \Delta$) Josephson junctions where the wire is contacted by a
superconductor at each end. For ballistic junctions the critical
current is quantized in units of $e\Delta/\hbar$ \cite{beenakker1991b}
and for disordered wires the sample-to-sample fluctuations of the
critical current have a universal magnitude of $e\Delta/\hbar$
\cite{beenakker1992}.  Another example is the Aharonov--Bohm effect in
a small ring connected to leads, see {\it e.g.} Ref. \cite{webb1988}.
Here, a magnetic flux $\phi$ enclosed by the ring causes conductance
oscillations as a function $\phi$ with a period $\phi_0=h/e$ (or
higher harmonics $\phi_0/n$). Within the scattering approach
\cite{buttiker1985} the effect can be understood from interference
between different Feynman paths circulating the enclosed flux
\cite{feynman2}. The effect is a clear signature of phase-coherence
and the damping of the higher harmonics can be attributed to a finite
phase-coherence length, see {\it e.g.} Ref. \cite{hansen2001}.

Not only the electronic degrees of freedom are subject to the bounds
of quantum mechanics -- also the phononic degrees are so. The latter
manifests itself in {\it e.g.} heat conductance quantized in units of
$\pi^2k^2 T/3h$ in suspended constrictions
\cite{angelescu1998,rego1998,schwab2000}. The ``universality'' is
somewhat weaker in the examples with supercurrent and heat transport
where the ``quantum unit'' includes either the superconducting pairing
potential $\Delta$ (material dependent) or the temperature $T$ whereas
for the conductance only fundamental constants are involved; Planck's
constant $h$ and the electron charge $e$.

Along with the phase-breaking length $\ell_\phi$ the other important
length scale is the Thouless length\index{length scales!, Thouless length} $L_T$. It
may be thought of as the thermal length over which phases
$\phi(\varepsilon)$ of the quasi-particle states can be considered
constant for $\varepsilon$ in a range of $kT$ around the Fermi level.
On longer length scales we can not ignore the energy dependence of the
acquired phase and upon thermal averaging the effect of the strong
phase dependence may be washed away.

One of the characteristic things of disordered systems in the
mesoscopic regime is that sample-to-sample fluctuations are
pronounced, though they may not always be universal. As the sample
dimensions and/or temperature are decreased the phase-breaking length
and Thouless length will at some point meet the sample dimensions. At
this point interference effects will modify the classical phenomena
and since the interference behavior depends strongly on the particular
potential landscape that the electrons move in there will be
dominating sample-to-sample fluctuations in the presence of even weak
disorder.

\section{The electron gas}

Most under graduate text books start with the non-interacting electron
gas, see {\it e.g.} Ref. \cite{ashcroft}. Surprisingly, a lot of phenomena
in metals and semiconductors can be understood from this
kind of na\"\i ve model. The Fermi liquid theory initiated by Landau
\cite{landau1957} explains how a complicated system of interacting
fermions may often respond to external perturbations just as if it was
a non-interacting electron gas. For an overview see Ref.~\cite{pines}.
When studying the excitations of the ground state at least two
important things emerge; {\it i)} the fermionic excitations are of
single-particle nature with a spectrum that is similar (within {\it
  e.g.}  some re-normalization of the mass to an effective mass) to
that of the non-interacting electron gas and {\it ii)} the life time
of the excitations diverges near the Fermi level. For those reasons
these long-lived excitations are often referred to as
quasi-particles.

A large part of the mesoscopic transport phenomena can successfully be
accounted for by the\index{scattering!, approach} ``scattering approach
to quantum transport'' \cite{landauer1957,landauer1970,buttiker1986}
--- often referred to as Landauer--B\"{u}ttiker formalism --- and
extensions within the same spirit. For an overview of the variety of
experiments see Ref.~\cite{beenakker1991a}. The formalism applies to
the non-inter\-acting electron gas so that its usefulness relies on
Fermi liquid theory. For a review of Landauer--B\"{u}ttiker formalism
see {\it e.g.}
Refs.~\cite{buttiker1996,imry1999,sohn1997,beenakker1997,imry,ferry,datta}.

If all electron systems were Fermi liquids we na\"{\i}vely only
had to know about the non-interacting electron gas and the interaction
corrections in terms of the small ratio, $F\ll 1$, between the interaction
energy and the kinetic energy \cite{pines}. However, there are various
examples of the break-down of Fermi liquid theory or non-Fermi liquid
behavior, but here we shall only mention a few cases which have also
been studied in the context of drag. When reducing the dimensionality
from three to two (by confining the electron gas in one direction)
there are already precursors of the break down like {\it e.g.}
logarithmic corrections to the particle-particle scattering rate,
$T^2\longrightarrow T^2 \ln T$ \cite{hodges1971,chaplik1971}.

The break down takes place in the strictly one-dimensional limit. Here
the electron gas forms a so-called Tomonaga--Luttinger liquid
\cite{tomonaga1950,luttinger1963,mahan} --- commonly referred to as a
Luttinger liquid. It is different from the Fermi liquid in the sense
that the excitations are bosonic collective modes rather than
quasi-particles. The Luttinger liquid has attracted much theoretical
interest both because of the analytical progress that can be made, but
also because of the existence of edge states in the quantum Hall
regime \cite{wen1992,grayson1998} and the discovery of single-walled
carbon nanotubes \cite{iijima1991} which are also believed to be
one-dimensional conductors \cite{dekker1999,bockrath1999}.
Experimentally, it is of course challenging to create a
one-dimensional system and the most convincing demonstrations of
Luttinger liquid effects are in the power-law dependence of the
density-of-states in tunneling studies of quantum--Hall edge states
\cite{grayson1998}, partly in tunneling studies of carbon
nanotubes \cite{bockrath1999}, whereas for Raman scattering studies in
long semiconductor wires some discussion apparently still remain
\cite{wang2000}. For not too long quantum wires, $L<\hbar v_F/kT$, attached to reservoirs
through open contacts the Fermi liquid is believed to be restored due
the strong proximity from the Fermi liquid reservoirs, though some
debate has existed \cite{yacoby1996,alekseev1998,lal2001}. Fermi liquid theory also breaks down in the extreme
low-density limit where the screening is reduced so that the energy
associated with the repulsive Coulomb interactions dominates over the
kinetic energy, {\it i.e.} $F>1$. In the ground state the electrons organize themselves
into a regular lattice, the so-called Wigner lattice, with bosonic
excitations which are phonon-like just as for the ordinary atomic
lattice in solids. The question of a possible metal-insulator
transition is as well a delicate problem.

Another interesting state is the quantum--Hall liquid of the two-dimensional
electrons gas in a strong magnetic field
\cite{vonklitzing1980,tsui1982,laughlin1983}. Depending on the filling
of the Landau levels the excitations can be so-called composite
fermions \cite{jain1989} with fractional charge --- a quasi-particle with
$1/\nu$ magnetic flux quanta attached at filling factor $\nu=1/3, 2/5,
4/7$ {\it etc}. Filling factor $\nu=1/2$ and other even denominators
are special cases where there is no quantized Hall effect and where
surprisingly the low-energy physics (near the Fermi level) is
basically that of a Fermi liquid in zero magnetic field
\cite{halperin1993}. 
The superconducting state of some metals at low temperatures is an
example where some other degree of freedom in the system is able to
mediate a particle-particle interaction which is effectively
attractive. When that happens two fermionic particles can form a
bosonic pair (the so-called Cooper pair) which is allowed to
condensate. In the conventional low-temperature superconductors ---
usually referred to as BCS superconductors \cite{bardeen1957} --- the
attractive interaction is mediated by the phonons of the atomic
lattice, but in the high-$T_c$ cuprates the situation seems to be
much more delicate and the mechanism is still to be understood. At
least for the BCS superconductors the excitations are known to be of
fermionic nature, though they are a kind of superpositions of the
electron-like and hole-like quasi-particles known from Fermi liquid
theory.

Finally, we will just mention the electron-hole pairs (excitons)
formed in semiconductors consisting of the pairing of an electron
excited to the conduction band and the oppositely charged hole left
behind in the valance band. Such pairs are bosonic and it has been
speculated that they might undergo condensation.

\section{Frictional drag}

\begin{figure}[h!]
\begin{center}
  \epsfig{file=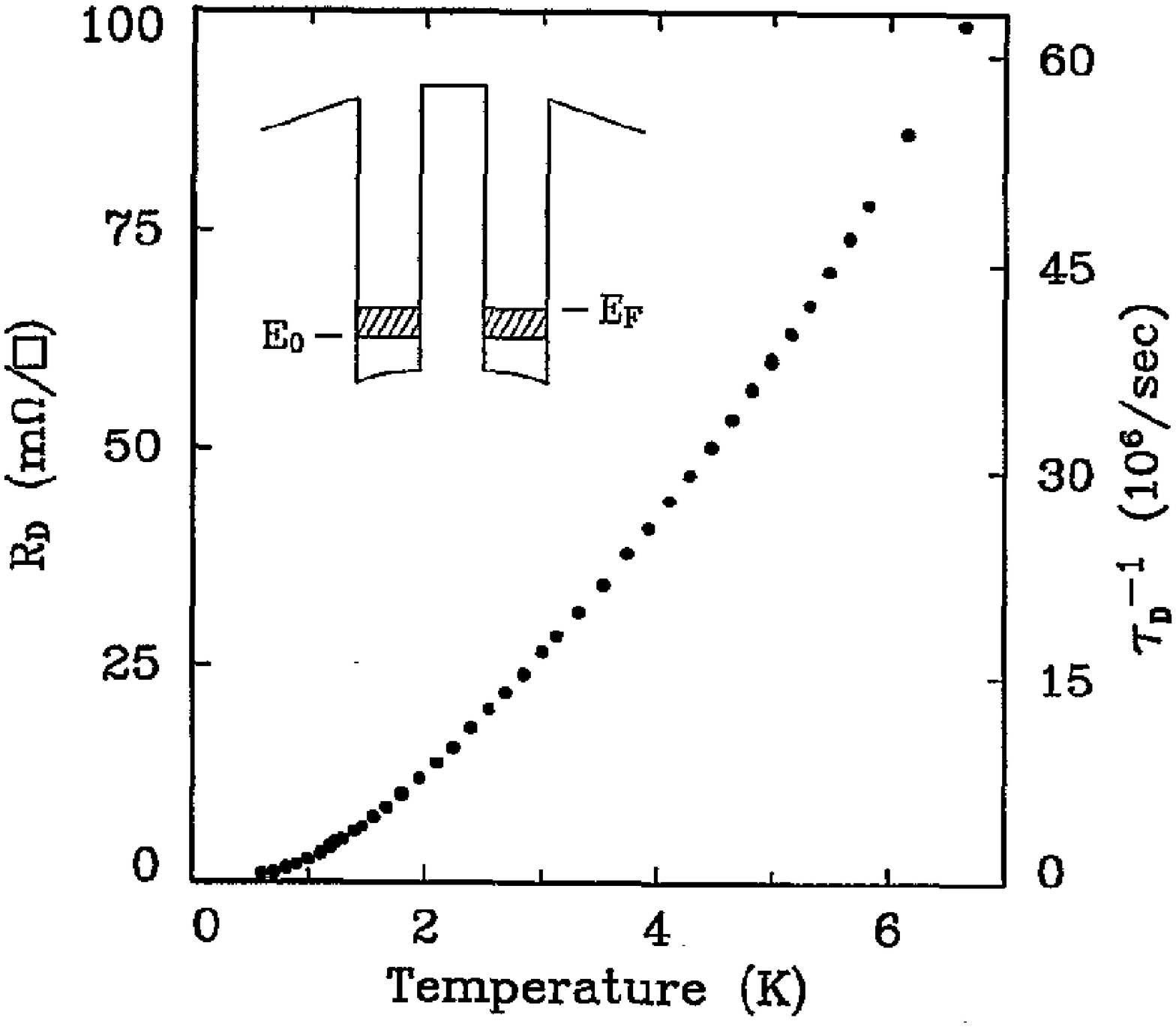,
    height=0.45\columnwidth,clip,angle=0}\epsfig{file=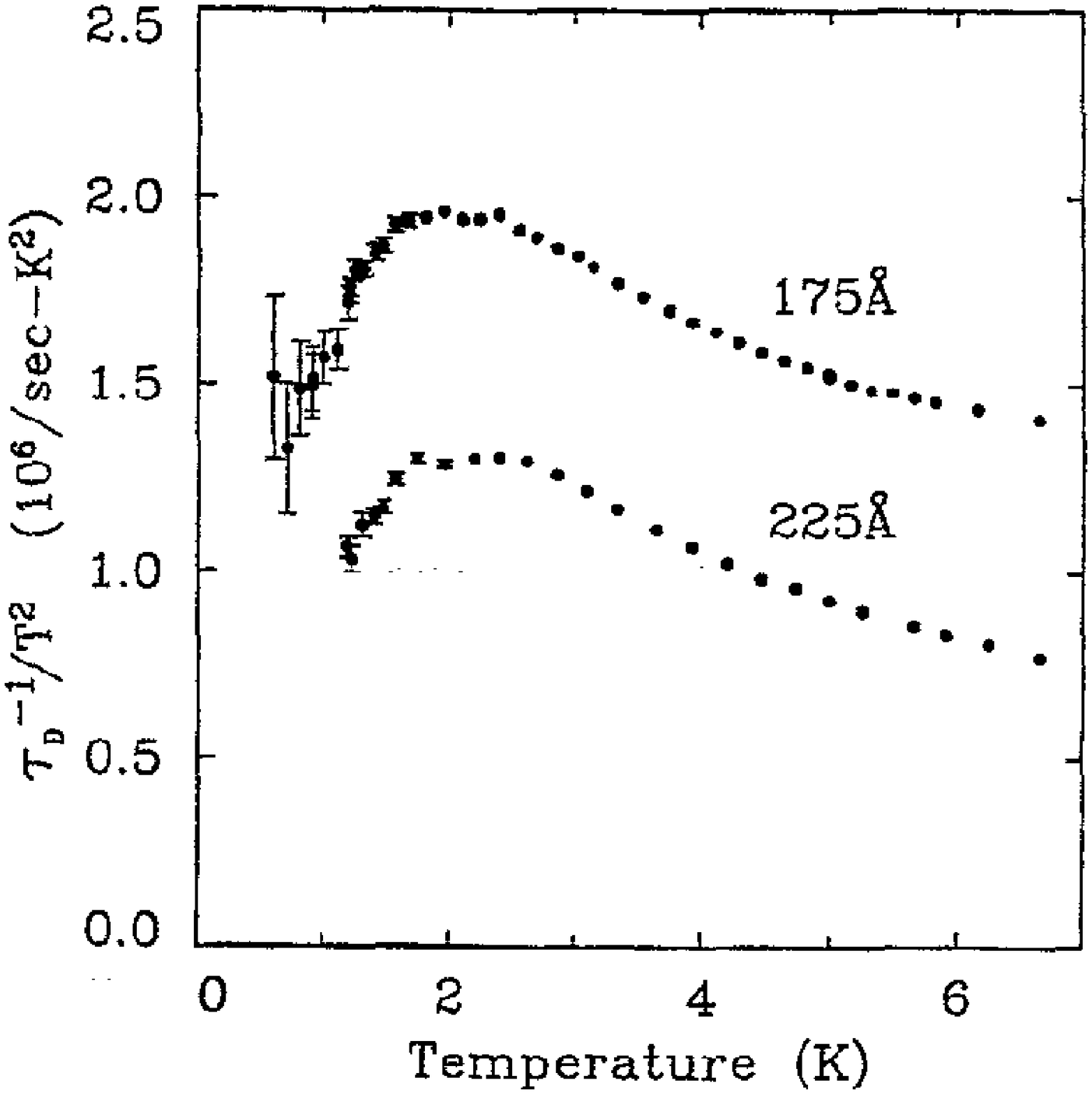,
    height=0.45\columnwidth,clip,angle=0}
\end{center}
\caption[Temperature dependence of drag]{Temperature dependence of drag in bi-layer system. The left figure shows the approximate $T^2$--dependence and the right figure shows the derivations from this in more detail. From Gramila {\it et al.} \cite{gramila1991}.}
\label{T2-gramila}
\end{figure}

When two (or more) electron systems are brought close together
many-body interactions between electrons belonging to each of the two
subsystems come into play. One of the consequences is the possible
momentum transfer between the subsystems even in the absence of
tunneling mediated charge transfer. Driving a current through one
electron system interaction mediated momentum transfer will show up as
an induced current in the other layer. The induced current is also
referred to as the drag current. The Coulomb drag effect was first
suggested in 1977 by Pogrebinski\u{\i} \cite{pogrebinskii1977} and in
1983 apparently independently by Price \cite{price1983}.  However, one
could imagine other situations such as phonon mediated drag
\cite{hubner1960,gramila1991,tso1993,bonsager1998}, drag mediated by
van der Waals interactions \cite{rojo1992}, and plasmon enhanced drag
\cite{flensberg1994,hill1997}. For Coulomb mediated interaction the
low-temperature behavior is given by a $T^2$--dependence similar to the
ordinary rate for carrier-carrier scattering, see {\it e.g.}  Ref.
\cite{smith}, though corrections emerge when reducing the
dimensionality from three to two \cite{hodges1971}.
Fig.~\ref{T2-gramila} shows the temperature dependence measured by
Gramila {\it et al.} \cite{gramila1991}.

The study of Coulomb drag is highly motivated by the fact that the
drag conductance directly depends on the Coulomb interaction in
contrast to {\it e.g} the two-probe conductance of an open system away
from the Coulomb blockade regime, see {\it e.g.} Ref.
\cite{beenakker1997}. In passing we note that Coulomb coupling also
has interest in other contexts, such as {\it e.g.} capacitive coupling
of a mesoscopic conductor to the environment \cite{martin2000,buttiker2000}, charge
pumping in quantum dots \cite{shutenko2000}, and also spin polarized
transport \cite{damico2000,flensberg2001}.

Below we will briefly mention a few examples from the development of
Cou\-lomb drag since the seminal works of Pogrebinski\u{\i}
\cite{pogrebinskii1977} and Price \cite{price1983}. For more details
on Coulomb drag in extended two-layer systems we refer to the review
by Rojo~\cite{rojo1999}.

By now drag has been studied both experimentally and theoretically for
quite many states of the two electron gases involved. For the bi-layer
system the theoretical framework has been Fermi liquid theory in zero
\cite{flensberg1995} and finite magnetic field
\cite{bonsager1996,khaetskii1999} though still in the integer quantum
Hall regime. In the fractional quantum Hall regime the framework has
been composite fermion theory \cite{ussishkin1998,narozhny2001}.

For quasi one-dimensional drag the studies have been based on
Boltzmann equation approaches \cite{hu1996,gurevich1998} within the
spirit of Fermi liquid theory and for strictly one-dimensional drag
between {\it e.g.} quantum--Hall edges, semiconductor quantum wires,
and carbon nanotubes there has been quite some activity based on
Luttinger liquid theory
\cite{nazarov1998,flensberg1998,komnik1998,ponomarenko2000,komnik2001}.
A subset of the Luttinger liquid studies has been on drag and
dissipation-less current in coupled Luttinger rings
\cite{shahbazyan1997,baker1999}.

In passing we mention work on drag between a superconductor and a
normal metal \cite{kamenev1995}, drag in electron-hole pair systems
\cite{vignale1996,hu2000}, and drag between ring-shaped Wigner
lattices \cite{baker} and between extended metal and Wigner crystals
\cite{braude}. Also the noise properties of Coulomb drag has been
studied \cite{gurevich2000a}.

The first experimental studies were on drag between silicon thin-films by
Hubner and Shockley \cite{hubner1960} and drag between a
two-dimensional and a three-dimen\-sional electron system by Solomon
{\it et al.} \cite{solomon1989}. The experimental studies of bi-layer
systems were initiated by Gramila {\it et al.}  \cite{gramila1991} in
the case of drag between two electron gases and by Sivan {\it et al.}
\cite{sivan1992} for drag between an electron gas and a hole gas. For
the subsequent studies the main part have been carried out in bi-layer
semiconductor systems. As far as we know the few exceptions are the
very recent studies of coupled quantum wires
\cite{kristensen1999,yamamoto2001,debray2001}.

While Coulomb drag of extended bi-layer systems in zero magnetic field
is fairly well-understood several open questions remain in the context
of finite magnetic field as well as reduced dimensionality --- here we
will only mention a few. Recent experiments in the fractional quantum
Hall regime, filling $\nu=1/2$, have revealed negative drag which could be interpreted
phenomenologically as drag mediated by a spin-dependent interaction
\cite{lok2001}, see also Ref. \cite{vonoppen2001}. Also the missing $2k_F$ response
\cite{zelakiewicz2000,zeitler} and the low-temperature behavior and
deviations from the $T^2$--dependence, see Fig.~\ref{T2-gramila}, have
given rise to continued discussion
\cite{gramila1991,tso1992,zheng1993,jauho1993,rojo1999}. The same
might apply to the Luttinger liquid interpretation of recent
experiments on quasi one-dimensional wires
\cite{yamamoto2001,debray2001}. Finally, there is the new field of
mesoscopic fluctuations and mesoscopic Coulomb drag which has only
been initiated by a few theoretical papers
\cite[E]{narozhny2000,narozhny2001} and which is
the focus of this thesis.

\begin{figure}[h!]
\begin{center}
  \epsfig{file=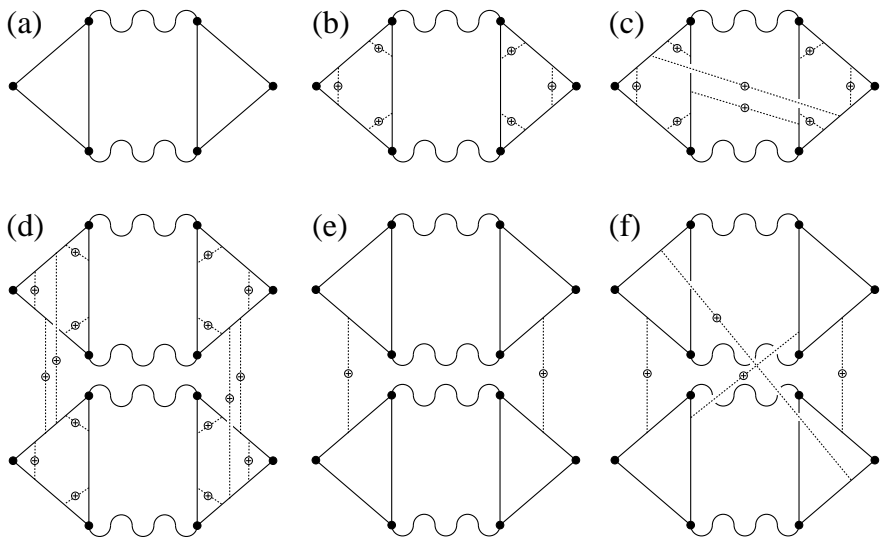,
    height=0.99\columnwidth,clip,angle=-90}
\end{center}
\caption[Diagrammatic representations of current-current correlation function]{Diagrammatic representations of current-current correlation functions with triangles representing the $\Delta_i$'s and the interaction lines representing the $U_{12}$'s. Diagram (a) is the bare diagram and diagrams (b) and (c) are the type of diagrams to be considered for the mean value. Upon disorder averaging the crossed diagram (c) is non-vanishing only for subsystems with mutually correlated disorder. Diagrams (d), (e), and (f) are the types of diagrams relevant for the fluctuations. Again, upon disorder averaging the crossed diagram (f) is non-vanishing only in the case of mutually correlated disorder.}
\label{triangles_collection}
\end{figure}

The initial theoretical work was mainly done in the framework of coupled Boltzmann equations
\cite{price1983,gramila1991,sivan1992,jauho1993} and first later on a
linear-response theory based on Kubo formalism was developed by
Kamenev and Oreg \cite{kamenev1995} and simultaneously by Flensberg
{\it et al.} \cite{flensberg1995}. Recently drag has also been studied in
the framework of non-equilibrium Green function formalism
\cite{raichev1999,wang2001}. In the Kubo formalism the DC
current-current correlation function $\Pi$ is calculated by aid of a
second order perturbation expansion in the interaction $U_{12}$. As a
result the drag conductivity,

 \begin{multline}\label{dragequation}
   \sigma_{21}({\boldsymbol r},{\boldsymbol r}')
   =\frac{e^2}{h}\left(-\frac{1}{2\hbar^2}\right) \iiiint{\rm
     d}{\boldsymbol r}_1 {\rm d}{\boldsymbol r}_2 {\rm d}{\boldsymbol
     r}_1' {\rm d}{\boldsymbol r}_2'\, U_{12}({\boldsymbol
     r}_1,{\boldsymbol r}_2) U_{12}({\boldsymbol
     r}_1',{\boldsymbol r}_2')\\
   \times{\mathscr P}\int_{-\infty}^\infty {\rm d}\omega\,
   \frac{\partial n_B(\omega)}{\partial \omega}
   {\Delta}_1({\boldsymbol r}',{\boldsymbol r}_1,{\boldsymbol
     r}_1',-\omega) {\Delta}_2({\boldsymbol r},{\boldsymbol
     r}_2,{\boldsymbol r}_2',\omega),
\end{multline}
is expressed in terms of two ``triangle'' functions
$\Delta_i=-\big<\hat{J}_i\hat{\rho}_i\hat\rho_i\big>$ and two
interaction potentials $U_{12}$. This can essentially be thought of
as two three-particle Green functions connected by two interaction
lines which as suggested in Ref.~\cite{flensberg1995} can be
illustrated by diagram (a) in Fig.~\ref{triangles_collection}. The
result stated in Eq.~(\ref{dragequation}) will be derived and
explained in more detail later in this thesis. At this point we note
that a very similar formalism is used in microscopic studies of
mechanical friction \cite{novotny1999,volokitin2001}.

Since there often is some degree of disorder present due to random
impurities {\it etc.} it is useful to know the statistical properties
of the drag conductance rather than its specific value for a specific
disorder configuration. Knowledge about the full distribution of the
drag conductance is of course the ultimate goal, but often one has to
be satisfied by the first moments like the mean value and the mean
fluctuations. These can be obtained with the aid of disorder ensemble
averaging techniques.  For the mean value it has become quite common
to consider mutually un-correlated ($\rm uc$) disorder potentials of
the two subsystems such that for the mean value \cite{flensberg1995}\index{disorder!, mutually un-correlated}

\begin{equation}
\big<\Delta_1\Delta_2\big>_{\rm uc} =\big<\Delta_1\big>\big<\Delta_2\big>.
\end{equation}
This corresponds to only considering diagrams of the type of diagram
(b) in Fig.~\ref{triangles_collection} where there are no impurity
lines crossing between the two triangle functions. This is motivated
by the situation in bi-layers where the main-source for disorder is
the ionized dopants located in the outside barriers --- the barrier
separating the wells is mainly un-doped. Disorder scattering in a given
well is then mainly due to disorder induced by the nearest by outside
barrier whereas for the other well the disorder is mainly induced by
the other outside barrier \cite{flensberg1995}.  However, recently it
was suggested by Gornyi {\it et al.} \cite{gornyi1999} to go beyond
this assumption and consider the case of mutually correlated ($\rm c$)
disorder represented by crossing diagrams like diagram (c). The
motivation is that disorder induced by the separating barrier will
give rise to a common disorder potential for the two electron systems.
This will especially be the case if the barrier is doped. In reality
the disorder is probably only partly correlated. Mutual correlation of
the disorder was found to increase the mean value compared to mutually
un-correlated disorder \cite{gornyi1999}.\index{disorder!, mutually correlated}

Calculating the fluctuating properties requires the evaluation of {\it two} of the bare
diagrams (a) connected by disorder lines. Studies in this spirit were recently initiated
by Narozhny and Aleiner \cite{narozhny2000} who considered mutually
un-correlated disorder\index{disorder!, mutually un-correlated}

\begin{equation}
\big<\Delta_1\Delta_1\Delta_2\Delta_2\big>_{\rm uc}
=\big<\Delta_1\Delta_1\big>\big<\Delta_2\Delta_2\big>,
\end{equation}
corresponding to diagrams of type (d). This has subsequently been
extended to the fractional quantum Hall regime with half-filled
($\nu=1/2$) Landau levels by Narozhny {\it et al.}
\cite{narozhny2001}. Finally, in Ref.~[E]
quasi-ballistic wires were considered where diagrams of the type (e)
in the presence of mutually un-correlated disorder give the
fluctuations to leading order in $1/k_F\ell$ with $\ell$ being the
mean free path. The extension of this limit to mutually correlated
disorder was studied in Ref.~[G]. Including crossed
diagrams of the same type as (e) it was found that the fluctuations
increase by a factor of $\sqrt{2}$ compared to mutually un-correlated
disorder.

\begin{figure}
  \begin{center}
\begin{minipage}[c]{0.65\textwidth}
 \epsfig{file=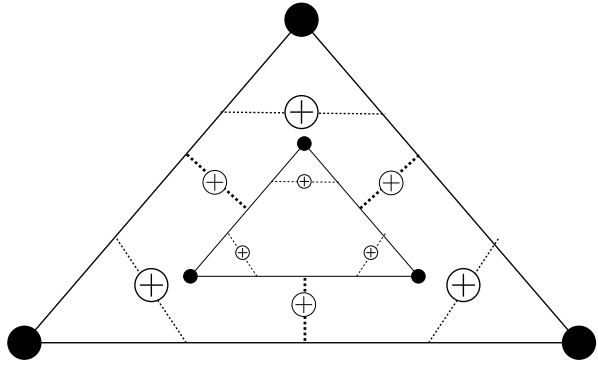,height=\columnwidth,clip,angle=-90}
\end{minipage}\hfill
\begin{minipage}[c]{0.31\textwidth}
\caption{Diagrammatic representation of the average of two triangle functions; $\big<\Delta_1\Delta_1\big>$, $\big<\Delta_2\Delta_2\big>$, and $\big<\Delta_1\Delta_2\big>$.}
\label{2triangles}
\end{minipage}
\end{center}
  \end{figure}

In general it is a quite difficult job to go beyond the assumption of
mutually un-correlated disorder for the fluctuations since it requires
the average of four triangle functions which includes crossing
diagrams of the type (f). Furthermore, in the two-dimensional case no
leading order term can be identified and a systematic infinite
summation over classes of diagrams has to be performed in order to get
a non-divergent behavior. Assuming un-correlated disorder one can on
the other hand do with calculating the average of two triangle
functions like illustrated in Fig.~\ref{2triangles}. In fact diagram
(c) can also be cast into that form and in that sense Gornyi {\it et
  al.} \cite{gornyi1999} were very close to also calculating
fluctuation properties.

The important thing demonstrated by Narozhny and Aleiner
\cite{narozhny2000} for diffusively disordered bi-layers was that at
temperatures smaller than the Thouless energy, {\it i.e.} samples
smaller than the Thouless length, the sample-to-sample fluctuations
become pronounced when compared to the mean value. Drag between
phase-coherent quasi-ballistic one-dimensional wires as well as
coupled chaotic quantum dots show the same behavior
[E].

Experimentally, Coulomb drag has not yet been brought into the
mesoscopic regime with $L\ll \min(\ell_\phi,L_T)$, though experiments
on coupled quantum wires are approaching this limit
\cite{kristensen1999,yamamoto2001,debray2001}.\index{length scales!, phase-breaking length}\index{length scales!, Thouless length} One of the reasons might be
that Coulomb drag increases with the sample size and decreases with
temperature and since the magnitude of Cou\-lomb drag is already small
in existing samples it is quite a challenge to move into the
mesoscopic regime. However, the idea of bringing together the fields
of Coulomb drag and mesoscopic physics is promising and a couple of
very recent reports
\cite[E]{narozhny2000,narozhny2001} clearly
demonstrate the potential gain of such experimental efforts: Though
the mesoscopic fluctuations are non-universal they are on the other
hand an extreme example of mesoscopic fluctuations since they can be
of the order of, or even exceed, the mean value!

\begin{figure}[h!]
\begin{center}
  \epsfig{file=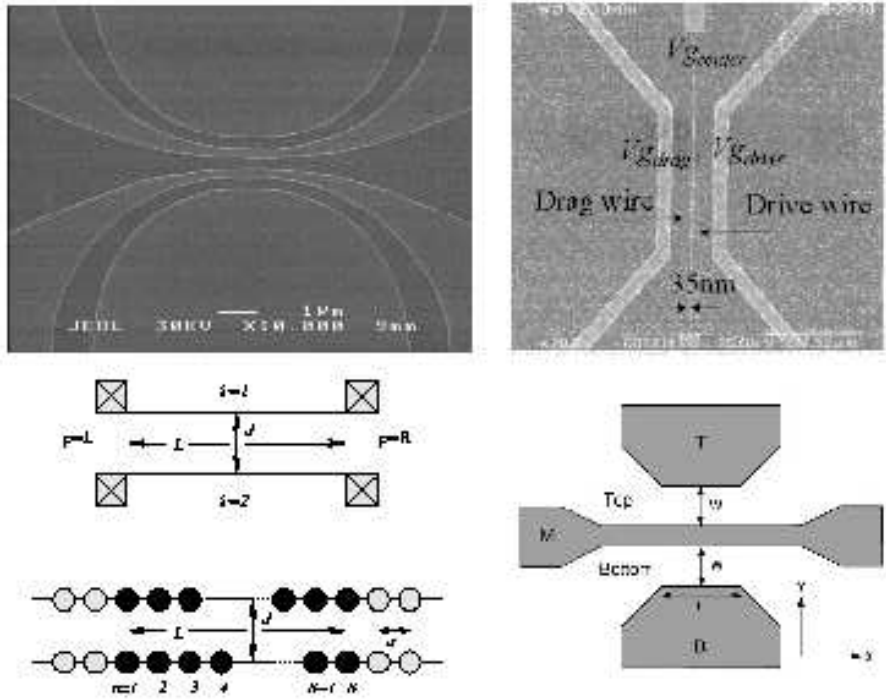, height=0.99\columnwidth,clip,angle=-90}
\end{center}
\caption[Coupled quantum wires]{Different geometries with coupling of two quantum wires. The upper left panel shows a scanning electron micro-graph (SEM) image of a GaAs sample with wires defined by removing parts of the 2DEG (the dark regions) by a shallow-etch technique \cite{kristensen1999}. The upper right panel shows a similar SEM image of a sample where the wires are defined by surface Schottky gates \cite{yamamoto2001}, and the same is the case for the sample sketched in the lower right corner \cite{debray2001}. The lower left corner shows the simplified geometry considered theoretically in Refs.~[E,\,F,\,G,\,H].\index{quantum!, wires}}
\label{SAMPLES_wires}
\end{figure}

Two interesting systems are coupled quantum wires and coupled quantum
dots. For coupled quantum wires there are at least two possibilities
based on 2DEGs. They can be formed in a bi-layer structure with one
wire above the other, each of the wires defined by depleting the
2DEGs. The other possibility is to have only a single 2DEG and then
form two in-plane wires by depleting the 2DEG, see
Fig.~\ref{SAMPLES_wires}. The first situation allows for studies of
both narrow wires with a few number of modes and also wide wires with
diffusive transport. The other situations is most suitable for narrow
wires with a width smaller or comparable to the separation of the
wires. For wider wires the Coulomb coupling will only be efficient
close to the nearby edges of the two wires due to screening.

Another very interesting possibility is to use Nature's own systems
such as {\it e.g.} two nearby carbon nanotubes. In fact, if for
multi-walled carbon nanotubes (which consist of several concentric
carbon nanotubes) one could imagine contacting different tubes
independently it would in principle allow for a study of drag between
concentric quantum wires. Whereas this might seem very difficult it is
definitely possible that only, say, the outermost wall is contacted
when performing a two-probe measurement. If tunneling between the
tubes is absent it means that the current in the other tubes is forced
to be zero. The electrons in the current-carrying tube will then be
subject to a Coulomb friction from the electrons in tubes that do not
carry a current and this will give rise to additional resistance.
Indeed this is very similar to the additional resistance due to
Coulomb drag between the spin-up and spin-down channels in
spin-polarized transport \cite{damico2000,flensberg2001}.

\begin{figure}[htb!]
\begin{center}
  \epsfig{file=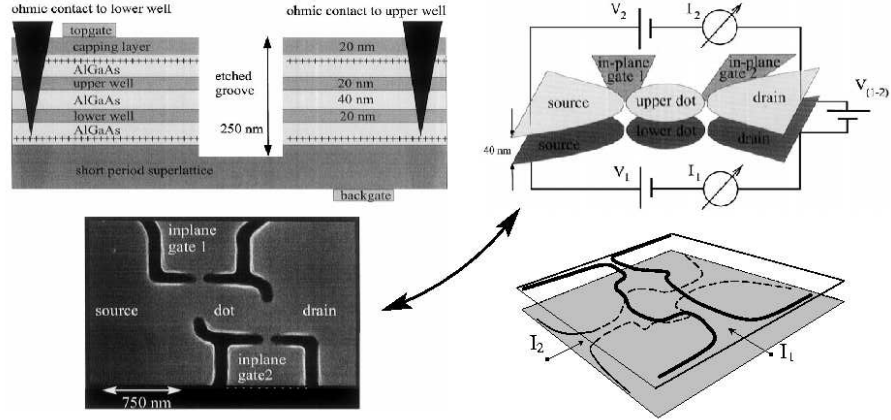, height=0.99\columnwidth,clip,angle=-90}
\end{center}
\caption[Coupled quantum dots]{Realization of Coulomb coupled quantum dots which are independently contacted \cite{wilhelm2000}. The lower right corner shows the generic geometry studied in Refs.~[E,\,F] with spatially separated contacts to avoid their otherwise dominating contributions to drag.\index{quantum!, dots}}
\label{SAMPLES_dots}
\end{figure}

The possibility of two Coulomb coupled chaotic quantum dots, see
Fig.~\ref{SAMPLES_dots}, with open contacts is an other interesting
possibility. For such systems drag will be zero on average since when
an electron in one dot is scattered by an electron in the other dot it
can with equal probability leave the dot through either of the two
leads. This suggests that the fluctuations will be finite. Such
systems are of course very difficult to realize, but the advances in
the fabrication and lithography of semiconductor structures has
reached a high level of sophistication and very recently Coulomb
coupled sub-micron quantum dots with independent electrical contacts
have been fabricated in a two-layer system \cite{wilhelm2000} even
though no drag measurements were performed.

\begin{table}[h!]
\begin{center}
\begin{tabular}{cccc}
\hline\\
Sample & Dimension & Separation & Temperature\\
\\
\hline\\
Quantum wires \cite{kristensen1999} & $L\sim 2\,{\rm \mu m}$ 
& $d\sim 200\,{\rm nm}$ &  $<4.2\, {\rm K}$\\
\\
Quantum wires \cite{yamamoto2001}& $L\sim 2\,{\rm \mu m}$
& $d\sim 35\,{\rm nm}$ & $<4.2\, {\rm K}$ \\
\\
Quantum wires \cite{debray2001}& $L\sim 2\,{\rm \mu m}$
& $d\sim 50\,{\rm nm}$ &  $0.2 \, -\, 1\, {\rm K}$\\
\\
Quantum dots \cite{wilhelm2000}& $\sqrt{\cal A}\sim 0.8\,{\rm \mu m}$
& $d\sim 40\,{\rm nm}$ & \\
\\
\hline
\end{tabular}
\end{center}
\caption[Properties of samples]{Properties of the quantum wire samples shown in Fig.~\ref{SAMPLES_wires} and the quantum dot sample shown in Fig.~\ref{SAMPLES_dots}. Note that Coulomb drag measurements have not been reported for the quantum dot sample.\index{quantum!, wires}\index{quantum!, dots} }
\label{TABLE_samples}
\end{table}

While the systems with coupled quantum wires
\cite{kristensen1999,yamamoto2001,debray2001} are examples which are
close to if not already in the mesoscopic regime ($L \sim\ell_\phi$ or
$L>\ell_\phi$) then the coupled quantum dot system \cite{wilhelm2000}
belongs to the mesoscopic regime ($\sqrt{\cal A}<\ell_\phi$) at not too low temperatures where there is
still sufficiently phase-space for the Coulomb drag to be measurable, see
Table~\ref{TABLE_samples}.

\section{This thesis}

The aim of this work is to formulate a linear-response theory of
Coulomb drag in coupled mesoscopic systems that otherwise behave as
Fermi liquids, {\it i.e.} non-interacting electrons. We closely follow
the work of Flensberg {\it et al.}  \cite{flensberg1995}, or
alternatively Kamenev and Oreg \cite{kamenev1995}, on translationally
invariant systems. Here, we will however generalize the formulation to
also include systems with broken translation symmetry and dimensions
smaller than both the phase-breaking length $\ell_\phi$ and the
Thouless length $L_T$.

\begin{figure}[b!]
  \begin{center}
\begin{minipage}[c]{0.60\textwidth}
\epsfig{file=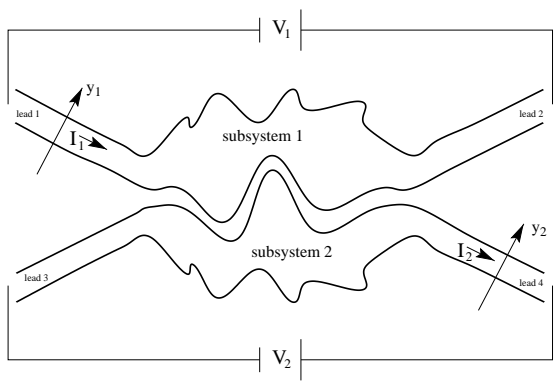, width=\columnwidth,clip}
\end{minipage}\hfill
\begin{minipage}[c]{0.36\textwidth}
\caption[Generic coupled two-subsystem geometry]{Generic coupled two-subsystem geometry.}
\label{generic}
\end{minipage}
\end{center}
  \end{figure}

This is also the approach in the very recent work on diffusive
bi-layer systems in zero magnetic field \cite{narozhny2000} and the
extension to half-filled Landau levels \cite{narozhny2001}. However,
here we consider arbitrarily shaped subsystems of non-interacting
electrons with an inter-subsystem Coulomb coupling of electrons, see
Fig.~\ref{generic}.

The thesis is organized as follows: Chapters \ref{chap:kubo},
\ref{chap:matsubara}, and \ref{chap:wick} contain a derivation of the
formal results for the drag conductance that form the starting point
of the calculations in
Refs.~[E,\,F,\,G,\,H].
Starting from Kubo formalism the drag conductance is calculated with
the aid of the Matsubara formalism. In chapter \ref{chap:numerical} a
discrete formulation is given which is suitable for numerical
implementations.  Chapter \ref{chap:wires} is on drag between quantum
wires and reports on the numerical and perturbative results in
Refs.~[E,\,G,\,H].
Chapter \ref{chap:dots} is on drag between chaotic quantum dots and
contains the results of
Refs.~[E,\,F] where the
statistical properties are obtain by use of random matrix theory.
Finally, Chapter \ref{chap:summary} gives a brief summary and
concluding remarks.

%% file: kubo.tex
\chapter{Linear response theory}\label{chap:kubo}

\section{Conductance matrix}
We consider two coupled electron systems in the linear-response limit
where the currents $I_i$ are related to the applied voltages $V_i$
through a $2\times 2$ conductance matrix,

\begin{equation}\label{G-matrix}
\begin{pmatrix}I_1\\I_2\end{pmatrix}
=\begin{pmatrix}G_{11}&G_{12}\\G_{21}&G_{22}\end{pmatrix}
\begin{pmatrix}V_1\\V_2\end{pmatrix}.
\end{equation}

In the phase-coherent regime one often assumes that the systems
consist of non-interacting particles and the diagonal parts can then
be calculated from the two-probe Landauer formula in terms of the
scattering matrices.\index{conductance!, drag $G_{21}$}\index{conductance!, Landauer $G_{ii}$} The aim of this
work is to study the off-diagonal elements in the case of inter-subsystem Coulomb interaction, but with electrons of each subsystem
being otherwise non-interacting. Furthermore, in order to study the
pure effect of Coulomb drag we assume that there is no tunneling, {\it
  i.e.} no charge transfer between the two electron systems.

We start from Kubo formalism which can as well be used to derive the
Landauer formula for the diagonal elements
\cite{fisher1981,sols1991,nockel1993}. The Kubo formalism has already
been used for studying Coulomb drag in extended ($L \gg \ell_\phi$)
two-layer systems \cite{kamenev1995,flensberg1995} and here we shall
closely follow the work of Flensberg {\it et al.}
\cite{flensberg1995}. The main difference is that for extended systems
the formalism becomes translationally invariant (after standard
impurity averaging is performed) and thus transformation to the
Fourier space is convenient. On the other hand this is not possible in
the case of mesoscopic systems ($\ll \ell_\phi$) where a formulation
in real space coordinates is required to deal with the broken
translation symmetry of these systems.

\begin{figure}
  \begin{center}
\begin{minipage}[c]{0.65\textwidth}
\epsfig{file=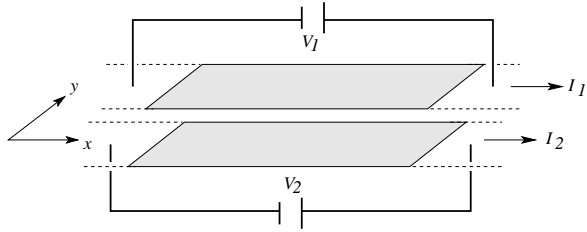, height=\columnwidth,angle=-90}
\end{minipage}\hfill
\begin{minipage}[c]{0.31\textwidth}
\caption[Bi-layer system]{Bi-layer system with ideal leads.}
\label{system_bilayer}
\end{minipage}
\end{center}
  \end{figure}

For simplicity our starting point is a bi-layer system as sketched in
Fig.~\ref{system_bilayer}, but the formalism is general valid and more exotic geometries like in Fig.~\ref{generic}, are in fact
included by adding some in-plane confinements. The relative
orientation of the two subsystems only enters the functional form of
the Coulomb interaction between them.

\section{From conductivity to conductance}

In linear response, the induced charge density currents $J_i$ are
related to the electric fields $E_i$ by the conductivity tensor
$\sigma$ through

\begin{equation}\label{fundamental}
J_i^\alpha({\boldsymbol r}_i,t)=\sum_{j,\beta} \int {\rm d}
  {\boldsymbol r}_j\,
  \sigma_{ij}^{\alpha\beta}({\boldsymbol r}_i,{\boldsymbol r}_j,\Omega)
E_j^\beta({\boldsymbol r}_j,t).
\end{equation}
Here, $E_i^{\alpha}({\boldsymbol r}_i,t)=
\Xi_{\alpha}e^{i({\boldsymbol q}\cdot{\boldsymbol r}_i-\Omega t)}$,
$\alpha$ and $\beta$ label the spatial directions, and $i$ and $j$
label the two subsystems.

For the study of drag we want to calculate, say, the DC current $I_2$
in subsystem 2 due to an applied voltage $V_1$ in system 1 so that we
can identify $G_{21}$. From Eq. (\ref{fundamental}) we get

\begin{equation}
I_2 = \int{\rm d} y_2 J_2^x({\boldsymbol r}_2)=G_{22} V_2
+ \sum_{\beta} \iint{\rm d} y_2  {\rm d}{\boldsymbol r}_1\,
  \sigma_{21}^{x\beta}({\boldsymbol r}_2,{\boldsymbol r}_1)
E_1^\beta({\boldsymbol r}_1),
\end{equation}
where ${\boldsymbol r}_2=(x_2,y_2)$ is a position in one of the leads
of subsystem 2 with $y_2$ being the transverse direction. In
fact, since we consider the DC limit we can choose the point any where
due to current conservation,

\begin{equation}
{\bf \nabla}\cdot{\boldsymbol J}_i=0.
\end{equation}
The electrical field is given by $E_1^\beta({\boldsymbol
  r}_1)=-{\boldsymbol \nabla}_{{\boldsymbol r}_1}^\beta
\Phi_1({\boldsymbol r}_1)$ with the bias $V_1$ defined by

\begin{equation}
V_1=\lim_{x_1\rightarrow -\infty} 
\Phi_1({\boldsymbol r}_1)-\lim_{x_1\rightarrow \infty} 
\Phi_1({\boldsymbol r}_1).
\end{equation}
By partial integration we get

\begin{multline}
I_2 =G_{22} V_2+ \iint{\rm d} y_2  {\rm d} {y}_1\,
  \sigma_{21}^{xx}({\boldsymbol r}_2,{\boldsymbol r}_1)V_1\\
+  \iint{\rm d} y_2  {\rm d}{\boldsymbol r}_1\,
\left[\sum_{\beta}{\boldsymbol \nabla}_{{\boldsymbol r}_1}^\beta 
\sigma_{21}^{x\beta}({\boldsymbol r}_2,{\boldsymbol r}_1)\right] 
\Phi_1({\boldsymbol r}_1),
\end{multline}
where in the second term both $x_1$ and $x_2$ are in the leads and we
have assumed the two leads of subsystem 1 to be identical. In the DC
limit the last term is zero due to current conservation and comparing
to Eq.~(\ref{G-matrix}) we identify the drag conductance to be given
by\index{conductance!, drag $G_{21}$}

\begin{equation}\label{sigma->G}
G_{21}=\iint{\rm d} y_2 {\rm d} {y}_1\,
\sigma_{21}^{xx}({\boldsymbol r}_2,{\boldsymbol r}_1),
\end{equation}
where ${\boldsymbol r}_1$ (${\boldsymbol r}_2$) is in one of the leads
of subsystem 1 (2). The advantage of this choice is that in the leads
the asymptotic form of the single-particle scattering states is given
in terms of plane waves and scattering matrices. In the following we
use Kubo formalism to establish a quantum mechanical description of
$\sigma_{21}$ in mesoscopic structures.

\section{Kubo formalism}
In the following we briefly review the Kubo formalism, but in the
case of coupling of two subsystems. First, we present the Hamiltonian
and include the perturbation by an external vector potential. Next, we
derive the Kubo result for the drag conductivity $\sigma_{21}$.

\subsection{Coupling of two subsystems}

We consider a coupling of the two subsystems described by the Hamiltonian
\begin{subequations}
\begin{equation}
\hat{\mathscr H}=\hat{\mathscr H}_1+\hat{\mathscr H}_2 +\hat{\mathscr H}_{12},
\end{equation}
where in absence of intra-subsystem interactions we have
\begin{equation}
\hat{\mathscr H}_i=\int{\rm d}{\boldsymbol r}_i\,
\hat{\psi}^\dag({\boldsymbol r}_i)\,\hat{H}_i({\boldsymbol r}_i)
\,\hat{\psi}_i({\boldsymbol r}_i),
\end{equation}
and for the inter-subsystem interactions 
\begin{equation}\label{H12}
\hat{\mathscr H}_{12}=\int{\rm d}{\boldsymbol r}_1\int{\rm d}{\boldsymbol r}_2\,\hat{\psi}^\dag({\boldsymbol r}_1)\hat{\psi}^\dag({\boldsymbol r}_2)\,U_{12}({\boldsymbol r}_1,{\boldsymbol r}_2)\,
\hat{\psi}({\boldsymbol r}_2)\hat{\psi}({\boldsymbol r}_1).
\end{equation}
\end{subequations}
Here, $\hat{\psi}$ are the field operators, $H_i$ is the Hamiltonian of subsystem
$i$, and $U_{ij}$ is the electron-electron interaction between
electrons in subsystem $i$ and electrons in subsystem $j$.

\subsection{Vector potential}

The effect of a vector potential may be included through an additional
term. This is seen by considering the kinetic part $\hat{T}=\frac{1}{2m}\left(-i\hbar
  {\boldsymbol\nabla}+e{\boldsymbol A}\right)^2$ of the
singe-particle Hamiltonian $\hat{H}= \hat{T} +V$,
\begin{multline}
\int {\rm d}{\boldsymbol r}\,\hat{\psi}^\dagger({\boldsymbol r})\,\hat{T}\,\hat\psi({\boldsymbol r})=\int {\rm d}{\boldsymbol r}\,\hat{\psi}^\dagger({\boldsymbol r})\left[\frac{1}{2m}\left(-i\hbar
  {\boldsymbol\nabla}\right)^2\right]\hat\psi({\boldsymbol r})\\+\int {\rm d}{\boldsymbol r}\,\Bigg\{\frac{\hbar
  e}{i2m}\left[\hat{\psi}^\dag({\boldsymbol r})\,{\boldsymbol \nabla}\cdot {\boldsymbol
  A}({\boldsymbol r},t)\,\hat\psi({\boldsymbol r})\,+\hat{\psi}^\dag({\boldsymbol r})\,{\boldsymbol
  A}({\boldsymbol r},t)\cdot {\boldsymbol \nabla} \hat\psi({\boldsymbol r}) \right]\\+\frac{e^2}{2m}{\boldsymbol A}({\boldsymbol r},t)\cdot {\boldsymbol
  A}({\boldsymbol r},t)\, \hat{\psi}^\dagger({\boldsymbol r})\,\hat{\psi}({\boldsymbol r})\Bigg\}.
\end{multline}
By partial integration we get
\begin{equation}
\int {\rm d}{\boldsymbol r}\,\hat{\psi}^\dag({\boldsymbol r})\, {\boldsymbol
  \nabla}\cdot {\boldsymbol
  A}({\boldsymbol r},t)\, \hat\psi({\boldsymbol r})\, =-\int {\rm d}{\boldsymbol r}\,{\boldsymbol
  A}({\boldsymbol r},t)\,\cdot\left[ {\boldsymbol
  \nabla}\hat{\psi}^\dag({\boldsymbol r})\right]\hat{\psi}({\boldsymbol r}),
\end{equation}
where the surface term at infinity has been omitted. As shown by Sols
\cite{sols1991} a more
careful analysis of the boundary term leads to the same result for the
Kubo formula, which we will derive in the subsequent section. This means that
\begin{multline}
\int {\rm d}{\boldsymbol r}\,\hat{\psi}^\dagger({\boldsymbol
  r})\,\hat{T}\,\hat\psi({\boldsymbol r})=\int {\rm
  d}{\boldsymbol r}\,\hat{\psi}^\dagger({\boldsymbol r})\,\hat{T}_{{\boldsymbol A}=0}\,\hat\psi({\boldsymbol r})\\-\int {\rm
  d}{\boldsymbol r}\, (-e){\boldsymbol
  A}({\boldsymbol r},t)\cdot\left\{\hat{\boldsymbol J}({\boldsymbol r})+\hat{\boldsymbol J}'({\boldsymbol r},t)\right\},
\end{multline}
where the particle current operator is given by $\hat{\boldsymbol J}({\boldsymbol r})
+\hat{\boldsymbol J}'({\boldsymbol r},t)$ with
\begin{subequations}
\begin{eqnarray}
\hat{\boldsymbol J}({\boldsymbol r})&=&\frac{\hbar}{2mi}\left\{\hat{\psi}^\dag({\boldsymbol r})\, {\boldsymbol
  \nabla} \hat{\psi}({\boldsymbol r})- \left[{\boldsymbol
  \nabla}\hat{\psi}^\dag({\boldsymbol r})\,\right]\hat{\psi}({\boldsymbol r})\right\},\\
\hat{\boldsymbol J}'({\boldsymbol r},t)&=&\frac{e}{m}{\boldsymbol
  A}({\boldsymbol r},t)\,\hat{\psi}^\dag({\boldsymbol r})\, \hat{\psi}({\boldsymbol r}).
\end{eqnarray}
\end{subequations}
The full Hamiltonian can therefore be written as $\hat{\mathscr
  H}+\hat{\mathscr H}'$ where
\begin{equation}\label{A-hamiltonian}
\hat{\mathscr H}' = -\sum_i \int{\rm
  d}{\boldsymbol r}_i\, (-e){\boldsymbol A}_i({\boldsymbol r}_i,t)\cdot\left\{ \hat{\boldsymbol J}_i({\boldsymbol r}_i)+ \hat{\boldsymbol J}_i'({\boldsymbol r}_i,t)\right\}.
\end{equation}
In the following we ``linearize'' this term in $\boldsymbol A$ and treat it as a perturbation.

\subsection{Conductivity}

In the Coulomb gauge where ${\boldsymbol \nabla}{\bf \cdot}{\boldsymbol A}=0$ the electric
  field is described in terms of a time-dependent vector potential
  ${\boldsymbol A}$ according to ${\boldsymbol E}=-\partial
  {\boldsymbol A}/\partial t$ corresponding to ${\boldsymbol
  A}=-(i/\Omega){\boldsymbol E}$. Eq. (\ref{A-hamiltonian}) can thus to
  first order in $\boldsymbol E$ (linear response) be written as

\begin{equation}
\hat{\mathscr H}' \simeq \frac{i(-e)}{\Omega}\sum_i \int{\rm
  d}{\boldsymbol r}_i\,  {\boldsymbol E}_i({\boldsymbol
  r}_i,t)\cdot \hat{\boldsymbol J}_i({\boldsymbol
  r}_i).
\end{equation}

In the following we consider the full Hamiltonian $\hat{\mathscr
  H}+\hat{\mathscr H}'$ where the perturbation $\hat{\mathscr H}'$ is
  switched on at time $t=-\infty$. After the perturbation is switched on,
  the system develops according to the Schr\"{o}dinger equation

\begin{equation}
\left[\hat{\mathscr
  H}+\hat{\mathscr H}'(t)\right] \left|\psi(t)\right>= i\hbar \frac{\partial }{\partial t} \left|\psi(t)\right>
\end{equation}
with
\begin{subequations}
\begin{eqnarray}
\left|\psi(t)\right>&=&e^{-i\hat{\mathscr H}t/\hbar}\hat{U}(t)\left|\psi(-\infty)\right>,\\
\hat{U}(t)&=&1+\frac{1}{i\hbar}\int_{-\infty}^t{\rm d}t'\,\hat{\mathscr
  H}'(t')+\ldots,
\end{eqnarray}
\end{subequations}
where from here on the time-dependent operators (with $t$ as an explicitly
    stated argument) are given in the
    Heisenberg picture\index{representation!, Heisenberg} where 

\begin{equation}
 \hat{\mathscr O}({\boldsymbol
    r},t) = e^{i\hat{\mathscr H}t/\hbar}\hat{\mathscr O }({\boldsymbol
    r})e^{-i\hat{\mathscr H}t/\hbar}.
\end{equation}

The total charge current can now be written as

\begin{eqnarray}
{\boldsymbol J}_i({\boldsymbol r}_i,t)&=&(-e)\big<\psi(t)\big|
   \hat{\boldsymbol J}_i({\boldsymbol r}_i)+ \hat{\boldsymbol
    J}_i'({\boldsymbol r}_i,t)\big|\psi(t)\big>\\
&=&(-e)\big<\psi(-\infty)\big|\hat{U}^\dag(t)
    \big\{\hat{\boldsymbol J}_i({\boldsymbol r}_i)+\hat{\boldsymbol J}_i'({\boldsymbol r}_i,t)\big\}\hat{U}(t)\big|\psi(-\infty)\big>\nonumber,
\end{eqnarray}
and in linear response we get

\begin{multline}
{\boldsymbol J}_i({\boldsymbol r}_i,t)
\simeq (-e)\big<\psi(-\infty)\big|\hat{\boldsymbol J}_{i}'({\boldsymbol
    r}_i,t)\big|\psi(-\infty)\big>+\frac{e^2}{\hbar\Omega}\sum_j \int{\rm
  d}{\boldsymbol r}_j\,\Big\{\\- \big<\psi(-\infty)\big|\big\{{\boldsymbol E}_j({\boldsymbol
  r}_j,t)\cdot\int_{-\infty}^t{\rm d}t'\, e^{i\Omega(t-t')} \hat{\boldsymbol J}_j({\boldsymbol
  r}_j,t')\big\} \hat{\boldsymbol J}_{i}({\boldsymbol
    r}_i,t)\big|\psi(-\infty)\big>
\\+\big<\psi(-\infty)\big| \hat{\boldsymbol J}_{i}({\boldsymbol
    r}_i,t) \big\{{\boldsymbol E}_j({\boldsymbol
  r}_j,t)\cdot\int_{-\infty}^t{\rm d}t'\, e^{i\Omega(t-t')} \hat{\boldsymbol J}_j({\boldsymbol
  r}_j,t')\big\}\big|\psi(-\infty)\big>\Big\}.
\end{multline}
Here, we have used that
\begin{eqnarray}
\hat{U}(t)
&\simeq& 1+\frac{1}{i\hbar}\int_{-\infty}^t{\rm d}t'\,\frac{ie}{\Omega}\sum_i \int{\rm
  d}{\boldsymbol r}_i\,  {\boldsymbol E}_i({\boldsymbol
  r}_i,t')\cdot \hat{\boldsymbol J}_i({\boldsymbol
  r}_i,t')\nonumber\\
&=& 1+\frac{1}{i\hbar}\sum_i \int{\rm
  d}{\boldsymbol r}_i\,  {\boldsymbol E}_i({\boldsymbol
  r}_i,t)\cdot \frac{ie}{\Omega}\int_{-\infty}^t{\rm d}t'\, e^{i\Omega(t-t')}\hat{\boldsymbol J}_i({\boldsymbol
  r}_i,t').
\end{eqnarray}
Let us now consider the current density in, say, the $\alpha$ direction
\begin{multline}
{J}_i^\alpha({\boldsymbol r}_i,t)
\simeq
\frac{ie^2}{m\Omega}\big<\psi(-\infty)\big|\hat{\psi}_i^\dag({\boldsymbol
      r}_i)\hat{\psi}_i({\boldsymbol
      r}_i)\big|\psi(-\infty)\big>{E}_i^\alpha({\boldsymbol r}_i,t)\\+ \sum_{j,\beta} \int{\rm
  d}{\boldsymbol r}_j\,\frac{ie^2}{\hbar\Omega}\int_{-\infty}^t{\rm d}t'\, e^{i\Omega(t-t')}\\\times (-i)\big<\psi(-\infty)\big|\big[ \hat{J}_i^{\alpha}({\boldsymbol
  r}_i,t)\;,\; \hat{J}_{j}^{\beta}({\boldsymbol
    r}_j,t')\big]_-\big|\psi(-\infty)\big>\cdot E_j^\beta({\boldsymbol
  r}_j,t),
\end{multline}
and comparing to Eq. (\ref{fundamental}), we identify $\sigma_{ij}^{\alpha\beta}$ and get

\begin{equation}\label{kubo}
 \sigma_{ij}^{\alpha\beta}({\boldsymbol r}_i,{\boldsymbol
 r}_j,\Omega)=\frac{ie^2}{m \Omega}\rho_i({\boldsymbol r}_i)\delta({\boldsymbol
 r}_i-{\boldsymbol
 r}_j)\delta_{ij}\delta_{\alpha\beta}+\frac{ie^2}{\hbar\Omega}\Pi_{ij}^{\alpha\beta,r}({\boldsymbol
 r}_i,{\boldsymbol r}_j,\Omega).
\end{equation}
Here,

\begin{equation}
\rho_i({\boldsymbol
      r}_i) =\big<\hat{\psi}_i^\dag({\boldsymbol
      r}_i)\hat{\psi}_i({\boldsymbol
      r}_i)\big>,
\end{equation}
is the unperturbed particle density
      and the retarded current-current correlation function is given by

\begin{subequations}
\begin{eqnarray}
\Pi_{ij}^{\alpha\beta,r}({\boldsymbol
 r}_i,{\boldsymbol
 r}_j,\Omega)&=&\int_{-\infty}^\infty{\rm
 d}(t-t')\,e^{i\Omega(t-t')}\Pi_{ij}^{\alpha\beta,r}({\boldsymbol
 r}_i,{\boldsymbol
 r}_j,t,t'),\\
\Pi_{ij}^{\alpha\beta,r}({\boldsymbol
 r}_i,{\boldsymbol
 r}_j,t,t')&=&-i\Theta(t-t')\big<\big[\hat{J}_i^\alpha({\boldsymbol r}_i,t),\hat{J}_j^\beta({\boldsymbol r}_j,t')\big]_-\big>,
\end{eqnarray}
\end{subequations}
where we have used that 
$$\int_{-\infty}^t {\rm d}t' \ldots =
\int_{-\infty}^\infty {\rm d}t'\, \Theta(t-t')\ldots=\int_{-\infty}^\infty {\rm d}(t-t')\, \Theta(t-t')\ldots.$$ 
Eq.~(\ref{kubo}) is the celebrated Kubo formula\index{Kubo formula} \cite{kubo1957,mahan}, but here derived in real space for the case of two subsystems so that it in principle applies to all the four elements in the conductance matrix in Eq.~(\ref{G-matrix}). For the diagonal elements the calculation leads to the Landauer formula \cite{fisher1981,sols1991,nockel1993}. For the off-diagonal elements we note that only the last term in Eq.~(\ref{kubo}) contributes so that we do not have to worry about the usual problem with the divergence of the diamagnetic term in the DC limit, see {\it e.g.} Ref.~\cite{haug}. 

\subsection{Finite temperatures}

At finite temperatures, $\big<\hat{A}\big>$ denotes the usual statistical average over the
grand-canonical ensemble of the quantum mechanical expectation value
of the operator $\hat{A}$, {\it i.e.}
\begin{equation}\label{thermal_average}
\big<\hat{A}\big>=\frac{\Tr\big<\psi(-\infty)\big|e^{-\beta \hat{\mathscr
        K}}\hat{A}\big|\psi(-\infty)\big>}{\Tr\big<\psi(-\infty)\big|e^{-\beta \hat{\mathscr K}}\big|\psi(-\infty)\big>}=e^{\beta\tilde\Omega} \Tr\big<\psi(-\infty)\big|e^{-\beta \hat{\mathscr K}}\hat{A}\big|\psi(-\infty)\big>,
\end{equation}
where $\beta=1/k_{\rm B}T$ (no confusion should be made with the
$\beta$ labeling the spatial direction!) and the ``Kamiltonian'' is given by

\begin{equation}
 \hat{\mathscr K}=\hat{\mathscr H}-\mu
\hat{N}.
\end{equation}
 For a discussion of the absence of
$\hat{\mathscr H}'$ in the Kamiltonian, see Ref. \cite{mahan}.
 The trace is taken over states $\left|\psi\right>$ with any
number $\big<\hat{N}\big>$ of particles and $\tilde\Omega$ is the grand thermodynamic
potential. We shall later make a perturbation expansion in $\hat{\mathscr H}_{12}$,
so that these states are
eigenstates corresponding to the unperturbed Hamiltonians of the two decoupled systems, {\it i.e.} $\left|\psi\right>=\left|\psi_1\right>\otimes
\left|\psi_2\right> $.

%% file: matsubara.tex
\chapter{Formal calculation of drag conductivity}\label{chap:matsubara}

We are only interested in the transconductivity which is given by the last term in Eq.~(\ref{kubo}),

\begin{equation}\label{sigma21}
\sigma_{21}^{\alpha\beta}(\Omega)=\frac{ie^2}{\hbar\Omega}\Pi_{21}^{\alpha\beta,r}({\boldsymbol
 r}_2,{\boldsymbol r}_1,\Omega).
\end{equation}
This is also the starting point of Ref.~\cite{flensberg1995}. In order to calculate it we apply the Matsubara formalism\index{Matsubara!, formalism}, imaginary-time formalism. Turning to the interaction representation we make a systematic perturbation expansion in $U_{12}$. After a Fourier transformation we utilize the generalized Lehmann representation to obtain an expression for the current-current correlation function.

\section{Matsubara formalism}

In the Matsubara formalism we write the current-current correlation
function as

\begin{equation}
\Pi_{21}^{\alpha\beta}({\boldsymbol
 r},{\boldsymbol
 r}',\tau,\tau')=-\big<\psi\big|\hat{T}_\tau\big\{\hat{J}_2^\alpha({\boldsymbol
 r},\tau
 )\,\hat{J}_1^\beta({\boldsymbol r}',\tau'
 )\big\}\big|\psi\big>,
\end{equation}
where $\hat{T}_\tau$ is the usual imaginary-time ordering
operator. No confusion with the kinetic energy operator $\hat{T}$ should be made. In the interaction representation we have that

\begin{equation}
\Pi_{21}^{\alpha\beta}({\boldsymbol
 r},{\boldsymbol
 r}',\tau,\tau')=-\frac{\big<\psi\big|\hat{T}_\tau\big\{ \hat{S}(\beta)\hat{J}_2^\alpha({\boldsymbol
 r},\tau
 )\,\hat{J}_1^\beta({\boldsymbol r}',\tau'
 )\big\}\big|\psi\big>}{\big<\psi\big|\hat{S}(\beta)\big|\psi\big>},
\end{equation}
and due to the usual cancellation of unconnected (unlinked) diagrams \cite{mahan} we get
\begin{equation}
\Pi_{21}^{\alpha\beta}({\boldsymbol
 r},{\boldsymbol
 r}',\tau,\tau')=-\big<\psi\big|\hat{T}_\tau\big\{ \hat{S}(\beta)\hat{J}_2^\alpha({\boldsymbol
 r},\tau
 )\,\hat{J}_1^\beta({\boldsymbol r}',\tau'
 )\big\}\big|\psi\big>,
\end{equation}
where only connected (linked) diagrams are to be considered. Here, $\left|\psi\right>=\left|\psi_1\right>\otimes
\left|\psi_2\right> $ and
\begin{eqnarray}
\hat{S}(\beta)&=& \hat{T}_\tau \left\{\exp\left[-\hbar^{-1}\int_0^{\hbar\beta}{\rm
    d}\tau_1\,\hat{\mathscr H}_{12}(\tau_1)\right]\right\}\nonumber\\
&\simeq& 1-  \hat{T}_\tau \left\{-\hbar^{-1}\int_0^{\hbar\beta}{\rm
    d}\tau_1\,\hat{\mathscr H}_{12}(\tau_1)\right\}\nonumber\\
&\quad&+\frac{1}{2}\hat{T}_\tau\left\{\hbar^{-2}\iint_0^{\hbar\beta}{\rm
    d}\tau_1{\rm
    d}\tau_2\,\hat{\mathscr H}_{12}(\tau_1)\hat{\mathscr
    H}_{12}(\tau_2)\right\}+\ldots.
\label{S_expansion}
\end{eqnarray}
In the interaction picture\index{representation!, interaction} Eq.~(\ref{H12}) becomes

\begin{equation}
\hat{\mathscr H}_{12}(\tau)
=\iint{\rm d}{\boldsymbol r}_1 {\rm d}{\boldsymbol
  r}_2\,\hat\rho_1({\boldsymbol r}_1,\tau) U_{12}({\boldsymbol r}_1,{\boldsymbol r}_2)\hat\rho_2({\boldsymbol r}_2,\tau),
\end{equation}
where $\hat\rho$ is the particle density operator.

\section{Perturbation expansion}

Using the expansion of $\hat{S}$, Eq.~(\ref{S_expansion}), we can now calculate $\Pi_{21}^{\alpha\beta}({\boldsymbol
 r},{\boldsymbol
 r}',\tau,\tau')$ to any order in $U_{12}$. The $0$th order contribution $\Pi_{21}^{\alpha\beta}({\boldsymbol
 r},{\boldsymbol
 r}',\tau,\tau')^{(0)}$ to the transconductivity is obviously vanishing. The $1$st order contribution $\Pi_{21}^{\alpha\beta}({\boldsymbol
 r},{\boldsymbol
 r}',\tau,\tau')^{(1)}$ also vanishes in the DC limit \cite{flensberg1995} so that the lowest-order contribution is of $2$nd order in $U_{12}$. We will not prove the vanishing of the $1$st order term, but just mention that application of Fermi's golden rule obviously would give a lowest-order contribution of 2nd order in $U_{12}$ too since it involves the absolute square of the matrix element.\index{Fermi's golden rule}

\subsection{Quadratic contribution}

\begin{figure}
  \begin{center}
\begin{minipage}[c]{0.65\textwidth}
 \epsfig{file=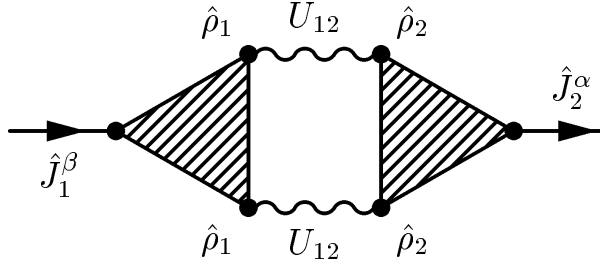, width=\textwidth,clip}
\end{minipage}\hfill
\begin{minipage}[c]{0.31\textwidth}
\caption[Diagrammatic representation of current-current correlation function]{Diagram corresponding to the current-current correlation function
  ${\Pi_{21}^{\alpha\beta}}^{(2)}$. The hatched triangles correspond
  to the functions $\Delta_1^\beta$ and $\Delta_2^\alpha$.}
\label{feynman}
\end{minipage}
\end{center}
  \end{figure}

The $2$nd
 order contribution becomes

\begin{multline}
\Pi_{21}^{\alpha\beta}({\boldsymbol
 r},{\boldsymbol
 r}',\tau,\tau')^{(2)}=-\frac{1}{2\hbar^2}\iint_0^{\hbar\beta}{\rm d}\tau_1
{\rm d}\tau_2\\
\times\iiiint{\rm d}{\boldsymbol r}_1
{\rm d}{\boldsymbol r}_2
{\rm d}{\boldsymbol r}_1'
{\rm d}{\boldsymbol r}_2'\,
 U_{12}({\boldsymbol r}_1,{\boldsymbol r}_2) U_{12}({\boldsymbol
 r}_1',{\boldsymbol r}_2')\\
\times\big<\psi\big|\hat{T}_\tau\big\{
 \hat{\rho}_1({\boldsymbol r}_1,\tau_1)\hat{\rho}_2({\boldsymbol
 r}_2,\tau_1)
\hat{\rho}_1({\boldsymbol r}_1',\tau_2)\hat{\rho}_2({\boldsymbol
 r}_2',\tau_2)
\hat{J}_2^\alpha({\boldsymbol
 r},\tau
 )\,\hat{J}_1^\beta({\boldsymbol r}',\tau'
 )
\big\}\big|\psi\big>.
\end{multline}
We can at no cost introduce an extra $\tau$-ordering operator
$\hat{T}_\tau$ and since $\left|\psi\right>=\left|\psi_1\right>\otimes
\left|\psi_2\right> $ we get

\begin{multline}
\Pi_{21}^{\alpha\beta}({\boldsymbol
 r},{\boldsymbol
 r}',\tau,\tau')^{(2)}=-\frac{1}{2\hbar^2}\iint_0^{\hbar\beta}{\rm d}\tau_1
{\rm d}\tau_2
\iiiint{\rm d}{\boldsymbol r}_1
{\rm d}{\boldsymbol r}_2
{\rm d}{\boldsymbol r}_1'
{\rm d}{\boldsymbol r}_2'\\
\times U_{12}({\boldsymbol r}_1,{\boldsymbol r}_2) U_{12}({\boldsymbol
 r}_1',{\boldsymbol r}_2')
\big<\psi_1\big|\hat{T}_\tau\big\{
 \hat{\rho}_1({\boldsymbol r}_1,\tau_1)
\hat{\rho}_1({\boldsymbol r}_1',\tau_2)\hat{J}_1^\beta({\boldsymbol r}',\tau'
 )
\big\}\big|\psi_1\big>\\
\times\big<\psi_2\big|\hat{T}_\tau\big\{
\hat{\rho}_2({\boldsymbol
 r}_2,\tau_1)
\hat{\rho}_2({\boldsymbol
 r}_2',\tau_2)
\hat{J}_2^\alpha({\boldsymbol
 r},\tau
 )
\big\}\big|\psi_2\big>.
\end{multline}
Introducing the following three-point correlation function, the so-called ``triangle'' function,
\begin{equation}\label{triangle_1}
{\Delta}_i^\alpha({\boldsymbol r},\tau,{\boldsymbol
  r}',\tau',{\boldsymbol r}'',\tau'')=-\big<\psi_i\big|\hat{T}_\tau\big\{  \hat{J}_i^\alpha({\boldsymbol
  r},\tau)\hat{\rho}_i({\boldsymbol
  r}',\tau') \hat{\rho}_i({\boldsymbol
  r}'',\tau'')\big\} \big|\psi_i\big>,
\end{equation}
we get

\begin{multline}\label{Pi_flensberg}
\Pi_{21}^{\alpha\beta}({\boldsymbol
 r},{\boldsymbol
 r}',\tau,\tau')^{(2)}=-\frac{1}{2\hbar^2}\iint_0^{\hbar\beta}{\rm d}\tau_1 {\rm d}\tau_2
\iiiint{\rm d}{\boldsymbol r}_1
{\rm d}{\boldsymbol r}_2
{\rm d}{\boldsymbol r}_1'
{\rm d}{\boldsymbol r}_2'\\
\times U_{12}({\boldsymbol r}_1,{\boldsymbol r}_2) U_{12}({\boldsymbol
 r}_1',{\boldsymbol r}_2')
{\Delta}_1^\beta({\boldsymbol r}',\tau',{\boldsymbol
  r}_1,\tau_1,{\boldsymbol r}_1',\tau_2)
{\Delta}_2^\alpha({\boldsymbol r},\tau,{\boldsymbol
  r}_2,\tau_1,{\boldsymbol r}_2',\tau_2),
\end{multline}
which is shown diagrammatically in Fig~\ref{feynman}. This result can also be found in Ref.~\cite{flensberg1995}.

\subsection{Three-particle Green function}

From Eq.~(\ref{triangle_1}) it follows that the triangle function can be expressed in terms of a three-particle Green function. We first rewrite the particle current operator

\begin{eqnarray}
\hat{\boldsymbol
J}(x)&=&\frac{\hbar}{2mi}\left\{\hat{\psi}^\dag(x)\,
{\boldsymbol
  \nabla}_{\boldsymbol r} \hat{\psi}(x)- \big[{\boldsymbol
  \nabla}_{\boldsymbol r}\hat{\psi}^\dag(x)\,\big]\hat{\psi}(x)\right\}\nonumber\\
  &=&\frac{\hbar}{2mi}\left\{{\boldsymbol
  \nabla}_{\boldsymbol r}-{\boldsymbol
  \nabla}_{\tilde{\boldsymbol r}}\right\}\hat{\psi}^\dag(\tilde{x})
  \hat{\psi}(x)\Big|_{\tilde{x}=x},
\end{eqnarray}
where $(x)=({\boldsymbol r},\tau)$, and thus

\begin{multline}
{\Delta}_i^\alpha(x,x',x'')=-\big<\hat{T}_\tau\big\{
\hat{J}_i^\alpha(x)\hat{\rho}_i(x') \hat{\rho}_i(x'')\big\}
\big>\\ 
=-\frac{\hbar}{2mi}\big\{{\boldsymbol \nabla}_{\boldsymbol
r}^\alpha-{\boldsymbol\nabla}_{\tilde{\boldsymbol r}}^\alpha\big\}
\big<\hat{T}_\tau\big\{\hat{\psi}_i^\dagger(\tilde{x})\hat{\psi}_i(x)\hat{\psi}_i^\dagger(x')\hat{\psi}_i(x')
\hat{\psi}_i^\dagger(x'') \hat{\psi}_i(x'')\big\} \big>
\Big|_{\tilde{x}=x}.
\end{multline}
Apart from the order of the field operators this is a three-particle Green function,
\begin{equation}
{\mathscr G}_i(x_1,x_2,x_3;y_3,y_2,y_1)=-\big<\hat{T}_\tau\big\{\hat{\psi}_i(x_1)\hat{\psi}_i(x_2)\hat{\psi}_i(x_3)\hat{\psi}_i^\dagger(y_3)
\hat{\psi}_i^\dagger(y_2) \hat{\psi}_i^\dagger(y_1)\big\}
\big>,
\end{equation}
so that we get the relation
\begin{equation}\label{three-particle}
{\Delta}_i^\alpha(x,x',x'')=\frac{\hbar}{2mi}\big\{\hat{\nabla}_{\boldsymbol
r}^\alpha-\hat{\nabla}_{\tilde{\boldsymbol r}}^\alpha\big\}
{\mathscr G}_i(x,x',x'';x'',x',\tilde{x}) \Big|_{\tilde{x}=x}.
\end{equation}
We note that in order to obtain the correct sequence of
field operators, $9$ interchanges are needed. This gives the additional factor
of $(-1)^9=-1$.

\section{Fourier transformation}
\subsection*{ --- in time, but not in space}

The next step will usually be to introduce a Fourier transformation in both time and space as in Ref.~\cite{flensberg1995}. However, as already discussed we want to be able to study also systems with broken translation symmetry and thus only Fourier transformation in time is convenient. The transformation is defined by

\begin{multline}\label{Fourier-time}
{\Delta}_i^\alpha({\boldsymbol r},\tau,{\boldsymbol
  r}',\tau',{\boldsymbol
  r}'',\tau'')=\frac{1}{(\hbar\beta)^2}\sum_{i\omega_m\,i\omega_n}
e^{-i\omega_m(\tau-\tau'')-i\omega_n(\tau'-\tau'')}\\\times{\Delta}_i^\alpha({\boldsymbol r},{\boldsymbol
  r}',{\boldsymbol
  r}'',i\omega_m,i\omega_n),
\end{multline}
with bosonic Matsubara frequencies,\index{Matsubara!, frequencies}
\begin{equation}
\omega_n=2n\pi/\hbar\beta,\qquad n=0,\pm1,\pm2,\ldots.
\end{equation}
Applying it to the current-current correlation function, Eq.~(\ref{Pi_flensberg}), we get

\begin{multline}
\Pi_{21}^{\alpha\beta}({\boldsymbol
 r},{\boldsymbol
 r}',\tau,\tau')^{(2)}=-\frac{1}{2\hbar^2} \frac{1}{(\hbar\beta)^4}\iint_0^{\hbar\beta}{\rm d}\tau_1
{\rm d}\tau_2\\
\times\iiiint{\rm d}{\boldsymbol r}_1
{\rm d}{\boldsymbol r}_2
{\rm d}{\boldsymbol r}_1'
{\rm d}{\boldsymbol r}_2'\,U_{12}({\boldsymbol r}_1,{\boldsymbol r}_2) U_{12}({\boldsymbol
 r}_1',{\boldsymbol r}_2')\\
\times \sum_{i\omega_m\,i\omega_n\,i\omega_{m'}\,i\omega_{n'}}\Big\{
e^{-i\omega_m\tau'-i\omega_{m'} \tau
 -i(\omega_n+\omega_{n'})(\tau_1-\tau_2)+i(\omega_m+\omega_{m'})\tau_2}
 \\
\times{\Delta}_1^\beta({\boldsymbol r}',{\boldsymbol
  r}_1,{\boldsymbol
  r}_1',i\omega_m,i\omega_n)\,{\Delta}_2^\alpha({\boldsymbol r},{\boldsymbol
  r}_2,{\boldsymbol
  r}_2',i\omega_{m'},i\omega_{n'})\Big\}.
\end{multline}
Using that $\int_0^{\hbar\beta}{\rm
  d}\tau\,e^{i(\omega_n-\omega_m)\tau}=\hbar\beta\delta_{nm}$ it is now easy to carry out two of the Matsubara sums\index{Matsubara!, sum} and get

\begin{multline}
\Pi_{21}^{\alpha\beta}({\boldsymbol
 r},{\boldsymbol
 r}',\tau,\tau')^{(2)}=
-\frac{1}{2\hbar^2} \frac{1}{(\hbar\beta)^2}
\iiiint{\rm d}{\boldsymbol r}_1
{\rm d}{\boldsymbol r}_2
{\rm d}{\boldsymbol r}_1'
{\rm d}{\boldsymbol r}_2'\,
 U_{12}({\boldsymbol r}_1,{\boldsymbol r}_2) U_{12}({\boldsymbol
 r}_1',{\boldsymbol r}_2')\\
\times \sum_{i\omega_m,i\omega_n}
e^{i\omega_m(\tau-\tau')}
{\Delta}_1^\beta({\boldsymbol r}',{\boldsymbol
  r}_1,{\boldsymbol
  r}_1',i\omega_m,i\omega_n)\,{\Delta}_2^\alpha({\boldsymbol r},{\boldsymbol
  r}_2,{\boldsymbol
  r}_2',-i\omega_{m},-i\omega_{n}).
\end{multline}
 Finally, we introduce the Fourier transform with respect to the ``frequency'' of the driving field,
\begin{equation}
\Pi_{21}^{\alpha\beta}({\boldsymbol r},{\boldsymbol
  r}',i\Omega_n)=\int_0^{\hbar\beta}{\rm d}(\tau-\tau')\,
  e^{i\Omega_n(\tau-\tau') }\Pi_{21}^{\alpha\beta}({\boldsymbol r},{\boldsymbol
  r}',\tau,\tau'),
\end{equation}
with bosonic Matsubara frequencies\index{Matsubara!, frequencies}
\begin{equation}
\Omega_n=2n\pi/\hbar\beta,\qquad n=0,\pm1,\pm2,\ldots.
\end{equation}
Applying this transformation we get
\begin{multline}
\Pi_{21}^{\alpha\beta}({\boldsymbol
 r},{\boldsymbol
 r}',i\Omega_n)^{(2)}
=-\frac{1}{2\hbar^2} \frac{1}{\hbar\beta}
\iiiint{\rm d}{\boldsymbol r}_1
{\rm d}{\boldsymbol r}_2
{\rm d}{\boldsymbol r}_1'
{\rm d}{\boldsymbol r}_2'\,U_{12}({\boldsymbol r}_1,{\boldsymbol r}_2) U_{12}({\boldsymbol
 r}_1',{\boldsymbol r}_2')\\
\times \sum_{i\omega_n}{\Delta}_1^\beta({\boldsymbol r}',{\boldsymbol
  r}_1,{\boldsymbol
  r}_1',-i\Omega_m,i\omega_n)\,{\Delta}_2^\alpha({\boldsymbol r},{\boldsymbol
  r}_2,{\boldsymbol
  r}_2',i\Omega_{m},-i\omega_{n}),
\end{multline}
which can rewritten as
\begin{multline}
\Pi_{21}^{\alpha\beta}({\boldsymbol
 r},{\boldsymbol
 r}',i\Omega_n)^{(2)}
=-\frac{1}{2\hbar^2} \frac{1}{\hbar\beta}
\iiiint{\rm d}{\boldsymbol r}_1
{\rm d}{\boldsymbol r}_2
{\rm d}{\boldsymbol r}_1'
{\rm d}{\boldsymbol r}_2'\,U_{12}({\boldsymbol r}_1,{\boldsymbol r}_2) U_{12}({\boldsymbol
 r}_1',{\boldsymbol r}_2')\\
\times \sum_{i\omega_n}{\Delta}_1^\beta({\boldsymbol r}',{\boldsymbol
  r}_1,{\boldsymbol
  r}_1',-i\Omega_n,-i\omega_n)\,{\Delta}_2^\alpha({\boldsymbol r},{\boldsymbol
  r}_2,{\boldsymbol
  r}_2',i\Omega_{n},i\omega_{n}),
\label{Pi(omega_n)}
\end{multline}
where we have reverted the sum, {\it i.e.} $i\omega_n\rightarrow -i\omega_n$.

\section{Generalized Lehmann representation}

The aim is now to obtain a formal result within the generalized Lehmann representation. First we find the resulting poles and subsequently we carry out the Matsubara summation by a contour integral in the complex plane. After obtaining the retarded correlation function by analytic continuation we finally take the DC limit. The derivation is in principle straight forward but contains many technical steps which may be skipped --- the final result is given in subsection \ref{subsectionDC}.

\subsection{Poles}                                             
We consider the reverse Fourier transform of Eq. (\ref{Fourier-time})

\begin{multline}\label{Fourier-time-reverse}
{\Delta}_i^\alpha({\boldsymbol r},{\boldsymbol
  r}',{\boldsymbol
  r}'',i\Omega_n,i\omega_n)=\iint_0^{\hbar\beta}{\rm
  d}(\tau-\tau'') {\rm d}(\tau'-\tau'')\,
\\\times e^{i\Omega_n(\tau-\tau'')+i\omega_n(\tau'-\tau'')}{\Delta}_i^\alpha({\boldsymbol r},\tau,{\boldsymbol
  r}',\tau',{\boldsymbol
  r}'',\tau''),
\end{multline}
so that the action of the imaginary-time ordering operator in Eq.~(\ref{triangle_1}) gives

\begin{multline}
{\Delta}_i^\alpha({\boldsymbol r},{\boldsymbol
  r}',{\boldsymbol
  r}'',i\Omega_n,i\omega_n)=
-
\int_0^{\hbar\beta}{\rm
  d}(\tau-\tau'')e^{i\Omega_n(\tau-\tau'')}\\\times\Bigg\{
  \int_0^{\tau-\tau''}{\rm d}(\tau'-\tau'')\, e^{i\omega_n(\tau'-\tau'')}
\big<\psi_i\big|\hat{J}_i^\alpha({\boldsymbol
  r},\tau)\hat{\rho}_i({\boldsymbol
  r}',\tau') \hat{\rho}_i({\boldsymbol
  r}'',\tau'') \big|\psi_i\big>\\+\int_{\tau-\tau''}^{\hbar\beta}{\rm d}(\tau'-\tau'')\,e^{i\omega_n(\tau'-\tau'')}
\big<\psi_i\big| \hat{\rho}_i({\boldsymbol
  r}',\tau')  \hat{J}_i^\alpha({\boldsymbol
  r},\tau)\hat{\rho}_i({\boldsymbol
  r}'',\tau'') \big|\psi_i\big>\Bigg\}.
\end{multline}
We now consider the evaluation within the generalized Lehmann
representation, see {\it e.g.} Ref. \cite{fetter}, where
we denote the eigenstates by
$\left|k\right>$ with\index{representation!, Lehmann} 
\begin{equation}
\hat{\mathscr
  K}_i\left|k\right>={\mathscr E}_k\left|k\right>,\qquad {\mathscr E}_k=\epsilon_k-\mu N_k.
\end{equation}
Using Eq.~(\ref{thermal_average}) this yields

\begin{multline}
{\Delta}_i^\alpha({\boldsymbol r},{\boldsymbol
  r}',{\boldsymbol
  r}'',i\Omega_n,i\omega_n)
=-e^{\beta\tilde\Omega} \int_0^{\hbar\beta}{\rm
  d}(\tau-\tau'')e^{i\Omega_n(\tau-\tau'')}\sum_{k}e^{-\beta {\mathscr E}_k}\\\times\Bigg\{
  \int_0^{\tau-\tau''}{\rm d}(\tau'-\tau'')\, e^{i\omega_n(\tau'-\tau'')}
\big<k\big|\hat{J}_i^\alpha({\boldsymbol
  r},\tau)\hat{\rho}_i({\boldsymbol
  r}',\tau') \hat{\rho}_i({\boldsymbol
  r}'',\tau'') \big|k\big>\\+\int_{\tau-\tau''}^{\hbar\beta}{\rm d}(\tau'-\tau'')\,e^{i\omega_n(\tau'-\tau'')}
\big<k\big| \hat{\rho}_i({\boldsymbol
  r}',\tau')  \hat{J}_i^\alpha({\boldsymbol
  r},\tau)\hat{\rho}_i({\boldsymbol
  r}'',\tau'') \big|k\big>\Bigg\},
\end{multline}
and since the $\tau$-dependence of an operator is given by 
\begin{equation}
\hat{\mathscr O}({\boldsymbol r},\tau)=e^{\hat{\mathscr
    K}\tau/\hbar}\hat{\mathscr O}({\boldsymbol r})e^{-\hat{\mathscr
    K}\tau/\hbar},
\end{equation}
we get

\begin{multline}
{\Delta}_i^\alpha({\boldsymbol r},{\boldsymbol
  r}',{\boldsymbol
  r}'',i\Omega_n,i\omega_n)
=-e^{\beta\tilde\Omega} \int_0^{\hbar\beta}{\rm
  d}(\tau-\tau'')e^{i\Omega_n(\tau-\tau'')}\sum_{k}e^{-\beta {\mathscr E}_k}\\\times\Bigg\{
  \int_0^{\tau-\tau''}{\rm d}(\tau'-\tau'')\, e^{i\omega_n(\tau'-\tau'')}e^{{\mathscr E}_k(\tau-\tau'')/\hbar}\\
\times \big<k\big|\hat{J}_i^\alpha({\boldsymbol r})e^{-\hat{\mathscr K}_i\tau/\hbar}e^{\hat{\mathscr K}_i\tau'/\hbar}\hat{\rho}_i({\boldsymbol
  r}')e^{-\hat{\mathscr K}_i\tau'/\hbar}e^{\hat{\mathscr K}_i\tau''/\hbar} \hat{\rho}_i({\boldsymbol
  r}'') \big|k\big>\\+\int_{\tau-\tau''}^{\hbar\beta}{\rm d}(\tau'-\tau'')\,e^{({\mathscr E}_k/\hbar+i\omega_n)(\tau'-\tau'')}\\
\times \big<k\big| \hat{\rho}_i({\boldsymbol
  r}')e^{-\hat{\mathscr K}_i\tau'/\hbar}e^{\hat{\mathscr K}_i\tau/\hbar}  \hat{J}_i^\alpha({\boldsymbol
  r})e^{-\hat{\mathscr K}_i\tau/\hbar}e^{\hat{\mathscr K}_i\tau''/\hbar} \hat{\rho}_i({\boldsymbol
  r}'') \big|k\big>\Bigg\}.
\end{multline}
Using the closure relation,
$\sum_m\big|m\big>\big<m\big|=1$, twice we now introduce the current and particle-density matrix elements and get

\begin{multline}
{\Delta}_i^\alpha({\boldsymbol r},{\boldsymbol
  r}',{\boldsymbol
  r}'',i\Omega_n,i\omega_n)
=-e^{\beta\tilde\Omega} \int_0^{\hbar\beta}{\rm
  d}(\tau-\tau'')e^{i\Omega_n(\tau-\tau'')}\sum_{kml}e^{-\beta {\mathscr E}_k}\\\times\Bigg\{e^{({\mathscr E}_k-{\mathscr E}_m)(\tau-\tau'')/\hbar}
  \int_0^{\tau-\tau''}{\rm d}(\tau'-\tau'')\, e^{({\mathscr E}_m/\hbar -{\mathscr E}_l/\hbar+i\omega_n)(\tau'-\tau'')}\\
\times \big<k\big|\hat{J}_i^\alpha({\boldsymbol r})\big|m\big>\big<m\big|\hat{\rho}_i({\boldsymbol
  r}')\big|l\big>\big<l\big| \hat{\rho}_i({\boldsymbol
  r}'') \big|k\big>\\+e^{({\mathscr E}_m-{\mathscr E}_l)(\tau-\tau'')/\hbar}\int_{\tau-\tau''}^{\hbar\beta}{\rm d}(\tau'-\tau'')\,e^{({\mathscr E}_k/\hbar-{\mathscr E}_m/\hbar+i\omega_n)(\tau'-\tau'')}\\
\times \big<k\big| \hat{\rho}_i({\boldsymbol
  r}')\big|m\big>\big<m\big|  \hat{J}_i^\alpha({\boldsymbol
  r})\big|l\big>\big<l\big| \hat{\rho}_i({\boldsymbol
  r}'') \big|k\big>\Bigg\}.
\end{multline}
Performing the inner integrals and using that the Matsubara frequencies are satisfying $e^{i\beta\hbar\Omega_n}=1$ the expression reduces to

\begin{multline}
{\Delta}_i^\alpha({\boldsymbol r},{\boldsymbol
  r}',{\boldsymbol
  r}'',i\Omega_n,i\omega_n)
=-e^{\beta\tilde\Omega} \int_0^{\hbar\beta}{\rm
  d}(\tau-\tau'')\sum_{kml}e^{-\beta {\mathscr E}_k}\\\times\Bigg\{
  \big[e^{(({\mathscr E}_k -{\mathscr E}_l)/\hbar+i\omega_n+i\Omega_n)(\tau-\tau'')}-e^{(({\mathscr E}_k-{\mathscr E}_m)/\hbar+i\Omega_n)(\tau-\tau'')}\big]\\
\times\frac{\big<k\big|\hat{J}_i^\alpha({\boldsymbol r})\big|m\big>\big<m\big|\hat{\rho}_i({\boldsymbol
  r}')\big|l\big>\big<l\big| \hat{\rho}_i({\boldsymbol
  r}'') \big|k\big>}{({\mathscr E}_m-{\mathscr E}_l)/\hbar+i\omega_n}\\+\big[e^{({\mathscr E}_k-{\mathscr E}_m)\beta}e^{(({\mathscr E}_m-{\mathscr E}_l)/\hbar +i\Omega_n)(\tau-\tau'')}-e^{(({\mathscr E}_k-{\mathscr E}_l)/\hbar+i\omega_n+i\Omega_n)(\tau-\tau'')}\big]\\
\times\frac{\big<k\big| \hat{\rho}_i({\boldsymbol
  r}')\big|m\big>\big<m\big|  \hat{J}_i^\alpha({\boldsymbol
  r})\big|l\big>\big<l\big| \hat{\rho}_i({\boldsymbol
  r}'') \big|k\big>}{({\mathscr E}_k-{\mathscr E}_m)/\hbar+i\omega_n}\Bigg\}.
\end{multline}
Proceeding in the same way with the remaining integral we get

\begin{multline}
{\Delta}_i^\alpha({\boldsymbol r},{\boldsymbol
  r}',{\boldsymbol
  r}'',i\Omega_n,i\omega_n)
=-e^{\beta\tilde\Omega} \sum_{kml}\\\times\Bigg\{
 e^{-\beta {\mathscr E}_k} \Bigg[\frac{e^{\beta({\mathscr E}_k -{\mathscr E}_l)}-1}{({\mathscr E}_k -{\mathscr E}_l)/\hbar+i\omega_n+i\Omega_n}-\frac{e^{\beta({\mathscr E}_k-{\mathscr E}_m)}-1}{({\mathscr E}_k-{\mathscr E}_m)/\hbar+i\Omega_n}\Bigg]\\
\times\frac{\big<k\big|\hat{J}_i^\alpha({\boldsymbol r})\big|m\big>\big<m\big|\hat{\rho}_i({\boldsymbol
  r}')\big|l\big>\big<l\big| \hat{\rho}_i({\boldsymbol
  r}'') \big|k\big>}{({\mathscr E}_m
  -{\mathscr E}_l)/\hbar+i\omega_n}\\
+e^{-\beta {\mathscr E}_k}\Bigg[e^{\beta({\mathscr E}_k-{\mathscr E}_m)}\frac{e^{\beta({\mathscr E}_m-{\mathscr E}_l)}-1}{({\mathscr E}_m-{\mathscr E}_l)/\hbar
  +i\Omega_n}-\frac{e^{\beta({\mathscr E}_k-{\mathscr E}_l)}-1}{({\mathscr E}_k-{\mathscr E}_l)/\hbar+i\omega_n+i\Omega_n}\Bigg]\\
\times
\frac{\big<k\big| \hat{\rho}_i({\boldsymbol
  r}')\big|m\big>\big<m\big|  \hat{J}_i^\alpha({\boldsymbol
  r})\big|l\big>\big<l\big| \hat{\rho}_i({\boldsymbol
  r}'') \big|k\big>}{({\mathscr E}_k-{\mathscr E}_m)/\hbar+i\omega_n}\Bigg\}.
\end{multline}
This can be further simplified due to the symmetries and interchanging $l$ and $k$ in the second term yields

\begin{multline}\label{triangle}
{\Delta}_i^\alpha({\boldsymbol r},{\boldsymbol
  r}',{\boldsymbol
  r}'',i\Omega_n,i\omega_n)
=-e^{\beta\tilde\Omega} \sum_{kml}\\\times\Bigg\{
  \Bigg[\frac{e^{-\beta {\mathscr E}_l}-e^{-\beta {\mathscr E}_k}}{({\mathscr E}_k  -{\mathscr E}_l)/\hbar+i\omega_n+i\Omega_n}+\frac{e^{-\beta {\mathscr E}_k}-e^{-\beta {\mathscr E}_m}}{({\mathscr E}_k-{\mathscr E}_m)/\hbar+i\Omega_n}\Bigg]\\
\times\frac{\big<k\big|\hat{J}_i^\alpha({\boldsymbol r})\big|m\big>\big<m\big|\hat{\rho}_i({\boldsymbol
  r}')\big|l\big>\big<l\big| \hat{\rho}_i({\boldsymbol
  r}'') \big|k\big>}{({\mathscr E}_m
  -{\mathscr E}_l)/\hbar+i\omega_n}\\
+\Bigg[\frac{e^{-\beta {\mathscr E}_k}-e^{-\beta {\mathscr E}_m}}{({\mathscr E}_m-{\mathscr E}_k)/\hbar
  +i\Omega_n}+\frac{e^{-\beta {\mathscr E}_l}-e^{-\beta {\mathscr E}_k}}{({\mathscr E}_l-{\mathscr E}_k)/\hbar+i\omega_n+i\Omega_n}\Bigg]\\
\times
\frac{\big<l\big| \hat{\rho}_i({\boldsymbol
  r}')\big|m\big>\big<m\big|  \hat{J}_i^\alpha({\boldsymbol
  r})\big|k\big>\big<k\big| \hat{\rho}_i({\boldsymbol
  r}'') \big|l\big>}{({\mathscr E}_l-{\mathscr E}_m)/\hbar+i\omega_n}\Bigg\}.
\end{multline}
This result corresponds to Eq.~(A2) of Ref.~\cite{flensberg1995}.
The triangle function is seen to be a {\it meromorphic} function in the
complex $z$-plane with simple poles at ${\rm Re}(z)={\mathscr E}_m-{\mathscr E}_k$ on
the lines corresponding to ${\rm Im}(z)=-\Omega_n$ and ${\rm
Im}(z)=0$ and it is analytic elsewhere. We also notice that at
${\rm
  Re}(z)={\mathscr E}_k-{\mathscr E}_l=0$ the pole at $i\omega_n=-i\Omega_n$ is removable. Since the operators are all Hermitian,
the result can be rewritten as

\begin{multline}\label{poles_Re}
{\Delta}_i^\alpha({\boldsymbol r},{\boldsymbol
  r}',{\boldsymbol
  r}'',i\Omega_n,i\omega_n)
=-e^{\beta\tilde\Omega} \sum_{kml}\\\times 2\Real\Bigg\{
  \Bigg[\frac{e^{-\beta {\mathscr E}_l}-e^{-\beta {\mathscr E}_k}}{({\mathscr E}_k
  -{\mathscr E}_l)/\hbar+i\omega_n+i\Omega_n}+\frac{e^{-\beta {\mathscr E}_k}-e^{-\beta {\mathscr E}_m}}{({\mathscr E}_k-{\mathscr E}_m)/\hbar+i\Omega_n}\Bigg]\\
\times\frac{\big<k\big|\hat{J}_i^\alpha({\boldsymbol
r})\big|m\big>\big<m\big|\hat{\rho}_i({\boldsymbol
  r}')\big|l\big>\big<l\big| \hat{\rho}_i({\boldsymbol
  r}'') \big|k\big>}{({\mathscr E}_m-{\mathscr E}_l)/\hbar+i\omega_n}\Bigg\},
\end{multline}
from which we see that for ${\mathscr E}_m={\mathscr E}_l$ the pole at $i\omega_n=0$ is removable too. We thus conclude that $\Delta_i^\alpha$ has no
poles coinciding with the Matsubara frequencies $i\omega_n=0$ and
$i\omega_n=-i\Omega_n$!\index{Matsubara!, frequencies}

For later use it is convenient to write $\Delta$ as a function of
the three variables $i\Omega_n$, $i\Omega_n+i\omega_n$, and
$i\omega_n$, {\it i.e.} 

\begin{equation}\label{newnotation}
{\Delta}_i^\alpha({\boldsymbol r},{\boldsymbol
  r}',{\boldsymbol
  r}'',i\Omega_n,i\omega_n)
\longrightarrow
{\Delta}_i^\alpha({\boldsymbol
r},{\boldsymbol
  r}',{\boldsymbol
  r}'',i\Omega_n,i\Omega_n+i\omega_n,i\omega_n),
\end{equation}
indicating the way in which $i\Omega_n$ and $i\omega_n$ enter the expression.

\subsection{Matsubara summation}

  \begin{figure}
  \begin{center}
\begin{minipage}[c]{0.70\textwidth}
 \epsfig{file=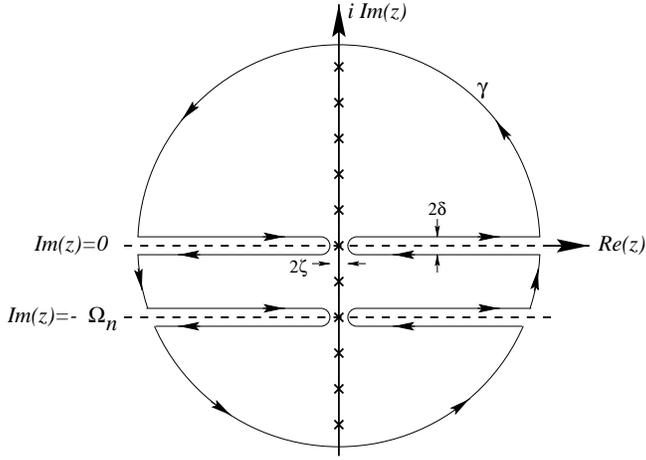, width=\textwidth,clip}
\end{minipage}\hfill
\begin{minipage}[c]{0.26\textwidth}
\caption[Complex contour for Matsubara summation]{Contour in complex plane for the evaluation of Matsubara
  sum. The Matsubara frequencies $i\omega_n$ are indicated by crosses
  ($\boldsymbol \times$) and the regions where $f(z)$ has poles by dashed lines.}
\label{contour}
\end{minipage}
\end{center}
  \end{figure}


For the evaluation of the Matsubara sum in Eq. (\ref{Pi(omega_n)}) we
consider the function

\begin{multline}\label{f}
f(i\Omega_n,i\Omega_n+z,z)= {\Delta}_1^\beta({\boldsymbol
r}',{\boldsymbol
  r}_1,{\boldsymbol
  r}_1',-i\Omega_n,-i\Omega_-,z,-z)\\\times{\Delta}_2^\alpha({\boldsymbol r},{\boldsymbol
  r}_2,{\boldsymbol
  r}_2',i\Omega_{n},i\Omega_n+z,z),
\end{multline}
and use the contour $\gamma$ shown in Fig. \ref{contour}. Here, $\delta$ and $\zeta$ are infinitesimal small and positive and the radius
$r$ of the arch is infinite. We consider the integral
\begin{equation}
I=\oint_{\gamma} \frac{{\rm
    d}z}{2\pi i}f(i\Omega_n,i\Omega_n+z,z) n_B(z)=\sum_{z_n} \Res(f\,n_B,z_n),
\end{equation}
where the Bose function,

\begin{equation}
n_B(z)=\big(e^{\hbar\beta z}-1\big)^{-1},
\end{equation}
has
simple poles at $z_n=i
(2\pi/\hbar\beta)n=i\omega_n $ so that we get a residue

\begin{equation}
\Res(f\,n_B,z_n)
=f(i\Omega_n,i\Omega_n+z_n,z_n)\,\Res(n_B,z_n)
=\frac{f(i\Omega_n,i\Omega_n+z_n,z_n)}{\hbar\beta}.
\end{equation}
On the other hand this means that, see Eqs. (\ref{Pi(omega_n)},\,\ref{f}), \index{Matsubara!, sum}
\begin{equation}
\frac{1}{\hbar\beta}\sum_{z_n}
f(i\Omega_n,i\Omega_n+z_n,z_n)=\oint_{\gamma} \frac{{\rm
    d}z}{2\pi i}f(i\Omega_n,i\Omega_n+z,z) n_B(z).
\end{equation}
The contribution to the contour integral from the arches,
\begin{equation}
 \lim_{r\rightarrow \infty}\int_0^{2\pi} \frac{{\rm
    d}\theta\,i r e^{i\theta}}{2\pi i}f(i\Omega_n,i\Omega_n+re^{i\theta},re^{i\theta}) n_B(re^{i\theta}),
\end{equation}
is vanishing since $f(i\Omega_n,i\Omega_n+z,z)$ vanishes in
infinity as $1/z^2$. The horizontal parts of the contour integral
can be expressed as Cauchy principal value integrals so that

\begin{multline}
\oint_{\gamma} \frac{{\rm
    d}z}{2\pi i}f(i\Omega_n,i\Omega_n+z,z) n_B(z)= {\mathscr P} \int_{-\infty}^{\infty} \frac{{\rm
    d}x}{2\pi i}f(i\Omega_n,i\Omega+x+i\delta,x+i\delta) n_B(x+i\delta)\\
+{\mathscr P}\int_{\infty}^{-\infty} \frac{{\rm
    d}x}{2\pi i}f(i\Omega_n,i\Omega_n+x-i\delta,x-i\delta) n_B(x-i\delta)\\
+{\mathscr P}\int_{-\infty}^{\infty} \frac{{\rm
    d}x}{2\pi i}f(i\Omega_n,x+i\delta,x-i\Omega_n+i\delta) n_{\rm
    B}(x-i\Omega_n+i\delta) \\
+{\mathscr P}\int_{\infty}^{-\infty} \frac{{\rm
    d}x}{2\pi i}f(i\Omega_n,x-i\delta,x-i\Omega_n-i\delta) n_{\rm
    B}(x-i\Omega_n-i\delta).
\end{multline}
Making the substitution $x\rightarrow \omega$ this means that
\begin{multline}
\frac{1}{\hbar\beta}\sum_{z_n} f(i\Omega_n,z_n)= \frac{1}{2\pi
i}{\mathscr P}\int_{-\infty}^\infty {\rm
    d}\omega\,n_B(\omega)\\\times\Big\{
 f(i\Omega_n,i\Omega_n+\omega,\omega+i\delta)
-f(i\Omega_n,i\Omega_n+\omega,\omega-i\delta)\\
+ f(i\Omega_n,\omega+i\delta,\omega-i\Omega_n)
-f(i\Omega_n,\omega-i\delta,\omega-i\Omega_n) \Big\},
\end{multline}
where we have let $\delta\rightarrow 0$ in terms with
$i\Omega_n$ and in the Bose functions $n_B$. Furthermore, we
have used that $n_{\rm
    B}(x-i\Omega_n)=n_{\rm
    B}(x)$.

\subsection{Analytic continuation}

We now extract the retarded part by analytic continuation
\cite{mahan,fetter}\index{analytic continuation}
\begin{multline}
\lim_{i\Omega_n\rightarrow
\Omega+i\delta}\frac{1}{\hbar\beta}\sum_{z_n} f(i\Omega_n,z_n)=
\frac{1}{2\pi i}{\mathscr P}\int_{-\infty}^\infty {\rm
    d}\omega\,n_B(\omega)\\\times\Big\{ 
f(\Omega+i\delta,\Omega+\omega+i\delta,\omega+i\delta)
-f(\Omega+i\delta,\Omega+\omega+i\delta,\omega-i\delta)\\
+ f(\Omega+i\delta,\omega+i\delta,\omega-\Omega-i\delta)
-f(\Omega+i\delta,\omega-i\delta,\omega-\Omega-i\delta) \Big\}.
\end{multline}
Shifting the variable $\omega\rightarrow \omega+\Omega$ in the
$3$rd and $4$th terms we get

\begin{multline}
\lim_{i\Omega_n\rightarrow
\Omega+i\delta}\frac{1}{\hbar\beta}\sum_{z_n} f(i\Omega_n,z_n)=
\frac{1}{2\pi i}{\mathscr P}\int_{-\infty}^\infty {\rm
    d}\omega\,\\\times\Big\{n_B(\omega)\big[ f(\Omega+i\delta,\Omega+\omega+i\delta,\omega+i\delta)
    -f(\Omega+i\delta,\Omega+\omega+i\delta,\omega-i\delta)\big]\\
     +n_B(\omega+\Omega)\big[ f(\Omega+i\delta,\Omega+\omega+i\delta,\omega-i\delta)
     -f(\Omega+i\delta,\Omega+\omega-i\delta,\omega-i\delta) \big]\Big\},
\end{multline}
and rearranging we have that
\begin{multline}
\lim_{i\Omega_n\rightarrow
\Omega+i\delta}\frac{1}{\hbar\beta}\sum_{z_n} f(i\Omega_n,z_n)=
\frac{1}{2\pi i}{\mathscr P}\int_{-\infty}^\infty {\rm
    d}\omega\,\\
\times\Big\{\left[n_B(\omega+\Omega)-n_{\rm
    B}(\omega)\right] f(\Omega+i\delta,\Omega+\omega+i\delta,\omega-i\delta)\\+n_{\rm
B}(\omega)
f(\Omega+i\delta,\Omega+\omega+i\delta,\omega+i\delta)
     -n_B(\omega+\Omega)f(\Omega+i\delta,\Omega+\omega-i\delta,\omega-i\delta)
     \Big\}.
\end{multline}
Using the notation of Ref.~\cite{flensberg1995} we now introduce the functions
\begin{equation}\label{+--+}
\Delta_1^\beta(\mp,\mp)\Delta_2^\alpha(\pm,\pm)=f(\pm,\pm)\equiv
f(\Omega+i\delta,\Omega+\omega\pm i\delta,\omega\pm i\delta),
\end{equation} 
and get
\begin{multline}
\lim_{i\Omega_n\rightarrow
\Omega+i\delta}\frac{1}{\hbar\beta}\sum_{z_n} f(i\Omega_n,z_n)=
\frac{1}{2\pi i}{\mathscr P}\int_{-\infty}^\infty {\rm
    d}\omega\,\\
\times\Big\{\left[n_B(\omega+\Omega)-n_B(\omega)\right]
\Delta_1^\beta(-,+)\Delta_2^\alpha(+,-)\\
+n_B(\omega) \Delta_1^\beta(-,-)\Delta_2^\alpha(+,+)
-n_B(\omega+\Omega)\Delta_1^\beta(+,+)\Delta_2^\alpha(-,-)
     \Big\},
\end{multline}
corresponding to Eq. (20) in Ref. \cite{flensberg1995}. 

\subsection{DC limit}\label{subsectionDC}
We now consider the DC limit,\index{conductance!, drag $G_{21}$}

\begin{equation}
\sigma_{21}^{\alpha\beta}({\boldsymbol r},{\boldsymbol
  r}')=\lim_{\Omega\rightarrow
  0}\sigma_{21}^{\alpha\beta}({\boldsymbol r},{\boldsymbol
  r}',\Omega).
\end{equation}
In Ref. \cite{flensberg1995} it was proven that 
\begin{equation}\lim_{\Omega\rightarrow 0}
\Delta_i^\alpha(+,+)=\lim_{\Omega\rightarrow 0}\Delta_i^\alpha(-,-)=0,
\end{equation} 
so that Eq. (\ref{sigma21}) now gives

\begin{multline}\label{sigma21dc+--+}
\sigma_{21}^{\alpha\beta}({\boldsymbol r},{\boldsymbol
  r}')=\frac{e^2}{h}\left(-\frac{1}{2\hbar^2}\right)
\iiiint{\rm d}{\boldsymbol r}_1
{\rm d}{\boldsymbol r}_2
{\rm d}{\boldsymbol r}_1'
{\rm d}{\boldsymbol r}_2'\,U_{12}({\boldsymbol r}_1,{\boldsymbol r}_2) U_{12}({\boldsymbol
 r}_1',{\boldsymbol r}_2')\\
\times\lim_{\Omega\rightarrow 0}{\mathscr P}\int_{-\infty}^\infty {\rm
    d}\omega\,\frac{n_B(\omega+\Omega)-n_{\rm
    B}(\omega)}{\Omega}\Delta_1^\beta(-,+)\Delta_2^\alpha(+,-),
\end{multline}
which simplifies to
\begin{multline}\label{sigma21dc}
\sigma_{21}^{\alpha\beta}({\boldsymbol r},{\boldsymbol r}')
=\frac{e^2}{h}\left(-\frac{1}{2\hbar^2}\right)
\iiiint{\rm d}{\boldsymbol r}_1
{\rm d}{\boldsymbol r}_2
{\rm d}{\boldsymbol r}_1'
{\rm d}{\boldsymbol r}_2'\,U_{12}({\boldsymbol r}_1,{\boldsymbol r}_2) U_{12}({\boldsymbol
 r}_1',{\boldsymbol r}_2')\\
\times{\mathscr P}\int_{-\infty}^\infty {\rm
    d}\omega\,\frac{\partial n_B(\omega)}{\partial \omega}{\Delta}_1^\beta({\boldsymbol
r}',{\boldsymbol
  r}_1,{\boldsymbol
  r}_1',0,-\omega-i\delta,-\omega+i\delta)\\\times{\Delta}_2^\alpha({\boldsymbol r},{\boldsymbol
  r}_2,{\boldsymbol
  r}_2',0,\omega+i\delta,\omega-i\delta).
\end{multline}
This is the generalization of Eq.~(22) in Ref.~\cite{flensberg1995} to also include the case of broken translation symmetry.

%% file: wick.tex
\chapter{Sub-systems of non-interacting particles}\label{chap:wick}

Until this point no assumptions have been made about the unperturbed
eigenstates of the subsystems; they could be complicated many-particle
states or more simple states. Assuming that we have Fermi liquids so
that we can consider sub-systems with quadratic Kamiltonians
\begin{equation}
\hat{\mathscr
K}=\sum_\lambda
\left(\varepsilon_\lambda-\mu\right)\hat{c}_{\lambda}^\dagger
\hat{c}_{\lambda}=\sum_\lambda {\mathscr E}_\lambda \hat{c}_{\lambda}^\dagger \hat{c}_{\lambda},
\end{equation}
we can immediately apply Wick's theorem to the three-particle Green
function, Eq. (\ref{three-particle}). After that we will make an
eigenstate expansion and follow the same line as in the Lehmann
representation. However, the general analysis of poles {\it etc.} can
be carried over to this problem. \index{representation!, Lehmann}

\section{Wick's theorem}

\begin{figure}
\begin{center}
\epsfig{file=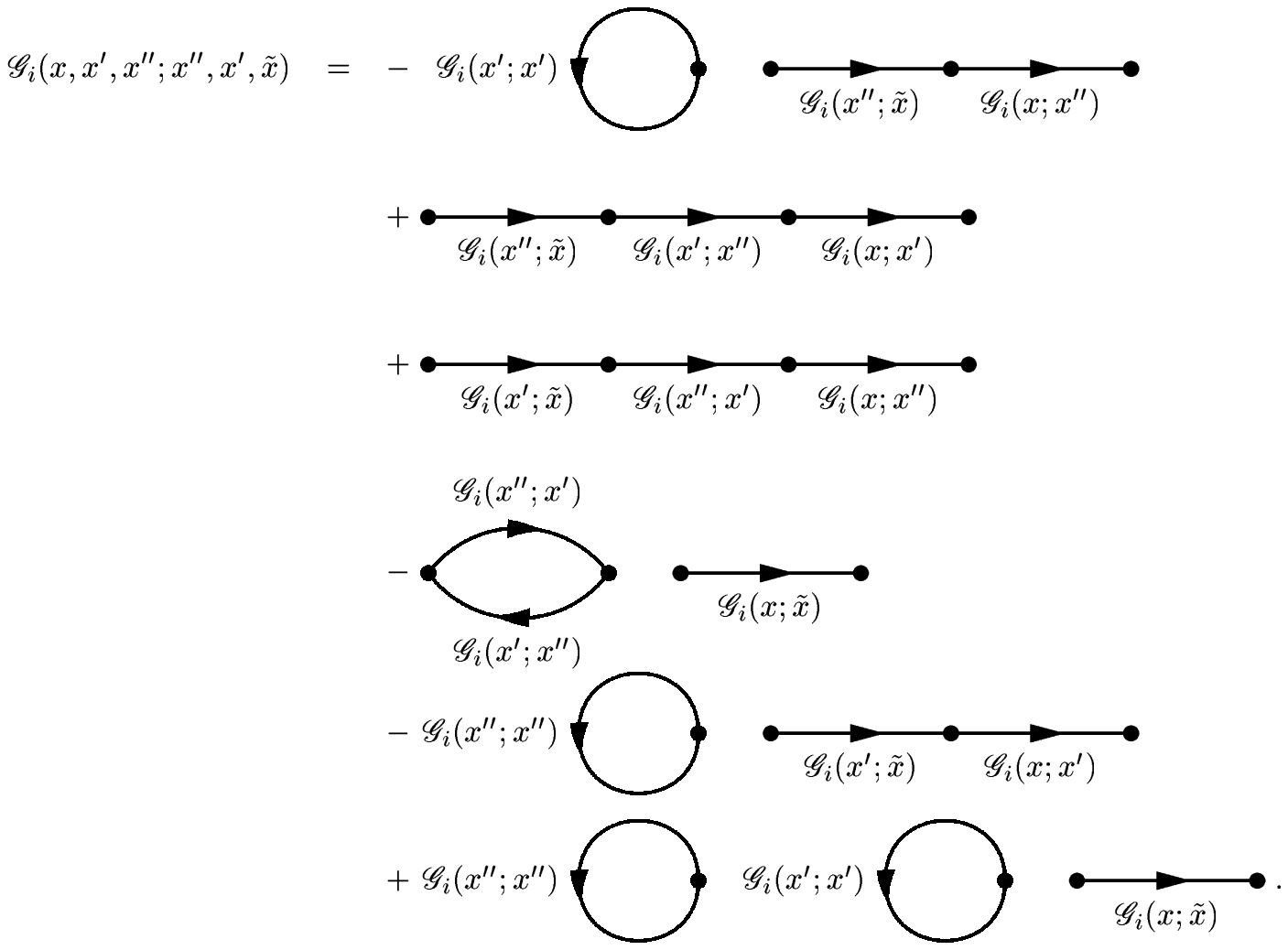, width=0.99\columnwidth}
\end{center}
\caption[Wick's theorem applied to three-body Green function]{Diagrammatic representation of Wick's theorem applied to the three-body Green function, see Eq.~(\ref{wick->G}).}
\label{3body}
\end{figure}

Applying Wick's theorem\index{Wick's theorem}~\cite{mahan}
\begin{subequations}
\begin{equation}
{\mathscr G}_i(x_1,x_2,x_3;y_3,y_2,y_1)=\left|
\begin{array}{ccc}
{\mathscr G}_i(x_1;y_1)&{\mathscr G}_i(x_1;y_2)&{\mathscr G}_i(x_1;y_3)\\
{\mathscr G}_i(x_2;y_1)&{\mathscr G}_i(x_2;y_2)&{\mathscr G}_i(x_2;y_3)\\
{\mathscr G}_i(x_3;y_1)&{\mathscr G}_i(x_3;y_2)&{\mathscr G}_i(x_3;y_3)
\end{array}\right|,
\end{equation}
where the single-particle Green function is given by

\begin{equation}
{\mathscr G}_i(x_1;y_1)=-\big<\hat{T}_\tau\big\{\hat{\psi}_i(x_1)\hat{\psi}_i^\dagger(y_1)\big\}
\big>,
\end{equation}
\end{subequations}
we get

\begin{multline}\label{wick->G}
{\mathscr G}_i(x,x',x'';x'',x',\tilde{x})=\\
-{\mathscr G}_i(x;x''){\mathscr G}_i(x';x'){\mathscr G}_i(x'';\tilde{x})
+{\mathscr G}_i(x;x'){\mathscr G}_i(x';x''){\mathscr G}_i(x'';\tilde{x})\\
+{\mathscr G}_i(x;x''){\mathscr G}_i(x'';x'){\mathscr G}_i(x';\tilde{x})
-{\mathscr G}_i(x;\tilde{x}){\mathscr G}_i(x';x''){\mathscr G}_i(x'';x')\\
-{\mathscr G}_i(x;x'){\mathscr G}_i(x';\tilde{x}){\mathscr G}_i(x'';x'')
+{\mathscr G}_i(x;\tilde{x}){\mathscr G}_i(x';x'){\mathscr G}_i(x'';x'').
\end{multline}
Only the $2$nd and $3$rd terms
correspond to connected diagrams, see Fig. \ref{3body}, so that Eq.~(\ref{three-particle}) becomes

\begin{multline}\label{DeltaGGG}
{\Delta}_i^\alpha(x,x',x'')=\frac{\hbar}{2mi}\left\{{\boldsymbol \nabla}_{\boldsymbol
r}^\alpha-{\boldsymbol\nabla}_{\tilde{\boldsymbol r}}^\alpha\right\}\\
\times\big[{\mathscr G}_i(x;x'){\mathscr G}_i(x';x''){\mathscr G}_i(x'';\tilde{x})+{\mathscr G}_i(x;x''){\mathscr G}_i(x'';x'){\mathscr G}_i(x';\tilde{x})\big]
\Big|_{\tilde{x}=x}.
\end{multline}
Indeed, since $\tilde{x}=x$ the sequence of the three single-particle propagators are connected so that they form a ``triangle'' as also indicated in Fig.~\ref{feynman}.

\section{Eigenstate expansion}

In the basis of the eigenstates of the Schr\"{o}dinger equation, suppressing the subsystem index $i$,
\begin{equation}
\hat{\mathscr H}({\boldsymbol r}) \phi_{\lambda} ({\boldsymbol
  r})=\varepsilon_\lambda\phi_{\lambda}({\boldsymbol
  r}),
\end{equation}
the single-particle Green function takes the form \cite{fetter}

\begin{equation}
{\mathscr G}(x_1;x_2)=\sum_{\lambda} \phi_{\lambda}^*({\boldsymbol
r}_2)\phi_{\lambda}({\boldsymbol r}_1){\mathscr G}_{\lambda}(\tau_1;\tau_2).
\end{equation}
Here, the new Green function is given by

\begin{multline}
{\mathscr G}_{\lambda}(\tau_1;\tau_2)=-e^{-\varepsilon_\lambda(\tau_1-\tau_2)/\hbar}\\\times
\left\{\Theta(\tau_1-\tau_2)\left[1-n_F({\mathscr E}_\lambda)\right]-\Theta(\tau_2-\tau_1)n_F({\mathscr E}_\lambda)\right\},
\end{multline}
and $n_F$ is the Fermi function.

\section{Matrix elements}\label{sectionME}
Introducing the Fourier transform
\begin{equation}
{\mathscr G}_\lambda(ik_n) = \int_0^{\hbar\beta}{\rm
  d}\left(\tau_1-\tau_2\right)\,e^{ik_n\left(\tau_1-\tau_2\right)}{\mathscr G}_{\lambda}(\tau_1;\tau_2)=\frac{1}{ik_n-{\mathscr E}_\lambda/\hbar},
\end{equation}
with fermionic Matsubara frequencies \index{Matsubara!, frequencies}
\begin{equation}
ik_n=(2n+1)\pi/\hbar\beta,\qquad n=0,\pm 1,\pm 2,\ldots,
\end{equation} 
we get
\begin{equation}
{\mathscr G}(x_1;x_2)=\sum_{\lambda} \phi_{\lambda}^*({\boldsymbol
r}_2)\phi_{\lambda}({\boldsymbol r}_1)\frac{1}{\hbar\beta}\sum_{ik_n}e^{-ik_n(\tau_1-\tau_2)}{\mathscr G}_\lambda(ik_n).
\end{equation}
Substituting into the first term of Eq. (\ref{DeltaGGG}) we get

\begin{multline}
\left\{{\boldsymbol\nabla}_{\boldsymbol
r}^\alpha-{\boldsymbol \nabla}_{\tilde{\boldsymbol r}}^\alpha\right\}{\mathscr G}(x;x'){\mathscr G}(x';x''){\mathscr G}(x'';\tilde{x})
\Big|_{\tilde{x}=x}=\frac{1}{(\hbar\beta)^3}\,\frac{2mi}{\hbar}\\\times\sum_{\lambda_1\lambda_2\lambda_3}
J_{\lambda_3\lambda_1}({\boldsymbol r}) \rho_{\lambda_1\lambda_2}({\boldsymbol r}')\rho_{\lambda_2\lambda_3}({\boldsymbol
  r}'')\\
\times\sum_{ik_n\,ik_n'\,ik_n''}e^{(ik_n''-ik_n)(\tau-\tau'')}e^{(ik_n-ik_n')(\tau'-\tau'')}{\mathscr G}_{\lambda_1}(ik_n){\mathscr G}_{\lambda_2}(ik_n'){\mathscr G}_{\lambda_3}(ik_n''),
\end{multline}
with the matrix elements\index{matrix!, current}\index{matrix!, particle-density}
\begin{subequations}
\begin{equation}
\rho_{\lambda'\lambda}({\boldsymbol
  r})=\phi_{\lambda'}^*({\boldsymbol r})\phi_{\lambda}({\boldsymbol r}),
\end{equation}
and
\begin{eqnarray}\label{Jmatrix}
 J_{\lambda'\lambda}({\boldsymbol r})&=&\frac{\hbar}{2mi}\left\{ \phi_{\lambda'}^*({\boldsymbol
    r}){\boldsymbol\nabla}_{\boldsymbol
    r}^\alpha\phi_{\lambda}({\boldsymbol r}) -
  \left[{\boldsymbol\nabla}_{\boldsymbol r}^\alpha \phi_{\lambda'}^*({\boldsymbol
    r}) \right]\phi_{\lambda}({\boldsymbol r}) \right\}\nonumber\\
&=& \frac{\hbar}{2mi}\lim_{\tilde{\boldsymbol r}\rightarrow {\boldsymbol r}}\big\{{\boldsymbol\nabla}_{\boldsymbol r}^\alpha -{\boldsymbol\nabla}_{\tilde{\boldsymbol r}}^\alpha\big\}\phi_{\lambda'}^*(\tilde{\boldsymbol r})\phi_{\lambda}({\boldsymbol r}).
\end{eqnarray}
\end{subequations}
For the current matrix element to be non-zero there is a special constraint on the two states involved in zero magnetic field. Applying the continuity equation ${\boldsymbol \nabla}{\boldsymbol \cdot} {\boldsymbol J}({\boldsymbol r})=0$ to the matrix element ${\boldsymbol J}_{\lambda'\lambda}$ and replacing the Laplacian by essentially $\hat{T}({\boldsymbol r})=\hat{\mathscr H}({\boldsymbol r})-V({\boldsymbol r})$ we get

\begin{equation}\label{JdeltaEE'}
0={\boldsymbol\nabla}{\boldsymbol\cdot} {\boldsymbol J}_{\lambda'\lambda}(\boldsymbol r)
= \left(\frac{\hbar}{2mi}\right)\frac{2m}{\hbar^2}({\mathscr E}_{\lambda'}-{\mathscr E}_{\lambda}) \rho_{\lambda'\lambda}({\boldsymbol r}) ,
\end{equation}
from which it is seen that the two states in general have to be degenerate, {\it i.e.} have the same energy. This will be useful later on.

Proceeding in the same way, the second term of Eq. (\ref{DeltaGGG}) becomes

\begin{multline}
\left\{{\boldsymbol \nabla}_{\boldsymbol
r}^\alpha-{\boldsymbol \nabla}_{\tilde{\boldsymbol r}}^\alpha\right\}{\mathscr G}(x;x''){\mathscr G}(x';\tilde{x}){\mathscr G}(x'';x')
\Big|_{\tilde{x}=x}=\frac{1}{(\hbar\beta)^3}\,\frac{2mi}{\hbar}\\
\times \sum_{\lambda_1\lambda_2\lambda_3}
J_{\lambda_2\lambda_1}({\boldsymbol r}) \rho_{\lambda_1\lambda_3}({\boldsymbol r}'')\rho_{\lambda_3\lambda_2}({\boldsymbol
  r}')
\\\times\sum_{ik_n\,ik_n'\,ik_n''}e^{(ik_n'-ik_n)(\tau-\tau'')}e^{(ik_n''-ik_n')(\tau'-\tau'')}{\mathscr G}_{\lambda_1}(ik_n){\mathscr G}_{\lambda_2}(ik_n'){\mathscr G}_{\lambda_3}(ik_n'').
\end{multline}
The next step is to Fourier transform with respect to time.

\section{Fourier transform}

We consider Eq.~(\ref{Fourier-time-reverse}) and from the first term of Eq.~(\ref{DeltaGGG}) we get


\begin{multline}
\iint_0^{\hbar\beta}{\rm
  d}(\tau-\tau''){\rm d}(\tau'-\tau'')\,
e^{i\Omega_n(\tau-\tau'')+i\omega_n(\tau'-\tau'')}\\
\times \left\{{\boldsymbol \nabla}_{\boldsymbol
r}^\alpha-{\boldsymbol \nabla}_{\tilde{\boldsymbol r}}^\alpha\right\}{\mathscr G}(x;x'){\mathscr G}(x';x''){\mathscr G}(x'';\tilde{x})
\Big|_{\tilde{x}=x}\\
=\frac{1}{\hbar\beta}\,\frac{2mi}{\hbar} \sum_{ik_n}K({\bf r},{\bf r}',{\bf r}'',ik_n,ik_n+i\omega_n,ik_n-i\Omega_n),
\end{multline}
where we have introduced the function

\begin{multline}\label{Kdeffinition}
K({\bf r},{\bf r}',{\bf r}'',ik_n,ik_n+i\omega_n,ik_n-i\Omega_n)= \sum_{\lambda_1\lambda_2\lambda_3}J_{\lambda_3\lambda_1}({\boldsymbol r}) \rho_{\lambda_1\lambda_2}({\boldsymbol r}')\rho_{\lambda_2\lambda_3}({\boldsymbol
  r}'')
\\\times\sum_{ik_n}{\mathscr G}_{\lambda_1}(ik_n){\mathscr G}_{\lambda_2}(ik_n+i\omega_n){\mathscr G}_{\lambda_3}(ik_n-i\Omega_n).
\end{multline}
The second term is treated similarly and putting things together we arrive at


\begin{multline}\label{Delta=K+K}
{\Delta}^\alpha({\boldsymbol r},{\boldsymbol
  r}',{\boldsymbol
  r}'',i\Omega_n,i\Omega_n+i\omega_n,i\omega_n)=\frac{1}{\hbar\beta}\\\times
  \sum_{ik_n}\Big\{K({\bf
  r},{\bf r}',{\bf r}'',ik_n,ik_n+i\omega_n,ik_n-i\Omega_n) \\+ K({\bf r},{\bf r}'',{\bf r}',ik_n,ik_n-i\omega_n-i\Omega_n,ik_n-i\Omega_n)\Big\}.
\end{multline}

\section{Matsubara summation}

\begin{figure}
  \begin{center}
\begin{minipage}[c]{0.65\textwidth}
 \epsfig{file=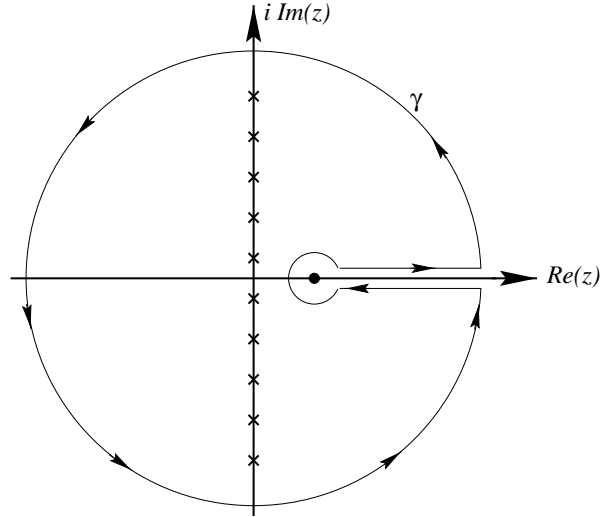, width=\textwidth,clip}
\end{minipage}\hfill
\begin{minipage}[c]{0.31\textwidth}
\caption[Complex contour for Matsubara summation]{Contours in complex plane for the evaluation of Matsubara
  sum $\frac{1}{\hbar\beta}\sum_{ik_n}{\mathscr
  G}_{\lambda_1}(ik_n)$. The Matsubara frequencies $ik_n$ are indicated by crosses
  ($\boldsymbol \times$) and the pole of ${\mathscr
  G}_{\lambda_1}(ik_n)$ by a full circle ($\bullet$).}
\label{contour-k}
\end{minipage}
\end{center}
  \end{figure}

For the evaluation of the Matsubara sum\index{Matsubara!, sum} in the function $K$ we use the identity

\begin{equation}
A^{-1}B^{-1}=(B-A)^{-1}(A^{-1}-B^{-1})
\end{equation}
so that

\begin{multline}
{\mathscr G}_{\lambda_1}(ik_n)
{\mathscr G}_{\lambda_2}(ik_n+i\omega_n)
{\mathscr G}_{\lambda_3}(ik_n-i\Omega_n)
=\frac{1}{({\mathscr E}_{\lambda_1}
-{\mathscr E}_{\lambda_3})/\hbar-i\Omega_n}
\\
\times{\mathscr G}_{\lambda_2}(ik_n+i\omega_n)\left[{\mathscr G}_{\lambda_1}(ik_n)- {\mathscr G}_{\lambda_3}(ik_n-i\Omega_n)\right].
\end{multline}
Repeating this ``trick'' we get

\begin{multline}
\frac{1}{\hbar\beta}\sum_{ik_n}{\mathscr G}_{\lambda_1}(ik_n)
{\mathscr G}_{\lambda_2}(ik_n+i\omega_n)
{\mathscr G}_{\lambda_3}(ik_n-i\Omega_n)
\\=\frac{1}{({\mathscr E}_{\lambda_1}-{\mathscr E}_{\lambda_3})\hbar-i\Omega_n}
\Bigg\{\frac{\frac{1}{\hbar\beta}\sum_{ik_n}[{\mathscr G}_{\lambda_2}(ik_n+i\omega_n) -{\mathscr G}_{\lambda_1}(ik_n)]}{({\mathscr E}_{\lambda_2}-{\mathscr E}_{\lambda_1})/\hbar-i\omega_n}
\\
+\frac{\frac{1}{\hbar\beta}\sum_{ik_n}[{\mathscr G}_{\lambda_3}(ik_n-i\Omega_n)-{\mathscr G}_{\lambda_2}(ik_n+i\omega_n)]}{({\mathscr E}_{\lambda_2}-{\mathscr E}_{\lambda_3})/\hbar-i\omega_n-i\Omega_n}
\Bigg\}.
\end{multline}
To evaluate $\frac{1}{\hbar\beta}\sum_{ik_n}{\mathscr
  G}_{\lambda_1}(ik_n)$ we
  consider the contour $\gamma$ in Fig. \ref{contour-k} and the integral

\begin{equation}
-\oint_\gamma \frac{{\rm d}z}{2\pi i}f(z) n_F(z) = -\sum_n {\rm
  Res}(f\,n_{\rm F},z_n)=\frac{1}{\hbar\beta} \sum_n f(z_n),
\end{equation}
where $f(z)={\mathscr
  G}_{\lambda_1}(z)$ and
\begin{equation}
n_F(z)=\left(e^{\hbar \beta z}+1\right)^{-1},
\end{equation}
is the Fermi function.

Since the contribution from the arches is zero and the two horizontal
parts cancel only the part encircling the pole
$\tilde{z}={\mathscr E}_{\lambda_1}/\hbar$ of $f(z)$
contributes. Therefore
\begin{equation}
\oint_\gamma \frac{{\rm d}z}{2\pi i}f(z) n_F(z) = - n_F(\tilde{z})\Res(f,\tilde{z})= - n_F(\tilde{z}),
\end{equation}
so that, removing the $\hbar$ in the definition of $n_F$,
\begin{equation}
\frac{1}{\hbar\beta}\sum_{ik_n}{\mathscr
  G}_{\lambda_1}(ik_n)= n_F({\mathscr E}_{\lambda_1}).
\end{equation}
Shifting the pole we in the same way find that 
\begin{equation}
\frac{1}{\hbar\beta}\sum_{ik_n}{\mathscr
  G}_{\lambda_2}(ik_n+i\omega_n)=n_F({\mathscr E}_{\lambda_2}),\qquad \frac{1}{\hbar\beta}\sum_{ik_n}{\mathscr
  G}_{\lambda_3}(ik_n-i\Omega_n)=n_F({\mathscr E}_{\lambda_3}).
\end{equation}
In this way we now get

\begin{multline}
\frac{1}{\hbar\beta}\sum_{ik_n}{\mathscr G}_{\lambda_1}(ik_n){\mathscr G}_{\lambda_2}(ik_n+i\omega_n){\mathscr G}_{\lambda_3}(ik_n-i\Omega_n)
=
\\
\frac{\chi_{\lambda_1\lambda_2}(i\omega_n)-\chi_{\lambda_3\lambda_2}(i\omega_n+i\Omega_n)}{({\mathscr E}_{\lambda_1}-{\mathscr E}_{\lambda_3})/\hbar-i\Omega_n},
\end{multline}
where we have introduced the susceptibility
\begin{equation}
\chi_{\lambda\lambda'}(i\omega_n)=\frac{n_F({\mathscr E}_{\lambda})-n_F({\mathscr E}_{\lambda'})}{i\omega_n+({\mathscr E}_{\lambda}-{\mathscr E}_{\lambda'})/\hbar}.
\end{equation}
Returning to Eq.~(\ref{Kdeffinition}) this means that
\begin{multline}
\frac{1}{\hbar\beta}\sum_{ik_n}K({\bf r},{\bf r}',{\bf r}'',ik_n,ik_n+i\omega_n,ik_n-i\Omega_n)=\\ \sum_{\lambda_1,\lambda_2,\lambda_3}
J_{\lambda_3,\lambda_1}({\boldsymbol r}) \rho_{\lambda_1,\lambda_2}({\boldsymbol r}')\rho_{\lambda_2,\lambda_3}({\boldsymbol
  r}'')
\frac{\chi_{\lambda_1,\lambda_2}(i\omega_n)-\chi_{\lambda_3,\lambda_2}(i\omega_n+i\Omega_n)}{({\mathscr E}_{\lambda_1}-{\mathscr E}_{\lambda_3})/\hbar-i\Omega_n},
\end{multline}
and treating the other term similarly Eq.~(\ref{Delta=K+K}) becomes
\begin{multline}
{\Delta}^\alpha({\boldsymbol r},{\boldsymbol
  r}',{\boldsymbol
  r}'',i\Omega_n,i\Omega_n+i\omega_n,i\omega_n)
= -\sum_{\lambda_1\lambda_2\lambda_3}\frac{1}{({\mathscr E}_{\lambda_1}-{\mathscr E}_{\lambda_3})/\hbar-i\Omega_n}\\\times \Bigg\{J_{\lambda_3\lambda_1}({\boldsymbol r}) \rho_{\lambda_1\lambda_2}({\boldsymbol r}')\rho_{\lambda_2\lambda_3}({\boldsymbol
  r}'')
\left[\chi_{\lambda_1\lambda_2}(i\omega_n)-\chi_{\lambda_1\lambda_2}(i\omega_n+i\Omega_n)\right]\\
-
\left[J_{\lambda_3\lambda_1}({\boldsymbol r}) \rho_{\lambda_1\lambda_2}({\boldsymbol r}')\rho_{\lambda_2\lambda_3}({\boldsymbol
  r}'')\right]^*
\left[\chi_{\lambda_1\lambda_2}(-i\omega_n)-\chi_{\lambda_1\lambda_2}(-i\omega_n-i\Omega_n)\right]
\Bigg\}.
\end{multline}
Here, we have used that $J_{\lambda_3\lambda_1}({\boldsymbol r})$ is non-zero for ${\mathscr E}_{\lambda_1}={\mathscr E}_{\lambda_3}$ only, see Eq.~(\ref{JdeltaEE'}).

\section{Analytic continuation}
At this stage we should in principle carry out a standard analytic continuation. However, this has already been done within the generalized Lehmann representation\index{representation!, Lehmann} and we can immediately take over the results obtained for $\Delta(-,+)$ and $\Delta(+-)$, see Eqs. (\ref{+--+},\,\ref{sigma21dc+--+},\,\ref{sigma21dc}). The first one \index{analytic continuation}

$${\Delta}(-,+)={\Delta}({\boldsymbol r},{\boldsymbol
  r}',{\boldsymbol
  r}'',i\delta,-\omega-i\delta,-\omega+i\delta),$$
becomes

\begin{multline}
{\Delta}(-,+)
= -\sum_{\lambda_1\lambda_2\lambda_3}\frac{1}{({\mathscr E}_{\lambda_1}-{\mathscr E}_{\lambda_3})/\hbar-i\delta}\\
\times \Bigg\{J_{\lambda_3\lambda_1}({\boldsymbol r}) \rho_{\lambda_1\lambda_2}({\boldsymbol r}')\rho_{\lambda_2\lambda_3}({\boldsymbol
  r}'')
\left[\chi_{\lambda_1\lambda_2}(-\omega+i\delta)-\chi_{\lambda_3\lambda_2}(-\omega-i\delta)\right]\\
-
\left[J_{\lambda_3\lambda_1}({\boldsymbol r}) \rho_{\lambda_1\lambda_2}({\boldsymbol r}')\rho_{\lambda_2\lambda_3}({\boldsymbol
  r}'')\right]^*
\left[\chi_{\lambda_3\lambda_2}(\omega-i\delta)-\chi_{\lambda_1\lambda_2}(\omega+i\delta)\right]
\Bigg\},
\end{multline}
or in terms of the retarded susceptibility
\begin{equation}
\chi_{\lambda\lambda'}^{r}(\omega)=\frac{n_{\rm
    FD}({\mathscr E}_{\lambda})-n_{\rm
    FD}({\mathscr E}_{\lambda'})}{\omega+({\mathscr E}_{\lambda}-{\mathscr E}_{\lambda'})/\hbar +i\delta},
\end{equation}
 we get

\begin{multline}
{\Delta}(-,+)
= -2\pi\hbar\sum_{\lambda_1\lambda_2\lambda_3} \Bigg\{J_{\lambda_3\lambda_1}({\boldsymbol r}) \rho_{\lambda_1\lambda_2}({\boldsymbol r}')\rho_{\lambda_2\lambda_3}({\boldsymbol
  r}'')
\Imag\left[\chi_{\lambda_2\lambda_1}^{r}(\omega)\right]\\
-
\left[J_{\lambda_3\lambda_1}({\boldsymbol r}) \rho_{\lambda_1\lambda_2}({\boldsymbol r}')\rho_{\lambda_2\lambda_3}({\boldsymbol
  r}'')\right]^*
\Imag\left[\chi_{\lambda_1\lambda_2}^{r}(\omega)\right]
\Bigg\}\delta({\mathscr E}_{\lambda_1}-{\mathscr E}_{\lambda_3}).
\end{multline}
Here, we have used that
${\mathscr E}_{\lambda_1}={\mathscr E}_{\lambda_3}$ to  replace
$[({\mathscr E}_{\lambda_1}-{\mathscr E}_{\lambda_3})/\hbar-i\delta]^{-1}$ by
$i\pi\hbar\delta({\mathscr E}_{\lambda_1}-{\mathscr E}_{\lambda_3}) $, see Eq.~(\ref{JdeltaEE'}). Since

\begin{equation}
\Imag[\chi_{\lambda\lambda'}^{r}(\omega)]
=-\pi\hbar\left[n_F({\mathscr E}_{\lambda})-n_F({\mathscr E}_{\lambda'})\right]
\delta(\hbar\omega+{\mathscr E}_{\lambda}-{\mathscr E}_{\lambda'}),
\end{equation}
we finally get

\begin{multline}
{\Delta}(-,+)
= 2\pi^2\hbar^2\sum_{\lambda_1\lambda_2\lambda_3} \Big\{J_{\lambda_3\lambda_1}({\boldsymbol r}) \rho_{\lambda_1\lambda_2}({\boldsymbol r}')\rho_{\lambda_2\lambda_3}({\boldsymbol
  r}'')
\delta({\mathscr E}_{\lambda_2}-{\mathscr E}_{\lambda_1}+\hbar\omega)\\
+
\left[J_{\lambda_3\lambda_1}({\boldsymbol r}) \rho_{\lambda_1\lambda_2}({\boldsymbol r}')\rho_{\lambda_2\lambda_3}({\boldsymbol
  r}'')\right]^*
\delta({\mathscr E}_{\lambda_2}-{\mathscr E}_{\lambda_1}-\hbar\omega)
\Big\}\\
\times\left[n_F({\mathscr E}_{\lambda_2})-n_F({\mathscr E}_{\lambda_1})\right]\delta({\mathscr E}_{\lambda_1}-{\mathscr E}_{\lambda_3}).
\end{multline}
In the same way we find that
\begin{equation}
{\Delta}(+,-)={\Delta}({\boldsymbol r},{\boldsymbol
  r}',{\boldsymbol
  r}'',i\delta,\omega+i\delta,\omega-i\delta)=-{\Delta}(-,+)\Big|_{\omega\rightarrow -\omega}.
\end{equation}

\section{DC drag conductance}

In the following we give the final result for the DC drag conductance, in terms of both the matrix elements introduced in Sec.~\ref{sectionME} and also in terms of spectral functions.

\subsection{Matrix element formulation}

From Eqs. (\ref{sigma->G},\,\ref{sigma21dc}) it now follows that\index{conductance!, drag $G_{21}$}

\begin{multline}\label{G21dc}
G_{21} =\frac{e^2}{h}
\iiiint{\rm d}{\boldsymbol r}_1'
{\rm d}{\boldsymbol r}_2'
{\rm d}{\boldsymbol r}_1''
{\rm d}{\boldsymbol r}_2''\,U_{12}({\boldsymbol r}_1',{\boldsymbol r}_2') U_{12}({\boldsymbol
 r}_1'',{\boldsymbol r}_2'')\\
\times{\mathscr P}\int_{-\infty}^\infty \hbar\,{\rm
    d}\omega\,\frac{{\Delta}_1(\omega,{\boldsymbol r}_1',{\boldsymbol r}_1'') {\Delta}_2(-\omega,{\boldsymbol r}_2',{\boldsymbol r}_2'')}{2kT\sinh^2(\hbar\omega/2kT)},
\end{multline}
where we have defined  ${\Delta}_i(\omega,{\bf r}_i',{\bf r}_i'')=-i\hbar^{-1}\int{\rm d}y_i\,{\Delta}(-,+)$ which is given by\index{matrix!, current}\index{matrix!, particle-density}

\begin{multline}\label{Deltai_matrix}
{\Delta}_i(\omega,{\bf r}_i',{\bf r}_i'')= -2i\pi^2\hbar\sum_{\lambda_1\lambda_2\lambda_3} \Bigg\{I_{\lambda_3\lambda_1} \rho_{\lambda_1\lambda_2}({\boldsymbol r}_i')\rho_{\lambda_2\lambda_3}({\boldsymbol
  r}_i'')\\
\times\delta({\mathscr E}_{\lambda_2}-{\mathscr E}_{\lambda_1}+\hbar\omega)
+
\left[I_{\lambda_3\lambda_1} \rho_{\lambda_1\lambda_2}({\boldsymbol r}_i')\rho_{\lambda_2\lambda_3}({\boldsymbol
  r}_i'')\right]^*
\delta({\mathscr E}_{\lambda_2}-{\mathscr E}_{\lambda_1}-\hbar\omega)
\Bigg\}\\
\times\left[n_F({\mathscr E}_{\lambda_2})-n_F({\mathscr E}_{\lambda_1})\right]\delta({\mathscr E}_{\lambda_1}-{\mathscr E}_{\lambda_3}).
\end{multline}
Here, $ I_{\lambda_3\lambda_1}\equiv\int{\rm d}y_i J_{\lambda_3\lambda_1}({\boldsymbol r}_i)$ has been introduced. There is no spatial dependence of the current matrix element $I_{\lambda_3\lambda_1}$ due two current conservation. To arrive at this result we have used that
\begin{equation}
\left(-\frac{1}{2}\right)\frac{\partial n_{B}(\omega)}{\partial \omega}=\frac{\hbar}{8kT\sinh^2(\hbar\omega/2kT)},
\end{equation}
and assumed spin-degeneracy\index{spin-degeneracy} so that
\begin{equation}
\sum_{\lambda_1\lambda_2\lambda_3}\longrightarrow 2\sum_{\lambda_1\lambda_2\lambda_3}.
\end{equation}

Eqs.~(\ref{G21dc},\,\ref{Deltai_matrix}) is the generalization of the existing theory of Coulomb drag to also include the mesoscopic regime with broken translation symmetry. This formulation constitutes the formal starting point in Refs.~[E,\,F,\,G,\,H]. We will end this section by writing Eq.~(\ref{Deltai_matrix}) in terms of spectral functions. 

\subsection{Spectral function formulation}

The aim is to write the triangle function in terms of the spectral function\index{spectral function}

\begin{multline}\label{spectralfunction}
{\mathscr A}_{\varepsilon}({\boldsymbol r}_1,{\boldsymbol r}_2)=i\left[{\mathscr G}_\varepsilon^r({\boldsymbol r}_1,{\boldsymbol r}_2) -{\mathscr G}_\varepsilon^a({\boldsymbol r}_1,{\boldsymbol r}_2)\right]\\=2\pi\sum_\lambda \phi_\lambda^*({\boldsymbol r}_1)\phi_\lambda({\boldsymbol r}_2)\delta(\varepsilon-\varepsilon_\lambda).
\end{multline}
To do that we first write out the matrix elements and re-arrange the order of the eigenfunctions
\begin{multline}
J_{\lambda_3\lambda_1}({\boldsymbol r}) \rho_{\lambda_1\lambda_2}({\boldsymbol r}')\rho_{\lambda_2\lambda_3}({\boldsymbol
  r}'')
=\left(\frac{\hbar}{2mi}\right)\lim_{\tilde{{\boldsymbol r}}\rightarrow {\boldsymbol r}}\big\{{\boldsymbol\nabla}_{\boldsymbol r}^\alpha -{\boldsymbol\nabla}_{\tilde{\boldsymbol r}}^\alpha\big\} \\
\times\phi_{\lambda_1}^*({\boldsymbol r}') \phi_{\lambda_1}({\boldsymbol r})\phi_{\lambda_2}^*({\boldsymbol r}'') \phi_{\lambda_2}({\boldsymbol r}')\phi_{\lambda_3}^*(\tilde{\boldsymbol r})\phi_{\lambda_3}({\boldsymbol r}''),
\end{multline}
and in Eq.~(\ref{Deltai_matrix}) we thus now have three pairs of eigenfunctions, three sums, but only two delta functions. We can at no cost get the third delta function by introducing $1=\int {\rm d}{\mathscr E}\,\delta({\mathscr E}-{\mathscr E}_{\lambda_2})$ into Eq.~(\ref{Deltai_matrix}). Putting things together this gives

\begin{multline}
{\Delta}_i(\omega,{\bf r}_i',{\bf r}_i'')
= -\frac{1}{4\pi}\frac{\hbar^2}{2m}\int{\rm d}y_i\int {\rm d}{\mathscr E}\lim_{\tilde{{\boldsymbol r}_i}\rightarrow {\boldsymbol r}_i}\big\{{\boldsymbol\nabla}_{{\boldsymbol r}_i}^x -{\boldsymbol\nabla}_{\tilde{\boldsymbol r}_i}^x\big\} \\
\times\Big\{{\mathscr A}_{{\mathscr E}+\hbar\omega}({\boldsymbol r}_i',{\boldsymbol r}_i){\mathscr A}_{{\mathscr E}}({\boldsymbol r}_i'',{\boldsymbol r}_i'){\mathscr A}_{{\mathscr E}+\hbar\omega}(\tilde{\boldsymbol r}_i,{\boldsymbol r}_i'')\left[n_F({\mathscr E})-n_F({\mathscr E}+\hbar\omega)\right]\\
-
\left[{\mathscr A}_{{\mathscr E}-\hbar\omega}({\boldsymbol r}_i',{\boldsymbol r}_i){\mathscr A}_{{\mathscr E}}({\boldsymbol r}_i'',{\boldsymbol r}_i'){\mathscr A}_{{\mathscr E}-\hbar\omega}(\tilde{\boldsymbol r}_i,{\boldsymbol r}_i'')\right]^*
\left[n_F({\mathscr E})-n_F({\mathscr E}-\hbar\omega)\right]
\Big\}.
\end{multline}
In the first term we make the shift ${\mathscr E}\rightarrow {\mathscr E}-\hbar\omega$ and in the second term  ${\mathscr E}\rightarrow {\mathscr E}+\hbar\omega$ and taking the limit $\tilde{\boldsymbol r}_i\rightarrow {\boldsymbol r}_i$ we get

\begin{multline}\label{Deltai_spectral}
{\Delta}_i(\omega,{\bf r}_i',{\bf r}_i'')
= \frac{1}{4\pi}\frac{\hbar^2}{2m}\int{\rm d}y_i\int {\rm d}{\mathscr E} \Big\{\left[n_F({\mathscr E}-\hbar\omega)-n_F({\mathscr E})\right]{\mathscr A}_{{\mathscr E}-\hbar\omega}({\boldsymbol r}_i'',{\boldsymbol r}_i')\\
\times[{\mathscr A}_{{\mathscr E}}({\boldsymbol r}_i',{\boldsymbol r}_i)\partial_{x_i} {\mathscr A}_{{\mathscr E}}({\boldsymbol r}_i,{\boldsymbol r}_i'')-{\mathscr A}_{{\mathscr E}}({\boldsymbol r}_i,{\boldsymbol r}_i'')\partial_{x_i} {\mathscr A}_{{\mathscr E}}({\boldsymbol r}_i',{\boldsymbol r}_i) ]-
\left[\omega\longrightarrow -\omega\right]^*
\Big\}.
\end{multline}
This result is the starting point of Ref.~[G].
It turns out to extremely useful for numerical implementations since efficient methods for obtaining the retarded Green function ${\mathscr G}^r$ in mesoscopic structures has already been developed. The original purpose of these methods was to obtain the diagonal part of the conductance matrix via the Fisher--Lee relation \cite{fisher1981} and other properties like the density-of-states which can be formulated in terms of scattering matrices \cite{gasparian1996}. For the diagonal conductance it is sufficient to know the amplitude of propagating from one boundary to the other (one lead to the other) whereas for the drag conductance the full Green function is needed since the drag is mediated by Coulomb interactions between particles located anywhere in the two subsystems. 

\section{Low temperature expansion}

At low temperatures $k_{\rm B}T\ll \mu$ we expect the same temperature dependence of the drag conductance as for the ordinary electron-electron scattering rate since the mechanism is the same. The available phase space for electron-electron scattering of course goes to zero at very low temperatures and thus the Coulomb drag conductance decreases with decreasing temperature and vanishes at $T=0$. At low temperatures the two Pauli blocking factors entering the electron-electron scattering rate gives rise to a $T^2$ dependence \cite{smith}. As as we shall see we also get at quadratic temperature dependence of the drag conductance.

We start by noting that the $\sinh^{-2}$ factor in Eq.~(\ref{G21dc}) cuts of the frequency integration and at low temperatures we can expand the triangle function to lowest order in $\omega$. This gives $\Delta\propto \omega$ and since in Eq.~(\ref{G21dc})

\begin{equation}\label{T^2}
{\mathscr P}\int_{-\infty}^\infty \hbar\,{\rm
    d}\omega\,\frac{(\hbar\omega)^2}{2kT\sinh^2(\hbar\omega/2kT)}=
\frac{(2\pi)^2}{3}(kT)^2,
\end{equation}
we get $G_{21}\propto T^2$. In the following we make the systematic expansion of the triangle function that gives rise to this result.

\subsection{Matrix element formulation}

First we note that

\begin{multline}
\lim_{T\rightarrow 0}\delta({\mathscr E}'-{\mathscr E}+\hbar\omega)[n_F({\mathscr E}-\hbar\omega)-n_F({\mathscr E})]
\simeq 
\hbar\omega\delta({\mathscr E}'-{\mathscr E})\\
\times\lim_{T\rightarrow 0} \left[-\frac{\partial n_F({\mathscr E})}{\partial {\mathscr E}}\right] +{\cal O}(\omega^2)
=\hbar\omega\delta({\mathscr E}'-{\mathscr E}) \delta({\mathscr E}) +{\cal O}(\omega^2)
,
\end{multline}
and substituting into Eq.~(\ref{Deltai_matrix}) we get\index{matrix!, current}\index{matrix!, particle-density}

\begin{multline}\label{Deltai_matrix_T^2}
{\Delta}_i(\omega,{\bf r}_i',{\bf r}_i'')= (2\pi)^2\hbar\,\hbar\omega\sum_{\lambda_1\lambda_2\lambda_3}\Imag\Big\{I_{\lambda_3\lambda_1} \rho_{\lambda_1\lambda_2}({\boldsymbol r}_i')\rho_{\lambda_2\lambda_3}({\boldsymbol
  r}_i'') \Big\}\\
\times\delta({\mathscr E}_{\lambda_1})\delta({\mathscr E}_{\lambda_2})\delta({\mathscr E}_{\lambda_3}).
\end{multline}
 As expected the low temperature behavior is governed by the states at the Fermi level where all the matrix elements are evaluated since $\delta({\mathscr E})=\delta(\varepsilon-\varepsilon_F)$. This expression is the starting point for the studies of chaotic quantum dots in Refs.~[E,\,F] and the perturbative studies of weakly disordered quantum wires in Refs.~[E,\,G,\,H].
 
\subsection{Spectral function formulation}

Similarly we note that
\begin{multline}
\lim_{T\rightarrow 0}{\mathscr A}_{{\mathscr E}-\hbar\omega}[n_F({\mathscr E}-\hbar\omega)-n_F({\mathscr E})]\\
\simeq \hbar\omega{\mathscr A}_{{\mathscr E}}\lim_{T\rightarrow 0} \left[-\frac{\partial n_F({\mathscr E})}{\partial {\mathscr E}}\right]+{\cal O}(\omega^2)=\hbar\omega{\mathscr A}_{{\mathscr E}}\delta({\mathscr E})+{\cal O}(\omega^2),
\end{multline}
and substituting into Eq.~(\ref{Deltai_spectral}) and performing the energy integral we now get\index{spectral function}

\begin{multline}\label{Deltai_spectral_T^2}
{\Delta}_i(\omega,{\bf r}_i',{\bf r}_i'')
\simeq \frac{1}{2\pi}\frac{\hbar^2}{2m}\hbar\omega \Real \Big\{{\mathscr A}_{\varepsilon_F}({\boldsymbol r}_i'',{\boldsymbol r}_i')\\
\times\int{\rm d}y_i[{\mathscr A}_{\varepsilon_F}({\boldsymbol r}_i',{\boldsymbol r}_i)\partial_{x_i} {\mathscr A}_{\varepsilon_F}({\boldsymbol r}_i,{\boldsymbol r}_i'')-{\mathscr A}_{\varepsilon_F}({\boldsymbol r}_i,{\boldsymbol r}_i'')\partial_{x_i} {\mathscr A}_{\varepsilon_F}({\boldsymbol r}_i',{\boldsymbol r}_i) ]
\Big\},
\end{multline}
with all spectral functions\index{spectral function} evaluated at the Fermi level. This expression is the starting point for the studies of quasi-ballistic 1D wires in Refs.~[E,\,G,\,H] where it is studied numerically by mapping it to a lattice.

\subsection{Scattering basis formulation}

We consider Eq.~(\ref{Deltai_matrix_T^2}) with the matrix elements calculated in the basis of scattering states,\index{scattering!, states}\index{scattering!, approach} 
\begin{equation}
(\psi_1^+,\psi_2^+,\ldots \psi_N^+,\ldots,\psi_1^-,\psi_2^-,\ldots \psi_N^-,\ldots),
\end{equation}
where $\psi_n^l$ is a scattering state incident in sub-band (mode) $n$ from either the left lead ($l=+$) or right lead ($l=-$), and $N$ is the number of propagating modes.

The current is conserved and thus we are free to evaluate the current matrix at any point we like. In particular we choose to do it within the leads where the asymptotic behavior of the scattering states are given in terms of plane waves and the scattering amplitudes entering the $2N\times 2N$ unitary scattering matrix $S$, which has the block form\index{scattering!, matrix}

\begin{equation}\label{Smatrix}
S=\begin{pmatrix}r& t'\\t&r'\end{pmatrix},
\end{equation}
with the blocks being $N\times N$ matrices. The states incident from the left lead have the form
\begin{subequations}
\label{psi_+-}
\begin{equation}
\psi_n^+({\boldsymbol r})=\left\{ 
\begin{array}{cc}
\phi_n^+({\boldsymbol r}) + \sum_{n'} r_{n'n} 
\phi_{n'}^{-}({\boldsymbol r})&, x<0,\\
\\
 \sum_{n'}  t_{n'n} \phi_{n'}^{+}({\boldsymbol r}) &, x>L,
\end{array}\right.
\end{equation}
and similarly for the states incident from the right 

\begin{equation}
\psi_{n}^{-}({\boldsymbol r})=\left\{ 
\begin{array}{cc}
\sum_{n'} t_{n'n}' \phi_{n'}^{-}({\boldsymbol r})&, x<0,\\
\\
\phi_{n}^{-}({\boldsymbol r}) + \sum_{n'}  r_{n'n}' 
\phi_{n'}^{+}({\boldsymbol r}) &, x>L.
\end{array}\right.
\end{equation}
\end{subequations}
Here
\begin{equation}\label{plane_waves}
\phi_n^\pm({\boldsymbol r})=\frac{\exp(\pm ik_n x)\Upsilon_n(y)}{\sqrt{k_n}},
\end{equation}
where the transverse wave functions are orthonormal
\begin{equation}
\int {\rm d}y\,\Upsilon_{m}^*(y)\Upsilon_{n}(y) = \delta_{mn}.
\end{equation}
The normalization in Eq.~(\ref{plane_waves}) makes all modes carry the same current.

 The previously introduced quantum number is expressed as $\lambda=(\eta,{\mathscr E})$ where $\eta=(l,n)$. Since 
\begin{equation}
{\mathscr E}=\hbar^2k_n^2/2m +\varepsilon_n -\mu,
\end{equation}
with $\varepsilon_n$ being the sub-band energy, the summations would normally be expressed as

\begin{equation}\nonumber
\sum_\lambda\longrightarrow \frac{1}{2\pi} \sum_{\eta} \int {\rm d}{\mathscr E}\, \left(\frac{\partial {\mathscr E}}{\partial k_n}\right)^{-1}\Theta({\mathscr E}+\mu-\varepsilon_n).
\end{equation}
However, the $k_n^{-1}=[\sqrt{k_n}\sqrt{k_n}]^{-1}$ factor in $\left(\partial {\mathscr E}/\partial k_n\right)^{-1} = m/\hbar^2k_n$ is already included in the normalization of the wave functions, see Eq.~(\ref{plane_waves}), and we get\index{scattering!, sum}

\begin{equation}\label{sum_scatteringstates}
\sum_\lambda\longrightarrow \frac{m}{2\pi\hbar^2} \sum_{\eta} \int {\rm d}{\mathscr E}\, \Theta({\mathscr E}+\mu-\varepsilon_n)\qquad [\;{\rm scattering\, states}!\;].
\end{equation}
From Eq.~(\ref{Deltai_matrix_T^2}) it now follows that, suppressing the subsystem index $i$,

\begin{equation}\label{Deltai_matrix_T^2_S}
{\Delta}(\omega,{\bf r}',{\bf r}'')= (2\pi)^2 \hbar \left(\frac{m}{2\pi\hbar^2}\right)^3\,\hbar\omega \Imag\Tr\big\{ I \rho({\boldsymbol r}')  \rho({\boldsymbol
  r}'')\big\},
\end{equation}
with all matrices evaluated at the Fermi level. Since the matrices are Hermitian we can now also write the result as

\begin{equation}\label{Deltai_matrix_T^2_commuatator}
{\Delta}(\omega,{\bf r}',{\bf r}'')= \frac{1}{2i}(2\pi)^2 \hbar \left(\frac{m}{2\pi\hbar^2}\right)^3
 \,\hbar\omega \Tr \big\{I\big[\rho({\boldsymbol r}') ;\rho({\boldsymbol
  r}'')\big]_-\big\},
\end{equation}
where the commutator of two matrices $M$ and $N$ is defined in the usual way $[M;N]_-=MN-NM$. 

\subsubsection[Current matrix]{Current matrix \\--- cancellation of velocity and density-of-states}
We now calculate the current matrix\index{matrix!, current}

\begin{equation}\label{Imatrixelements}
I=\begin{pmatrix}I_{++}&I_{+-}\\I_{-+}&I_{--}\end{pmatrix}, \quad\big\{I_{ll'}\big\}_{nn'} =\frac{\hbar}{2mi}\int {\rm d}y\,\lim_{\tilde{\boldsymbol r}\rightarrow {\boldsymbol r}}\big\{\partial_x -\partial_{\tilde{x}}\big\}
\{\psi_{n}^l(\tilde{\boldsymbol r}) \}^* \psi_{n'}^{l'}({\boldsymbol r}).
\end{equation}
Here, it is most convenient, but not necessary, to consider the right lead ($x>L$) for $I_{++}$ and the left lead ($x<0$) for  $I_{--}$. This gives $I_{++} =(\hbar/m)\,  t^\dagger t $ and similarly $
I_{--} =-(\hbar/m)\,  {t'}^\dagger t'$. Though slightly more complicated the off-diagonal matrices $I_{-+}=I_{+-}^\dagger$ follow in the same way so that

\begin{equation}
I=\frac{\hbar}{m}\begin{pmatrix}
t^\dagger t & -r^\dagger t'\\
-{t'}^\dagger r  & -{t'}^\dagger t'
\end{pmatrix}.
\end{equation}
Note that the dimension of $I$ changes to that of $\hbar/m$ with our deffiniton of scattering states. By a straight forward matrix multiplication and use of the unitarity of the scattering matrix it is easy to show that the matrix can be written in the compact form\index{matrix!, current}\index{matrix!, scattering}

\begin{equation}\label{I_S}
I=\frac{\hbar}{2m} \left[\tau^3- S^\dagger \tau^3 S\right],
\end{equation}
where $\tau^3$ is a block matrix with the form of the $3$rd Pauli matrix

\begin{equation}
\tau^3=\begin{pmatrix}
1 & 0\\
0 & -1
\end{pmatrix},
\end{equation}
where the blocks are $N\times N$ matrices.
As seen the velocity $\hbar k_n/m$ is canceled by the density-of-states, {\it i.e.} $[\sqrt{k_n}\sqrt{k_n}]^{-1}$ from the normalization of the wave function, see Eq.~(\ref{plane_waves}). This is  a central point in the derivation of the Landauer--B\"{u}ttiker formalism for the conductance -- essentially the diagonal parts of $I$ -- but of course also in calculation of the off-diagonal elements which are important in the context of Coulomb drag.

The compact notation in Eq.~(\ref{I_S}) will be useful when studying drag between chaotic quantum dots since in that case standard random matrix theory \cite{beenakker1997} applies immediately to the scattering matrix $S$ and thereby to $I$.

\subsubsection[Particle-density matrix]{Particle-density matrix \\--- van Hove singularities}

For the particle-density matrix\index{matrix!, particle-density} $\rho$ it is obvious that there is no cancellation of the density-of-states, {\it i.e.} 

\begin{equation}
\big\{\rho_{ll'}({\boldsymbol r})\big\}_{nn'}=\{\psi_{n}^l({\boldsymbol r}) \}^* \psi_{n'}^{l'}({\boldsymbol r}),
\end{equation}
will have a singular behavior --- van Hove singularities --- if the Fermi level is at the onset of mode $n$ or $n'$ due to the normalizations $1/\sqrt{k_n}$ and $1/\sqrt{k_{n'}}$ of the wave functions, see Eq.~(\ref{plane_waves}). In fact this will be the case for any mode for an arbitrary scattering matrix that allows for mode-mixing, see the the sum over $n'$ in Eqs.~(\ref{psi_+-}). The apparent divergence of the drag at the onset of new modes is of course a failure of the low-temperature expansion and the built-in thermal averaging in a small window around the Fermi level in Eq.~(\ref{Deltai_matrix}) removes this artifact. Though non-divergent there however will still be a peak in the drag conductance at the onset of new modes. This was also found in a theoretical study of drag between ballistic wires with a multiple number of sub-bands \cite{gurevich1998}, but here it is proven for an arbitrary scattering potential.

\subsubsection[Capacitive and point-like coupling]{Capacitive and point-like coupling \\--- no contribution to Coulomb drag}

Some very general consequences follow directly from Eq.~( \ref{Deltai_matrix_T^2_commuatator}). First we notice that the general property ${\Delta}(\omega,{\bf r}',{\bf r}')=0$ follows directly from the commutator. However, even more importantly it is seen that if we in Eq.~(\ref{G21dc}) consider a capacitive (constant) coupling,
\begin{equation}
U_{12}({\boldsymbol r}_1',{\boldsymbol r}_2')=U_{12}={\rm constant},
\end{equation} 
then $G_{21}=0$ since it is easily realized that\index{coupling!, capacitive}

\begin{multline}
\iint{\rm d}{\boldsymbol r}'{\rm d}{\boldsymbol r}'' {\Delta}(\omega,{\boldsymbol r}',{\boldsymbol r}'')\propto \hbar\omega \Tr\big\{ I\big[\int{\rm d}{\boldsymbol r}' \rho({\boldsymbol r}')\, ;\,\int{\rm d}{\boldsymbol r}'' \rho({\boldsymbol
  r}'')\big]_-\big\}=0.
\end{multline}
This conclusion also follows in concrete calculations on simple systems like {\it e.g.} ballistic wires, but for an arbitrary disorder potential this property is difficult to show explicitly without Eq.~( \ref{Deltai_matrix_T^2_commuatator}). 
The physical reason is that the electrons do not repel (or attract) each other, {\it i.e.} when two electrons interact they might stay very close to each other as well as far apart. Both situations are equally favorable with respect to the interaction energy.

If the two subsystems interact in a single point ${\boldsymbol R}$, {\it i.e.} 
\begin{equation}
U_{12}({\boldsymbol r}_1',{\boldsymbol r}_2')\propto \delta({\boldsymbol r}_1'-{\boldsymbol R})\delta({\boldsymbol r}_2'-{\boldsymbol R}),
\end{equation}
it again follows directly from Eqs. (\ref{G21dc},\,\ref{Deltai_matrix_T^2_commuatator}) that $G_{21}=0$. We note that this does not exclude neither a short ranged interaction like  
\begin{equation}
U_{12}({\boldsymbol r}_1',{\boldsymbol r}_2')\propto \delta({\boldsymbol r}_1'-{\boldsymbol r}_2'),
\end{equation}
from contributing to Coulomb drag nor a set of point-like couplings.\index{coupling!, point-like} \index{coupling!, short-range}

%% file: numerical.tex
\chapter{The discrete problem}\label{chap:numerical}

\begin{figure}
\begin{center}
\epsfig{file=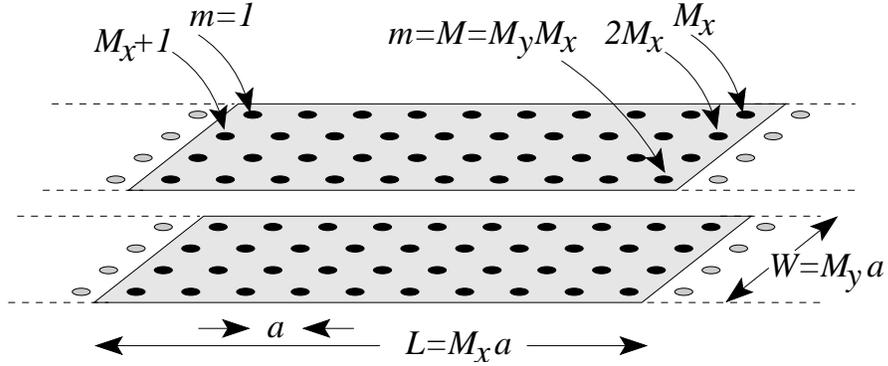, height=0.99\columnwidth,clip,angle=-90}
\end{center}
\caption[Lattice formulation]{Bi-layer system  of width $W$ and length $L$ with each 2DEG represented by an $N_x\times N_y$ square lattice with lattice constant $a$.}
\label{fig:lattice}
\end{figure}

The numerical method that we will develop in the following is --- as is usual the case --- based on a mapping of the continuous problem onto a discrete lattice problem. For each subsystem we choose a finite set of spatial points where the needed functions and quantities are calculated and evaluated. In general the set of points may form a regular, say, square lattice for two-dimensional problems or it might have some irregular and even three-dimensional form. In this work the lattice only has a mathematical meaning and the lattice constant is to be chosen small enough that the convergence towards the continuous problem is ensured. However, the lattice could also be the underlying physical lattice of the atoms like it would be the case if we were going to make a detailed study of drag between molecular wires like {\it e.g.} carbon nanotubes. In what follows we will consider two square lattices as illustrated in Fig.~\ref{fig:lattice} which is the discrete version of Fig.~\ref{system_bilayer}. The way we number the lattice points is not crucial and here we will for simplicity choose the one indicated in Fig.~\ref{fig:lattice} though at the end the overall formulation will not depend on the particular way of numbering.

\section{Trace formula for drag conductance}

We consider Eq.~(\ref{G21dc}), where both the triangle function $\Delta$ and the interaction $U_{12}$ are functions of two spatial variables and can thus be represented as two-dimensional matrices with elements

\begin{equation}
\{U_{12}\}_{m_1m_2}=U_{12}( {\boldsymbol r}_1(m_1),{\boldsymbol r}_2(m_2)),
\end{equation}

\begin{equation}
\{\Delta_{i}(\omega)\}_{m_im_i'}=\Delta_{i}( \omega,{\boldsymbol r}_i(m_i),{\boldsymbol r}_i(m_i')).
\end{equation}
Approximating the spatial integrals by sums over the lattice sites,\index{finite differences approximation}
$$\int d{\bf r}\longrightarrow (V/M)\sum_m,\qquad V_{1D}=L,\qquad V_{2D}=WL,$$
we get\index{conductance!, drag $G_{21}$}

\begin{equation}\label{G21dc_lattice}
G_{21} =\frac{e^2}{h}
\times{\mathscr P}\int_{-\infty}^\infty \hbar\,{\rm
    d}\omega\,\frac{\Tr \big\{U_{21}{\Delta}_1(\omega) U_{12} {\Delta}_2(\omega)\big\}}{2kT\sinh^2(\hbar\omega/2kT)},
\end{equation}
where we have used that ${\Delta}_i(-\omega,{\boldsymbol r}_i',{\boldsymbol r}_i'')=\Delta_i(\omega,{\boldsymbol r}_i'',{\boldsymbol r}_i')$ and introduced $U_{21}=U_{12}^T$. The ``missing'' factor $(V/M)^4$ will be incorporated into the spectral function so that ${\mathscr A}^{-1}$ will have the dimension of energy.

For simplicity we will assume the same number of sites in the two subsystems so that all matrices are $M\times M$ matrices instead of the general case with $\Delta_i$'s being $M_i\times M_i$ matrices and $U_{12}$ being an $M_1\times M_2$ matrix.

The one-dimensional case ($M_y=1$) is most illustrative and the generalization to a finite number of modes will be given afterwards.

\section{One-dimensional formulation}

For the one-dimensional problem we label the points by $m$ from $1$ to $M$ and formulate the problem in terms of $M\times M$ matrices.

\subsection{Coulomb coupling}

For the Coulomb coupling the interaction between site $m$ in one wire and site $m'$ in the other wire is given by
\begin{equation}
\big\{U_{12}\big\}_{mm'}=U_{12}(x_1'\rightarrow ma, x_2'\rightarrow m'a).
\end{equation}
In most cases $U_{12}$ will be symmetric, but {\it e.g.} a spatially dependent screening would break this symmetry. 

\subsection{Triangle function}

For the triangle function we similarly have that

\begin{equation}
\big\{\Delta_i(\omega)\big\}_{mm'}=\Delta_i(\omega,x_i'\rightarrow ma, x_i'\rightarrow m'a).
\end{equation}
It turns out to be most convenient to work with spectral functions, Eq.~(\ref{Deltai_spectral}), since the spectral function $\mathscr A$ is easy to obtain numerically whereas the eigenfunctions only follow implicitly.

Similarly to the transformation of integrals into sums we now need to transform the spatial derivative into finite differences, see {\it e.g. } Refs.~\cite{ferry,datta}. For some function $F$ this means that\index{finite differences approximation}

\begin{equation}\label{gradient_approx}
\partial_{x} F(x=ma)\longrightarrow \frac{F((m+1)a)-F(ma)}{a},
\end{equation}
which of course only becomes a good approximation when the lattice constant $a$ becomes sufficiently small. We will return to a more detailed criterion for that later.

Due to current conservation the point at which the derivative is taken can be anywhere. Applying Eq.~(\ref{gradient_approx}) to Eq.~(\ref{Deltai_spectral}) at $x=\tilde{m}a$ we get
\begin{multline}\label{Deltamatrix1D}
\big\{\Delta(\omega)\big\}_{mm'}
= \frac{1}{4\pi} \frac{\hbar^2}{2ma^2}\int{\rm d}\varepsilon 
\big\{{\mathscr A}_{\varepsilon-\hbar\omega}\big\}_{m'm} \\
\times
\big[ \big\{{\mathscr A}_{\varepsilon}\big\}_{\tilde{m} m'} 
\big\{{\mathscr A}_{\varepsilon}\big\}_{m,\tilde{m}+1} 
- \big\{{\mathscr A}_{\varepsilon}\big\}_{m\tilde{m}} 
\big\{{\mathscr A}_{\varepsilon}\big\}_{\tilde{m}+1,m'} \big]\\
\times
\big[n_F(\varepsilon-\hbar\omega)
-n_F(\varepsilon)\big]-\big(\;\omega\rightarrow-\omega\;\big)^*.
\end{multline}
In the square brackets two out of four terms have canceled when introducing the finite differences approximation.
Here, the spectral function matrix has matrix elements\index{spectral function}
\begin{equation}
\big\{{\mathscr A}_\varepsilon\big\}_{mm'} 
\equiv a\times {\mathscr A}_\varepsilon(x\rightarrow ma,x'\rightarrow m'a).
\end{equation}
The normalization by the lattice constant $a$ gives ${\mathscr A}^{-1}$ the dimension of energy --- it is a matter of choice which turns out to be convenient.

Due to current conservation we are free to sum over $\tilde m$ and divide by the number of lattice points corresponding to integrating along the wire and dividing by the length. This sum can be incorporated by a matrix product. Introducing the matrix $\Lambda$ with elements
\begin{equation}\label{Lambda}
\Lambda_{mm'}=\frac{1}{M-1}\sum_{\tilde{m}=1}^{M-1}
\delta_{m,\tilde{m}+1}\delta_{\tilde{m}m'}
-\delta_{m\tilde{m}}\delta_{\tilde{m}+1,m'}
= \frac{\pm\delta_{m,m'\pm 1}}{M-1},
\end{equation}
we get
\begin{multline}
\Delta(\omega)= \frac{1}{4\pi} \frac{\hbar^2}{2ma^2}
\int{\rm d}\varepsilon \big\{{\mathscr A}_{\varepsilon-\hbar\omega}\big\}^T
\otimes\big[ {\mathscr A}_{\varepsilon}\Lambda  {\mathscr A}_{\varepsilon} \big]
\\\times
\big[n_F(\varepsilon-\hbar\omega)-n_F(\varepsilon)\big]
-\big(\;\omega\rightarrow-\omega\;\big)^*,
\end{multline}
where $\otimes$ denotes an element-by-element multiplication, $\{X
\otimes Y\}_{nm} = X_{nm}Y_{nm}$. In Eq.~(\ref{Lambda}) we only sum over the first $M-1$ sites since Eq.~(\ref{gradient_approx}) can not be applied to the site $M$ if we insist on having $M\times M$ matrices.

\subsection{Spectral function}
The spectral function $\mathscr A$ is obtained from the retarded Green function ${\mathscr G}^r$ since from Eq.~(\ref{spectralfunction})

\begin{equation}
{\mathscr A}_{\varepsilon}=i\left[{\mathscr G}_\varepsilon^r -{\mathscr G}_\varepsilon^a\right]=i\left[{\mathscr G}_\varepsilon^r -\{{\mathscr G}_\varepsilon^r\}^\dagger\right].
\end{equation}
The problem is thus reduced to that of finding a method which provides the retarded Green function for any given scattering potential.
 
\subsection{Hamiltonian}

\begin{figure}
  \begin{center}
\begin{minipage}[c]{0.65\textwidth}
 \epsfig{file=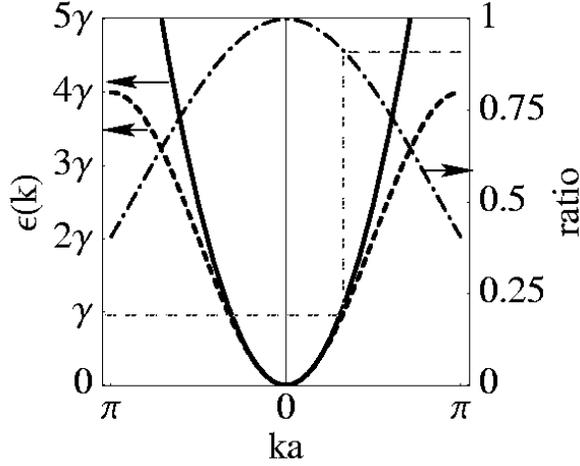, height=\textwidth,clip,angle=-90}
\end{minipage}\hfill
\begin{minipage}[c]{0.31\textwidth}
\caption[One-dimensional dispersion relations]{Plot of the parabolic (full line) and cosine (dashed line) dispersion relations, Eqs.~(\ref{cosine_band},\,\ref{parabolic}). The dash-dotted line shows the ratio of the two dispersions. For a quarter-filled band, $\varepsilon=\gamma$, the ratio is $9/\pi^2\simeq 0.91$.}
\label{Eq}
\end{minipage}
\end{center}
  \end{figure}

To get the Green function we first map the continuous Hamiltonian 
\begin{equation}
\hat{\mathscr H}=-\frac{\hbar^2}{2m}\partial_x^2 +V(x),
\end{equation}
onto a lattice. We do that by applying the finite differences approximation to the Laplacian. Eq.~(\ref{gradient_approx}) is not a unique definition and we might as well have considered the other neighboring point $m-1$. Using both neighboring points for the Laplacian we get a symmetric expression \cite{ferry,datta}\index{finite differences approximation}

\begin{equation}\label{laplacian}
\partial_{x}^2 F(x=ma)\longrightarrow \frac{F((m+1)a)+F((m-1)a)-2F(ma)}{a^2}.
\end{equation}
This means that

\begin{equation}
\hat{\mathscr H} \psi(x=ma)\longrightarrow -\gamma\Big[\psi_{m+1}+\psi_{m-1}-2\psi_m\Big]+V_m\psi_m,
\end{equation}
where $\gamma=\hbar^2/2ma^2$, $V_m=V(ma)$, and $\psi_m=\psi(ma)$. In matrix notation this means that

\begin{equation}
{\mathscr H}\psi =\varepsilon \psi,
\end{equation}
where the Hamiltonian has the matrix elements

\begin{equation}\label{Hmatrix}
{\mathscr H}_{mm'}= -\gamma\delta_{m,m'\pm 1}+(V_m+2\gamma)\delta_{mm'}.
\end{equation}
Interestingly, this is the form of a nearest neighbor tight binding Hamiltonian with hopping element $\gamma$ and a constant shift by $2\gamma$ of the on-site energies from $V_m$ to $V_m+2\gamma$. For $V_m=0$ the solution is the usual cosine band shifted by $2\gamma$,

\begin{equation}\label{cosine_band}
\varepsilon(k)=2\gamma(1-\cos ka),
\end{equation}
so that $\varepsilon(k=0)=0$. Expanding the dispersion relation near $k=0$ gives
\begin{equation}\label{parabolic}
\varepsilon(k)\simeq \gamma(ka)^2 = \frac{\hbar^2 k^2}{2m},
\end{equation}
and for energies $\varepsilon <\gamma$ the finite differences approximation provides a very accurate description of the continuous problem, see Fig.~\ref{Eq}. Making the grid finer --- $a$ smaller --- improves the accuracy and/or allows for a treatment of higher energies.\index{finite differences approximation}

\subsection{Retarded Green function}

Since we are considering an open system the Hamiltonian, Eq.~(\ref{Hmatrix}), is in principle represented by an infinite matrix. This also means that the retarded Green function

\begin{equation}\label{Gformal}
{\mathscr G}_\varepsilon^r=[\varepsilon-{\mathscr H}+i\delta]^{-1},
\end{equation}
involves the inversion of an infinite matrix. If we just truncate the Hamiltonian ${\mathscr H}\rightarrow {\mathscr H}_{\rm c}$ with ${\mathscr H}_{\rm c}$ being an $M\times M$ matrix it would correspond to a closed system -- the wire would no longer be connected to the outside through the leads. To deal with this we consider the retarded Green function of the closed system, suppressing the energy sub-script,

\begin{equation}
{\mathscr G}_{\rm c}^r=[\varepsilon-{\mathscr H}_{\rm c}+i\delta]^{-1},
\end{equation}
and the Dyson equation for the full retarded Green function

\begin{equation}
{\mathscr G}^r= {\mathscr G}_{\rm c}^r +  {\mathscr G}_{\rm c}^r\Sigma^r{\mathscr G}^r.
\end{equation}
The latter can formally be solved to give

\begin{equation}\label{GfromDyson}
{\mathscr G}^r= \big[ \{{\mathscr G}_{\rm c}^r\}^{-1}-\Sigma^r\big]^{-1}=\big[ \varepsilon-{\mathscr H}_{\rm c}-\Sigma^r+i\delta\big]^{-1},
\end{equation}
which leaves us with the job of obtaining the retarded self-energy $\Sigma^r$ due to coupling to the leads. Normally this involves approximations but for this particular problem where the electrons of each subsystem are non-interacting this is not necessary. To obtain the retarded self-energy we write the full retarded Green function as a $3\times 3$ block-matrix 

\begin{equation}\label{FullG}
{\mathscr G}^r=\begin{pmatrix}{\mathscr G}_{11}^r & {\mathscr G}_{12}^r&{\mathscr G}_{13}^r\\
{\mathscr G}_{21}^r & {\mathscr G}_{22}^r&{\mathscr G}_{23}^r\\
{\mathscr G}_{31}^r & {\mathscr G}_{32}^r&{\mathscr G}_{33}^r\end{pmatrix},
\end{equation}
where each of the three blocks along the diagonal corresponds to the left lead, the wire, and the right lead, respectively. For the Hamiltonian we can of course do the same 

\begin{equation}\label{FullH}
{\mathscr H}=\begin{pmatrix}
{\mathscr H}_{\rm L}&h &0\\
h^\dagger&{\mathscr H}_{\rm c}&h \\
0&h^\dagger &{\mathscr H}_{\rm R}\end{pmatrix},\qquad h=\begin{pmatrix}
\vdots& \vdots\\
0     & 0     & \ldots\\
-\gamma     & 0    & \ldots
\end{pmatrix},
\end{equation}
where ${\mathscr H}_{\rm L}$ and ${\mathscr H}_{\rm R}$ are the Hamiltonians of the semi-infinite leads and the coupling of the leads to the wire is described by $h$. Combining Eqs.~(\ref{Gformal},\,\ref{FullG},\,\ref{FullH}) we get

\begin{equation}
\begin{pmatrix}{\mathscr G}_{11}^r & {\mathscr G}_{12}^r&{\mathscr G}_{13}^r\\
{\mathscr G}_{21}^r & {\mathscr G}_{22}^r&{\mathscr G}_{23}^r\\
{\mathscr G}_{31}^r & {\mathscr G}_{32}^r&{\mathscr G}_{33}^r\end{pmatrix}
= \begin{pmatrix}
\{{\mathscr G}_{\rm L}^r\}^{-1} &-h &0\\
-h^\dagger&\{{\mathscr G}_{\rm c}^r\}^{-1}&-h \\
0&-h^\dagger & \{{\mathscr G}_{\rm R}^r\}^{-1}
 \end{pmatrix}^{-1}.
\end{equation}
From this result it is straight forward to obtain the sub-matrix ${\mathscr G}_{22}^r$ which is the full retarded Green function inside the wire

\begin{equation}
{\mathscr G}_{22}^r=\{ \{{\mathscr G}_{\rm c}^r\}^{-1} -h^\dagger {\mathscr G}_{\rm L}^rh-h {\mathscr G}_{\rm R}^rh^\dagger 
  \}^{-1}.
\end{equation}
Comparing to Eq.~(\ref{GfromDyson}) we read of the self-energies $\Sigma^r=\Sigma_{\rm L}^r+\Sigma_{\rm R}^r$ due to coupling to the leads

\begin{equation}
\{\Sigma_{\rm L}^r\}_{mm'}= \delta_{1m}\delta_{1m'}\gamma^2 \{{\mathscr G}_{\rm L}^r\}_{00} ,\qquad     \{\Sigma_{\rm R}^r\}_{mm'}= \delta_{Mm}\delta_{Mm'} \gamma^2\{{\mathscr G}_{\rm R}^r\}_{M+1,M+1},
\end{equation}
which are given in terms of the diagonal part of the retarded Green functions of the semi-infinite leads taken at their respective end-points.

For a semi-infinite lead (for convenience it starts at site $1$) the states are those of a semi-infinite tight-binding chain,
\begin{equation}
\psi_m^k=\frac{e^{ikma}-e^{-ikma}}{\sqrt{2}}=i\sqrt{2}\sin kma.
\end{equation}
By a suitable contour integration it follows that
\begin{equation}
\{{\mathscr G}_{\rm L,R}^r\}_{11} = \sum_k \frac{\{\psi_1^k\}^*\psi_1^k}{\varepsilon-\varepsilon(k)+i\delta}= -\frac{\exp ika}{\gamma},
\end{equation}
and thereby

\begin{equation}
\{\Sigma_{\rm L}^r\}_{mm'}= -\delta_{1m}\delta_{1m'}\gamma \exp ika ,\qquad     \{\Sigma_{\rm R}^r\}_{mm'}= -\delta_{Mm}\delta_{Mm'} \gamma\exp ika.
\end{equation}
Here, $k(\varepsilon)$ is given by the dispersion relation $\varepsilon(k)$ in Eq.~(\ref{cosine_band}).

\subsection{Low-temperature limit}
We will end by summarizing the results for the low-temperature limit where\index{spectral function}

\begin{equation}
\Delta_i(\omega)\simeq \frac{1}{2\pi} \hbar\omega\gamma \tilde\Delta_i,\qquad
\tilde\Delta_i\equiv
 \Real\Big\{ \big\{{\mathscr A}_{\varepsilon_F}^i\big\}^T
\otimes\big\{ {\mathscr A}_{\varepsilon_F}^i\Lambda  {\mathscr A}_{\varepsilon_F}^i \big\}\Big\}.
\end{equation}
Performing the $\omega$ integration in Eq.~(\ref{G21dc_lattice}) we get\index{conductance!, drag $G_{21}$}

\begin{equation}\label{G21lattice}
G_{21} =\frac{e^2}{h}(kT)^2 \frac{\gamma^2}{3} \Tr\big\{ U_{21}\tilde{\Delta}_1 U_{12} \tilde{\Delta}_2\big\}.
\end{equation}

This is the main result and it forms the
basis for all numerical work reported in Refs.~[E,\,F,\,G,\,H]. The generalization to several transverse channels is straight forward and it provides a very efficient method to investigate drag in different geometries and/or for different disorder configurations.

We note that starting from the Fisher--Lee relation\index{Fisher--Lee relation} \cite{fisher1981} the Landauer conductance $G_{ii}=\partial I_i/\partial V_i$ of the
individual wires can be expressed in a similar trace-form, see {\it e.g.} Refs.~\cite{datta,meir1992},\index{conductance!, Landauer $G_{ii}$}

\begin{equation}\label{Giilattice}
G_{ii}=\frac{2e^2}{h}{\rm Tr}
\big[\Gamma_{\rm \scriptscriptstyle L}^i
{\mathscr G}_{\varepsilon_F}^i 
\Gamma_{\rm \scriptscriptstyle R}^i
\big\{{\mathscr G}_{\varepsilon_F}^i\big\}^\dagger \big],
\end{equation} 
where the leads are described by
\begin{equation}
\Gamma_{\scriptscriptstyle p}^i
=i\big[\Sigma_{\scriptscriptstyle p}^i -
  \big\{\Sigma_{\scriptscriptstyle p}^i\big\}^\dagger\big].
\end{equation}

\section{Two-dimensional formulation}

In the presence of transverse degrees of freedom we can return to Eq.~(\ref{Deltamatrix1D}) and add the transverse integral that was absent in the 1D calculation. The net effect is to modify the matrix $\Lambda$ by a sum over the transverse direction and with the numbering in Fig.~\ref{fig:lattice} we get

\begin{multline}\label{Lambda2D}
\Lambda_{mm'}^{\rm 2D}=\frac{1}{M_y(M_x-1)}\sum_{n=1}^{N_y} \sum_{\tilde{m}=1+(n-1)M_x}^{(M_x-1)+(n-1) M_x} 
\delta_{m,\tilde{m}+1}\delta_{\tilde{m}m'}
-\delta_{m\tilde{m}}\delta_{\tilde{m}+1,m'}.
\end{multline}
However, this just means that $\Lambda^{\rm 2D}$ is an $N_y\times N_y$ block-matrix with $\Lambda$ in each block along the diagonal

\begin{equation}
\Lambda^{\rm 2D}=\diag (\overbrace{\Lambda,\Lambda,\Lambda,\Lambda\ldots}^{N_y \,{\rm elements}}).
\end{equation}
The real difference between the one- and two-dimensional systems is thus at the level of obtaining the retarded Green function where one need to modify the self-energies, see {\it e.g.} \cite{ferry,datta}. Choosing a way of numbering different from that in Fig.~(\ref{fig:lattice}) does not change the overall structure but only that of $U_{12}$ and $\Lambda$.

\section{Implementation}

The matrix formulation is readily implemented on a computer and the
accuracy can be increased simply by having more lattice points $N$ for a given size of the sample. The retarded Green function can be
obtained either by the suggestive direct matrix inversion, see {\it
  e.g.} Ref.~\cite{datta2000}, or by a recursive method, see {\it
  e.g.}  Ref.~\cite{baranger1991}. The numerics to be presented here are
based on the former.

For disordered systems the discretized set of equations are often considered a model system where the lattice constant is just set to unity and the important parameter is the band-filling, see {\it e.g.} Ref.~\cite{kramer1993}. Alternatively, one can also view Eqs.~(\ref{G21lattice},\,\ref{Giilattice}) as formulas for a tight-binding system where
$\gamma$ is a 'hopping matrix element' between different orbitals and
the local potential $V_m+2\gamma$ is the energy of the orbital
localized at site $m$.

For the implementation it has been useful to carry out tests for both
the Landauer conductance $G_{ii}$ and the drag conductance $G_{21}$.

For the Landauer conductance it is obvious to test the
numerics for various geometries which can also be solved analytically,
{\it e.g.} ballistic regime, single barrier, and double barrier
tunneling. For disordered systems the outcome of the Anderson model\index{disorder!, Anderson model}
can be directly compared to the analytic scaling results of Abrikosov
\cite{abrikosov1981} by changing the ratio of the mean free path
$\ell$ to the length of the wire $L$ from the localized regime
($\ell/L\ll 1$) to the de-localized regime ($\ell/L\gg 1$).

For the transconductance $G_{21}$ it is
difficult to obtain analytical results. However, drag between one-dimensional ballistic wires is an exception where numerics can be compared to semi-analytical
results where spatial integrals are evaluated numerically by some other method. Another test is the predicted peaking of $G_{21}$ at the
on-set of modes in either of the two wires. Especially this can be seen for systems with resonance transmission, {\it i.e.} peaks in $G_{ii}$.

%% file: wires.tex
\chapter{One-dimensional wires}\label{chap:wires}

In the following we consider one-dimensional wires which is relavant
to the situation where the width is comparable to the Fermi wave
length so that only a single mode is propagating at the Fermi level.\index{length scales!, Fermi wave length}
We first consider the ballistic regime and afterwards we study the
fluctuations of Coulomb drag in the quasi-ballistic regime. First we
however make a unitary transformation by which the current matrix $I$
becomes diagonal.\index{quantum!, wires}

\section{Unitary transformation}

We start from Eq.~(\ref{Deltai_matrix_T^2_S}) where in one dimension all matrices $I$ and $\rho$ are $2\times 2$ matrices. In the presence of time-reversal symmetry ($S=S^T$) we write the $2\times 2$ unitary scattering matrix as\index{scattering!, matrix}\index{matrix!, scattering}

\begin{equation}
  S=\begin{pmatrix}r& t\\t & r'\end{pmatrix}
  =\begin{pmatrix}\sqrt{\cal R}e^{i\theta}& \sqrt{\cal T}e^{i\phi}\\
  \sqrt{\cal T}e^{i\phi} &-\sqrt{\cal R}e^{i(2\phi-\theta)} \end{pmatrix},
  \end{equation}
where ${\cal R}=1-{\cal T}=|r|^2=|r'|^2$ has been introduced. The current matrix, Eq.~(\ref{I_S}) can now be written as\index{matrix!, current}

\begin{equation}
  I=\frac{\hbar}{2m}(\tau^3 -S^\dagger \tau^3 S)=\frac{\hbar}{2m} \begin{pmatrix}{\cal T} & -\sqrt{\cal RT}e^{i(\phi-\theta)}\\
  -\sqrt{\cal RT}e^{-i(\phi-\theta)}&-{\cal T} \end{pmatrix}.
\end{equation}
Next we employ a
  unitary transformation
  \begin{equation}
  {\cal U}=\begin{pmatrix}u & - v\\v^* & u^*\end{pmatrix},\qquad|u|^2+|v|^2=1,
  \end{equation}
that satisfies 

\begin{equation}
{\cal U}J {\cal U}^\dagger  =\frac{\hbar}{m}\sqrt{{\cal T}}\tau^3,
\end{equation}
by choosing 
\begin{equation}
|u|^2=\frac{1}{2}(1+\sqrt{\cal T}),\qquad
|v|^2=\frac{1}{2}(1-\sqrt{\cal T}),\qquad
vu^*=\frac{1}{2}\sqrt{\cal R}e^{i(\phi-\theta)}.
\end{equation}
We are free to choose the phases as 
\begin{equation}\label{phase}
v=|v|e^{i(\phi-\theta)/2},\qquad u=|u|e^{-i(\phi-\theta)/2}.
\end{equation}
This means that
\begin{equation}
\Tr\big\{ I \rho(x') \rho(x'')\big\}=\frac{\hbar}{m}\sqrt{\cal T}\Tr\big\{ \tau^3 \tilde\rho(x') \tilde\rho(x'')\big\},\qquad  \tilde\rho(x)={\cal U} \rho(x){\cal U}^\dagger,
\end{equation}
where we have inserted $1={\cal U}^\dagger {\cal U}$ three times and made use of the cyclic invariance of the trace. Since the diagonal elements of $\rho$ are real this also means that

\begin{equation}
\Imag \Tr \big\{I \rho(x') \rho(x'')\big\}=\frac{2\hbar}{m}\sqrt{\cal T} \Imag \tilde\rho_{+-}(x')\tilde\rho_{-+}(x'').
\end{equation}
From Eq.~(\ref{Deltai_matrix_T^2_S}) it now follows that in the new basis

\begin{equation}\label{Delta_new_basis}
{\Delta}(\omega,x',x'')= (2\pi)^2 \hbar \left(\frac{m}{2\pi\hbar^2}\right)^3\,\hbar\omega  \frac{2\hbar}{m}\sqrt{\cal T} \Imag \tilde\rho_{+-}(x')\tilde\rho_{-+}(x''),
\end{equation}
and with the choice of relative phase in Eq.~(\ref{phase}) we have that

\begin{multline}\label{rho_tilde}
\tilde\rho_{+-}(x)=\tilde\rho_{-+}^*(x)= \frac{e^{i(\phi-\theta)}}{2}\Big[\sqrt{\cal R}
  \big[\rho_{--}(x)-\rho_{++}(x)\big]\\
+(1+\sqrt{\cal T})\rho_{+-}(x)-(1-\sqrt{\cal T})\rho_{+-}^*(x)\Big],
\end{multline}
though the phase-factor $\exp i(\phi-\theta)$ has no importance in the context of Eq.~(\ref{Delta_new_basis}). In the following we will apply Eq.~(\ref{Delta_new_basis}) in our studies of both ballistic and disordered wires.

\section{Ballistic regime}

In the ballistic regime we have eigenstates given by
\begin{equation}\label{psi_ballistic1D}
\psi^\pm(x)= k^{-1/2}\exp[\pm i k x],\qquad {\mathscr E}=\hbar^2k^2/2m-\mu,
\end{equation}
where the normalization is that of scattering states.
In order to calculate the triangle function there are several possibilities: starting from Eq.~(\ref{psi_ballistic1D}) we may either {\it i)} calculated $I$ and $\rho$ and then use Eq.~(\ref{Deltai_matrix}) or {\it ii)} calculate the spectral function ${\mathscr A}_{\mathscr E}$ and use Eq.~(\ref{Deltai_spectral}). It can be checked that both routes of course give the same result.
 Here, we will only consider the low-temperature regime and use the approach of scattering states, Eq.~(\ref{Deltai_matrix_T^2_S}), as also suggested by the normalization in Eq.~(\ref{psi_ballistic1D}).

For the ballistic regime ${\cal T}=1$ and it is thus easy to utilize the unitary transformation.
 Eq.~(\ref{rho_tilde}) gives, omitting the phase-factor,

\begin{equation}
\tilde\rho_{+-}(x)=\tilde\rho_{-+}^*(x)= \rho_{+-}(x)=k_F^{-1}e^{i 2k_F x},
\end{equation}
and thereby

\begin{equation}\label{Imagtilde}
\Imag \tilde\rho_{+-}(x')\tilde\rho_{-+}(x'')=k_F^{-2}\sin 2k_F (x''-x').
\end{equation}
Using Eqs.~(\ref{Delta_new_basis},\,\ref{Imagtilde}) we get

\begin{equation}\label{Delta_sin}
{\Delta}(\omega,x',x'')= \frac{1}{4\pi}\frac{\hbar\omega}{\varepsilon_F} \frac{2m}{\hbar^2 }\sin 2k_F (x''-x'),
\end{equation}
and from Eqs.~(\ref{G21dc},\,\ref{T^2}) it now follows that in the low-temperature limit\index{conductance!, drag $G_{21}$}

\begin{multline}\label{G21dc_wires}
G_{21} =\frac{e^2}{h}
\frac{1}{12}\frac{(kT)^2}{\varepsilon_F^{\scriptscriptstyle (1)}\varepsilon_F^{\scriptscriptstyle (2)}}
\big(k_F^{\scriptscriptstyle (1)}k_F^{\scriptscriptstyle (2)}\big)^2
\iiiint{\rm d}x_1'
{\rm d}x_2'
{\rm d}x_1''
{\rm d}x_2''\,\\
\times \frac{U_{12}(x_1',x_2') U_{12}(x_1'',x_2'')}{\varepsilon_F^{\scriptscriptstyle (1)}\varepsilon_F^{\scriptscriptstyle (2)}}  \sin 2k_F^{\scriptscriptstyle (1)} (x_1''-x_1') \sin 2k_F^{\scriptscriptstyle (2)} (x_2''-x_2').
\end{multline}
Here, we have allowed for different Fermi levels of the two wires --- the drag is largest when they are equal.

\subsection{Matching Fermi levels}
For matching Fermi levels we use

\begin{multline}
\sin 2k_F (x_1''-x_1') \sin 2k_F (x_2''-x_2')\\
= \frac{1}{4}
\Big[-e^{i 2k_F (x_1''+x_2'')}e^{-i 2k_F (x_1'+x_2')} 
+ e^{-i2k_F (x_1'' -x_2'')}e^{i 2k_F (x_1'  -x_2')}   \\ 
+e^{i 2k_F (x_1'' -x_2'')}e^{-i 2k_F (x_1'-x_2')}
- e^{-i2k_F (x_1''+x_2'')}e^{i2k_F (x_1'+x_2')}\big]
\end{multline}
to introduce the Fourier-like transform
\begin{equation}
U_{12}^{\pm}(q)=k_F^2
\iint_0^L{\rm d}x_1'
{\rm d}x_2'
\,U_{12}(x_1',x_2') e^{i q (x_1'\pm x_2')} .
\end{equation}
The drag conductance can then be written as\index{conductance!, drag $G_{21}$}
\begin{equation}\label{1D:ballistic}
G_{21} =\frac{e^2}{h}
\frac{1}{12}\frac{(kT)^2}{\varepsilon_F^2}\cdot\frac{\big|U_{12}^{-}(2k_F)\big|^2-\big|U_{12}^{+}(2k_F)\big|^2}{2\varepsilon_F^2}.
\end{equation}
In the following we consider the long-wire limit.
\subsection{Long-wire limit}

For $k_FL\gg 1$ the contribution from $U_{12}^{+}(2k_F)$ goes to zero due to the rapidly oscillating exponential and we get

\begin{equation}\label{longe-wire}
G_{21} \simeq \frac{e^2}{h}
\frac{1}{12}\frac{(kT)^2}{\varepsilon_F^2}\cdot\frac{\big|U_{12}(2k_F)\big|^2}{2\varepsilon_F^2},
\end{equation}
in terms of the usual Fourier transform of the interaction potential $U_{12}(q)\equiv U_{12}^{-}(q)$. 

This relation is well-known, see {\it e.g.} Ref. \cite{hu1996}, and is easy to understand: In long translationally invariant wires the electrons can only undergo back-scattering or forward-scattering. The back-scattering from $k\sim +k_F$ to $k\sim -k_F$ and {\it vice versa} corresponds to a transferred momentum of $q\sim \pm 2k_F$ to the other layer whereas forward-scattering with $q\sim 0$ does obviously not contribute to drag since no momentum is transferred. In Fig.~\ref{feynman} each of the two interaction lines will carry momentum $2k_F$ (opposite to each other) and each line contributes a factor $U_{12}(2k_F)$ to the drag so that $G_{21}\propto U_{12}^2(2k_F)$.

\subsection{Long-range Coulomb coupling}

\begin{figure}
\begin{center}
\epsfig{file=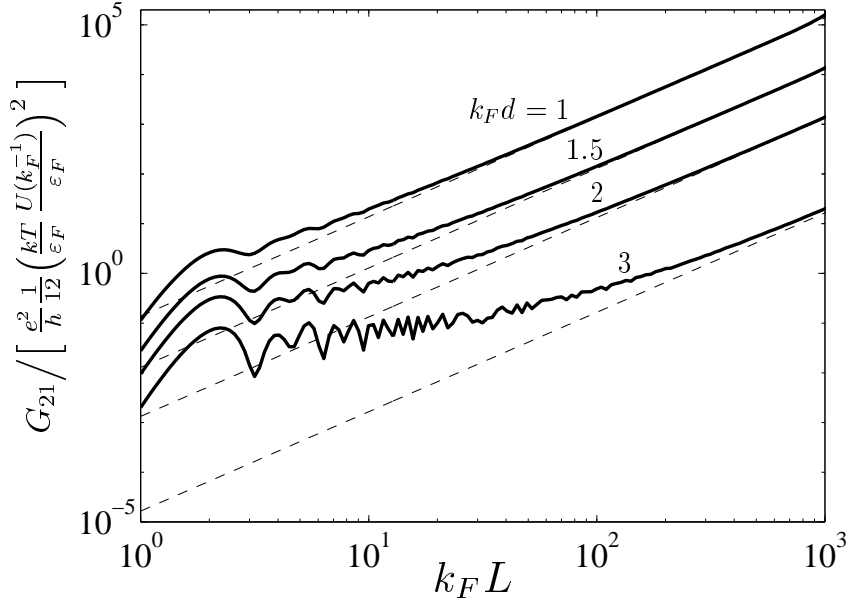, width=0.99\columnwidth,clip}
\end{center}
\caption[Drag conductance of ballistic wires]{Plot of $G_{21}$ for two ballistic wires, Eq.~(\ref{1D:ballistic}), as a function of the length $k_FL$ for different values of the separation $k_Fd$ in the case of long-range Coulomb coupling. The dashed curves show the asymptotic result, Eq.~(\ref{1D:ballistic_assymp}). Note that $U(k_F^{-1})=U(d)\,k_Fd$.}
\label{fig:ballistic}
\end{figure}

We consider a long-range Coulomb coupling\index{coupling!, long-range}
\begin{equation}
U_{12}(x_1',x_2')=U\big(\big[(x_1'-x_2')^ 2+d^2\big]^{1/2}\big),\qquad U(r)=\frac{e^2}{4\pi\epsilon_0\epsilon_r r},
\end{equation}
where $d$ is the wire separation. In this case the Fourier transform of the interaction becomes

\begin{equation}
U_{12}(q)=U(d)\,(k_Fd)\,k_F
\iint_0^L{\rm d}x
{\rm d}y
\frac{e^{i q (x- y)}}{\sqrt{(x_1'-x_2')^ 2+d^2}}
\end{equation}
which for long wires, $k_FL\gg 1$, can be approximated by
\begin{equation}
U_{12}(q)\simeq U(d)\,(k_Fd)\,( k_F L)
\int_{-\infty}^\infty{\rm d}z
\frac{e^{i q z}}{\sqrt{z^2+d^2}}.
\end{equation}
The integral can be performed analytically, see Eq.~(3.754.2) in Ref.~\cite{gradshteyn}, with the result
\begin{equation}\label{K0}
U_{12}(q)\simeq  U(d)\,(k_FL)\, (2k_Fd) \, K_0(2k_Fd)
\end{equation}
where $K_0$ is a modified Bessel function of the second kind of order
zero. This means that\index{conductance!, drag $G_{21}$}

\begin{equation}\label{1D:ballistic_assymp}
G_{21}\simeq \frac{e^2}{h} \frac{1}{6} \, \left(\frac{kT}{\varepsilon_F}\,
\frac{ U(d)}{\varepsilon_F}\,(k_FL)\, (k_Fd) \, K_0(2k_Fd)\right)^2 .
\end{equation}
Fig.~\ref{fig:ballistic}
shows a numerical evaluation of Eq.~(\ref{1D:ballistic}) along with
the asymptotic result, Eq.~(\ref{1D:ballistic_assymp}). The long wire limit was also (implicitly) studied in
Ref.~\cite{gurevich1998}, but with no justification of the $K_0$ approximation. As seen it gives the right asymptotic behavior, but under-estimates the drag conductance by even orders of magnitude for short wires. The oscillations for small values of $k_FL$ are due to the $\sin$ functions in the triangle functions, see Eq.~(\ref{Delta_sin}), and they are not a result of the numerical method that we have used. The reason that their magnitude appears so dramatic is due to the log-scale plot.

\subsection{Screened Coulomb coupling}

\begin{figure}
  \begin{center}
\begin{minipage}[c]{0.65\textwidth}
 \epsfig{file=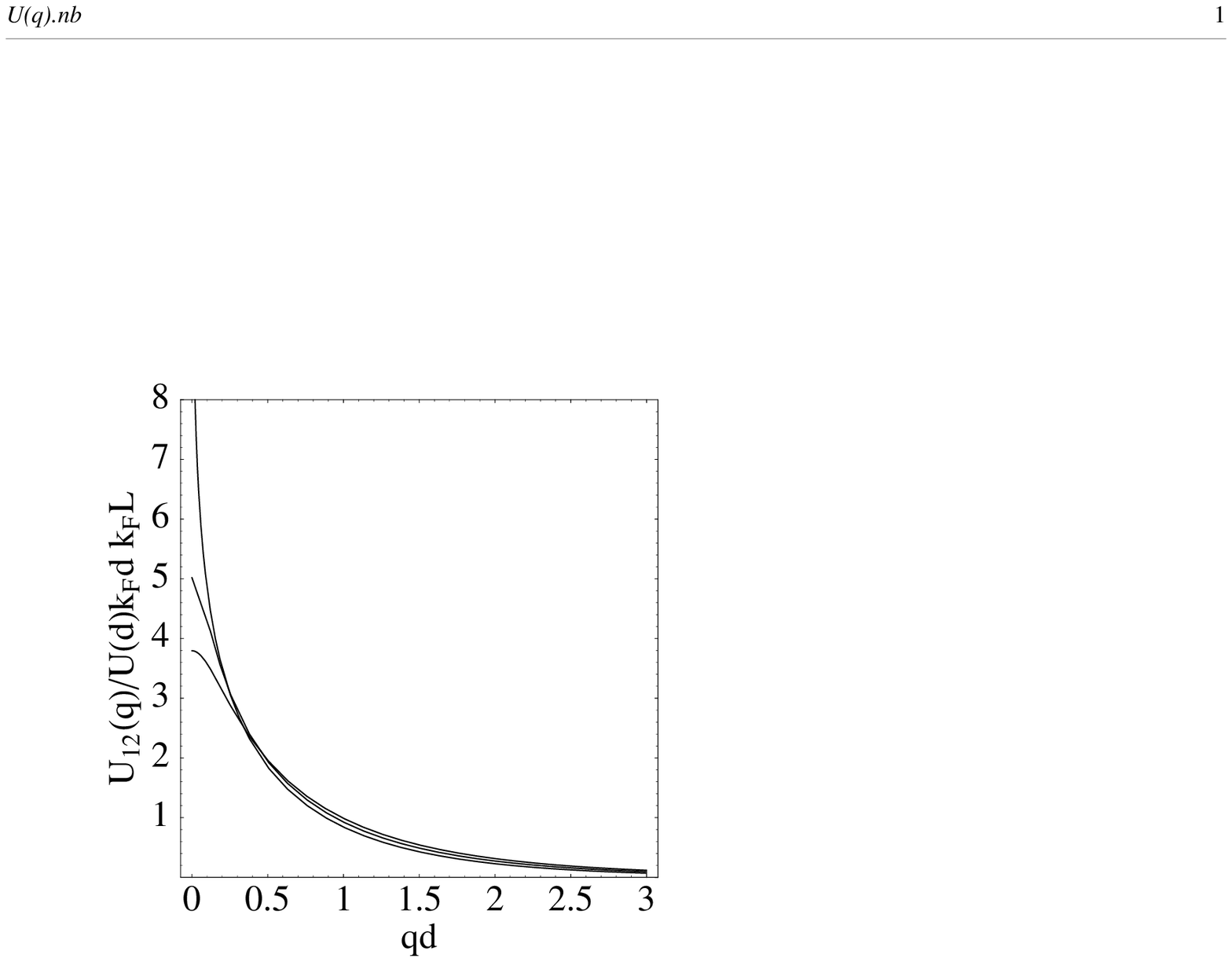, width=\textwidth,clip}
\end{minipage}\hfill
\begin{minipage}[c]{0.31\textwidth}
\caption[Fourier transformed interaction]{Plot of $U_{12}(q)$. The top curve is for long-range interaction, Eq.~(\ref{K0}), and the two other are for screened interaction, Eq.~(\ref{Uqscreened}), with $r_s/d=10$ and  $r_s/d=5$ from above. For $r_s\rightarrow \infty$ Eq.~(\ref{Uqscreened}) approaches the result of Eq.~(\ref{K0}).}
\label{fig:Uq}
\end{minipage}
\end{center}
  \end{figure}

In the following we study screened Coulomb interaction where\index{coupling!, screened}

\begin{equation}
U_{12}(x_1',x_2')=\bar{U}\big(\big[(x_1'-x_2')^ 2+d^2\big]^{1/2}\big),\qquad \bar{U}(r)=U(r)\exp(-r/r_s),
\end{equation}
with $r_s$ being the screening length.\index{length scales!, screening length} For $d\gg r_s$ drag is strongly suppressed due to screening. In the other limit $d\ll r_s$ we may neglect $d$ in the exponential screening and in the long-wire limit get

\begin{multline}\label{Uqscreened}
U_{12}(q)\simeq U(d)\,(k_FL)\, (k_F d)
\int_{-\infty}^\infty{\rm d}z
\frac{e^{i q z- |z|/r_s}}{\sqrt{z^2+d^2}}\\=U(d)\,(k_FL)\, (k_F d)
\pi\Real\{H_0(iqd+d/r_s)-Y_0(iqd+d/r_s)\},
\end{multline}
where $H_n$ is the Struve function of order $n$ and $Y_n$ is the Bessel function of the second kind of order $n$. This means that\index{conductance!, drag $G_{21}$}

\begin{multline}\label{G21_screened}
G_{21}\simeq \frac{e^2}{h} \frac{1}{24} \, \Bigg(\frac{kT}{\varepsilon_F}\,
\frac{ U(d)}{\varepsilon_F}\,(k_FL)\, (k_Fd) \\\times\pi\Real\{H_0(i2k_Fd+d/r_s)-Y_0(i2k_Fd+d/r_s)\} \Bigg)^2 .
\end{multline}
In Fig.~\ref{fig:Uq} we compare $U_{12}(q)$ to the case of long-range interaction. For moderate screening ($r_s> d$) the effect of screening is mainly to modify the short wave length behavior and since drag probes $U_{12}(2k_F)$ the effect of screening is not dramatic. However, as we shall see later the fluctuations of drag probes $U_{12}(q\sim 0)$ and then screening will make a more pronounced difference.

\subsection{Short-range Coulomb coupling}

We consider a point-like coupling \index{coupling!, short-range}

\begin{equation}
U_{12}(x_1,x_2)= {\mathscr U}(d) \delta[k_F(x_1-x_2)].
\end{equation}
This gives
\begin{equation}
U_{12}(q)=k_F L\, {\mathscr U}(d),
\end{equation}
so that in the long wire limit $k_FL\gg 1$\index{conductance!, drag $G_{21}$}
\begin{equation}\label{G21_point}
G_{21} =\frac{e^2}{h}
\frac{1}{12}\frac{1}{2}\left(\frac{kT}{\varepsilon_F}\cdot\frac{{\mathscr U}(d)}{\varepsilon_F}\,k_FL\right)^2.
\end{equation}

All three models for the interaction have the common feature that $G_{21}\propto L^2$ in the long-wire limit, see Eqs.~(\ref{1D:ballistic_assymp},\,\ref{G21_screened},\,\ref{G21_point}).

\section[Quasi-ballistic regime --- analytics]{Quasi-ballistic regime --- analytics}

We consider wires with a mean free path $\ell$ much longer than than the length of the wires, $\ell \gg L$, so that they behave as almost ideal quantum wires with the Landauer conductance of the two wires characterized by\index{conductance!, Landauer $G_{ii}$}

\begin{equation}
\big<G_{ii}\big>\sim G_{ii}(\ell\rightarrow\infty) = \frac{2e^2}{h}, \qquad \big<(\delta G_{ii})^2\big>\sim 0,\qquad \ell \gg L.
\end{equation}
Though the mesoscopic fluctuations of the Landauer conductance $G_{ii}$ are vanishing in the quasi-ballistic regime $L\ll \ell$ it turns out that the fluctuations of the drag conductance $G_{21}$ are finite and they can even exceed the mean value\index{conductance!, drag $G_{21}$}

\begin{equation}
\big<G_{21}\big>\sim G_{21}(\ell\rightarrow\infty) , \qquad \big<(\delta G_{21})^2\big>^{1/2}\sim  \big<G_{21}\big>,\qquad \ell \gg L.
\end{equation}
In the following sections we derive this relation and study it both perturbatively and numerically.

 We consider long wires, $k_FL\gg 1$, in the low-temperature regime. We are going to consider the magnitude of the fluctuations relative to the mean value and thus we for simplicity use the short-hand notation

  \begin{equation}
  G_{21}={\cal C}\int  U_{12}(x,y)U_{12}(x'y') \Delta_1(x,x')\Delta_2(y',y),
  \end{equation}
  with, see Eq.~(\ref{Delta_new_basis}), 
  \begin{equation}
  \Delta(x,x')\equiv {\rm Im}\big\{\tilde\rho_{+-}(x)\tilde\rho_{-+}(x')\big\}.
  \end{equation}
Here, $\int$ means an integral over all spatial degrees of freedom --- in this case $x$, $x'$, $y$, and $y'$. The frequency integration has been
  performed to give the $T^2$--dependence and all prefactors are
  included in ${\cal C}$, {\it i.e.} ${\cal C}\propto (e^2/h)(kT)^2$.

\subsection{Ballistic regime}
The notation is most easily illustrated by re-considering the ballistic regime. Since we only have a single sub-band we can safely incorporate the $k_F^{-1/2}$ normalization, see Eq.~(\ref{psi_ballistic1D}), into the prefactor $\cal C$. With this notation we get
  \begin{equation}
  \bar\Delta(x,x')= {\rm Im}\big\{e^{-i2k_Fx}e^{i2k_Fx'} \big\}=\sin2k_F(x'-x),
  \end{equation}
which can be compared to Eq.~(\ref{Imagtilde}). The over-line of the triangle function is used to indicate the ballistic regime. Eq.~(\ref{longe-wire}) now becomes

\begin{equation}\label{2kF}
  G_{21}(\infty) \equiv \lim_{\ell\rightarrow \infty}G_{21}(\ell)= {\cal C} \frac{1}{2} U_{12}^2(2k_F),
\end{equation}
with
  
\begin{equation}
  U_{12}(q)=\iint_0^Ldx_1dx_2\,e^{iq(x_1-x_2)}U_{12}(x_1,x_2),
\end{equation}
being the Fourier transform of the interaction.

\subsection{Mutually un-correlated disorder}

In the presence of disorder we consider the correction to the triangle function,\index{disorder!, mutually un-correlated} 

\begin{equation}
\Delta(x,x')=\bar\Delta(x,x')+\delta\Delta(x,x'),
\end{equation}
and if the disorder potentials of the two wires are mutually un-correlated ({\rm uc}) we can write
\begin{equation}
\big<\Delta_1(x,x')\Delta_2(y,y')\big>_{\rm uc} =\big<\Delta_1(x,x')\big>\big<\Delta_2(y,y')\big>,
\end{equation}
as
\begin{multline}\label{DeltaDelta_uc}
\big<\Delta_1(x,x')\Delta_2(y,y')\big>_{\rm uc}= \bar\Delta_1(x,x')\bar\Delta_2(y,y')
+\bar\Delta_1(x,x')\big< \delta\Delta_2(y,y')\big>\\
+\big<\delta\Delta_1(x,x')\big>\bar\Delta_2(y,y')+\big<\delta\Delta_1(x,x')\big>\big< \delta\Delta_2(y,y')\big>.
\end{multline}
In the quasi-ballistic regime we can make an expansion in the disorder strength. Diagrammatically we have that the average of a triangle function with only a single impurity line attached vanishes during disorder averaging since for the averages over
  disorder we assume a potential with zero mean and a short-range correlation\index{disorder!, short-range correlated}
\begin{equation}\label{disorder_uc}
\big<V_i(x)\big>=0,\qquad \big<V_i(x)V_j(x')\big>_{\rm uc}=\delta_{ij}V_0^2\delta(x-x'),\qquad \ell = (\hbar v_F/V_0)^2.
\end{equation}
Here, the back-scattering mean free path\index{length scales!, mean free path} $\ell$ is related to the disorder strength $V_0$ within the Born approximation. This result is easy to show with the aid of Fermi's golden rule and we will do this calculation in the discrete case at a later point.\label{Fermi's golden rule}

This means that all corrections to averages $\big<\Delta_i\big>$ and $\big<\Delta_i\Delta_i\big>$ will be of second order in the disorder strength, {\it i.e.} to lowest order in $1/k_F\ell$. Diagrammatically this means that there should be impurity lines connected to two points on the triangle function(s).

\subsection{Mean value}

For the mean value we note that
\begin{equation}
\big<\delta\Delta(x,x')\big>\ll \bar\Delta(x,x'),\qquad L\ll \ell,
\end{equation}
so that the corrections to the mean value are small, {\it i.e.}

\begin{equation}\label{mean2kF}
\big<G_{21}\big>_{\rm uc}\simeq  G_{21}(\infty) +{\cal O}(1/k_F\ell)= {\cal C} \frac{1}{2} U_{12}^2(2k_F)+{\cal O}(1/k_F\ell).
\end{equation}
To lowest order in $1/k_F\ell$ the effect of disorder is just to broaden the distribution of the drag conductance without shifting it. We now calculate the magnitude of these fluctuations as a function of the mean free path.

\subsection{Fluctuations}
For the fluctuations,

\begin{equation}
\big<(\delta G_{21})^2\big>=\big<G_{21}^2\big>-\big<G_{21}\big>^2,
\end{equation}
 the presence of even weak disorder makes a big difference because the broken translation invariance allows for transferred momentum different from $2k_F$.

  \begin{figure}
  \begin{center}
 \epsfig{file=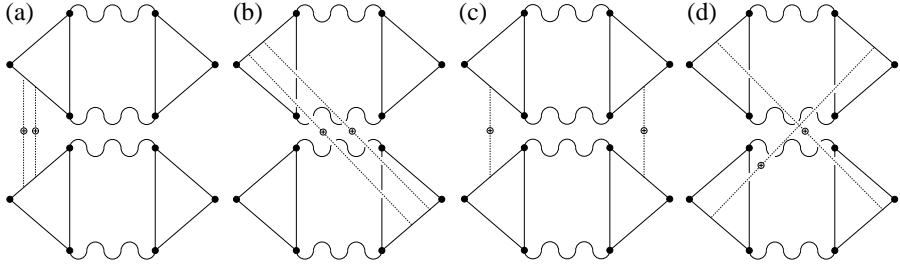, height=0.99\columnwidth,clip,angle=-90}
  \end{center}
  \caption[Diagrams for the drag conductance fluctuations]{Examples of the lowest order connected diagrams in the diagrammatic
  expansion for the fluctuations giving rise to a
  $\big<(\delta G_{21})^2\big>^{1/2}\propto V_0^2 \propto 1/k_F\ell$ dependence.
  Due to momentum conservation diagrams with only one impurity line do not
  contribute significantly to the drag fluctuations since they have to carry momentum $q=0$ corresponding to forward-scattering. The diagrams (a) and (c)
  are relevant for both correlated and un-correlated disorder whereas the diagrams (b) and (d) are relevant for correlated disorder only. }
  \label{fig:fluctuations_diagrams}
  \end{figure}

Fig.~\ref{fig:fluctuations_diagrams} shows different examples of diagrams with two impurity lines. The impurity lines carry
  momentum $2k_F$ corresponding to disorder induced back-scattering within the wires. Diagrams with a single impurity line are in principle also possible, but only if the line carries momentum $q=0$ corresponding to forward-scattering \cite{cheianov}. However, forward-scattering in the disorder channel will not contribute significantly to the fluctuations for a sufficiently long-range interaction and for a short-range interaction\index{coupling!, short-range} where forward-scattering becomes important the fluctuations are substantially suppressed when compared to the mean value. 

The main difference between diagram (a) and (c) is that in (a)
  the interaction lines in each sub-diagram must carry the same
  momentum whereas in (c) one of them can carry {\it e.g.} zero
  momentum while the other carries $2k_F$. Each impurity line is characterized by $\big<{\cal R}\big>\simeq L/\ell$ and the interaction lines by $U_{12}(q)$. For the fluctuations the contribution from diagrams of type (a) can be estimated as

\begin{equation}\label{estimate}
(a):\qquad\big<(\delta G_{21})^2\big>_{\rm uc} \sim {\cal C}^2\big<{\cal R}_1\big>^2 U_{12}^4(2k_F),
  \end{equation}
whereas for diagrams of type (c) we get
\begin{equation}\label{estimatewires}
(c):\qquad\big<(\delta G_{21})^2\big>_{\rm uc} \sim {\cal C}^2\big<{\cal R}_1\big> \big<{\cal R}_2\big> U_{12}^2(2k_F)U_{12}^2(0).
  \end{equation}
Since $U_{12}(0)\gg
  U_{12}(2k_F)$ for a long-range interaction\index{coupling!, long-range} this means that diagrams of type (c)
  give the major contribution to the fluctuations. This can be
  tested numerically by noting that the diagrams of type (a) are
  relevant to the case where wire $1$ is disordered and wire $2$ is
  either disordered or ballistic whereas diagrams of type (c) are
  only relevant to the case of both wires being disordered. Indeed, by
  numerically calculating the fluctuations for a system where one of
  the wires is ballistic and the other disordered we have found a very
  dramatic reduction of the fluctuations compared to the case of both wires being disordered.

To show the result in Eq.~(\ref{estimatewires}) in more detail we note that in the quasi-ballistic regime

\begin{multline}\label{fluc1}
\big<(\delta G_{21})^2\big>_{\rm uc} \simeq
  {\cal C}^2 \int  U_{12}(x,y)U_{12}(x',y') U_{12}(\bar{x},\bar{y})
U_{12}(\bar{x}',\bar{y}') \\\times \big<\delta\Delta_1(x,x')\delta
  \Delta_1(\bar{x},\bar{x}')\big>
\big<\delta\Delta_2(y',y)\delta\Delta_2(\bar{y}',\bar{y})\big>,
\end{multline}
corresponding to diagram (c) in Fig.~\ref{fig:fluctuations_diagrams} where all four triangle functions have an impurity line attached, {\it i.e.} two points for each of the averages.

\subsection{Lippmann--Schwinger equation}

The task is thus to calculate the correlation
  $\big<\delta\Delta_i(x,x')\delta\Delta_i(\bar{x},\bar{x}')\big>$
  to lowest order (second order) in the disorder strength. For quasi-ballistic wires, where ${\cal T}=1-{\cal R}\sim 1$, Eqs.~(\ref{Delta_new_basis},\,\ref{rho_tilde}) give

\begin{equation}
\Delta(x,x')\approx {\rm Im}\big\{\rho_{+-}(x)\rho_{-+}(x')\big\},
\end{equation}
and formally we thus have that
  \begin{multline}
  \delta\Delta(x,x')=  {\rm Im}\big\{\delta\rho_{+-}(x)\,\rho_{-+}(x')+\rho_{+-}(x)\,\delta\rho_{-+}(x')\big\}\\
={\rm Im}\big\{\delta \rho_{+-}(x)e^{i2k_Fx'}+
  e^{-i2k_Fx}\delta \rho_{+-}^*(x') \big\}.
  \end{multline}

  Using the Lippmann--Schwinger equation\index{Lippmann--Schwinger equation}, see {\it e.g.} Ref.~\cite{merzbacher}, we have to
  lowest order in the disorder strength that

\begin{equation}
  \psi_\pm(x)\simeq e^{\pm i k_Fx}+\int_0^L {\rm d}\chi\,{\mathscr G}_0^r(x,\chi)V(\chi)
  e^{\pm i k_F\chi},
\end{equation}
  where 
\begin{equation}
{\mathscr G}_0^r(x,x')=(i\hbar v_F)^{-1}e^{ik_F|x'-x|},
\end{equation}
is the unperturbed retarded Green function, {\it i.e.} that of a ballistic wire. Introducing $\delta \rho_{+-}(x)\equiv u(x)+\nu(x)$ this means that
\begin{subequations}
\begin{eqnarray}
u(x)
  &=&\frac{1}{i\hbar v_F}\Big[ \int_0^x {\rm d}\chi\,V(\chi)e^{-i 2k_F\chi}-\int_x^L {\rm d}\chi\,V(\chi)e^{-i2 k_F\chi}\Big],\\
\nu(x) &=&- \frac{1}{i\hbar v_F}\Big[ \int_0^x {\rm d}\chi\,
  V(\chi)- \int_x^L {\rm d}\chi\,V(\chi)\Big]e^{-i2 k_Fx}.
\end{eqnarray}
\end{subequations}
The disorder potential $V(x)$ is ``self-averaging'' in the sence that
\begin{equation}
\int_a^b{\rm d}x\, V(x)\sim (a-b) \big<V\big>=0
\end{equation}
and thus the last term gives a small contribution
  \begin{multline}
  \delta\Delta^{(\nu)}(x,x')=-\frac{1}{i\hbar v_F}\Bigg( \int_0^x {\rm d}\chi\,V(\chi)-
  \int_x^L {\rm d}\chi\,V(\chi)\\+ \int_0^{x'} {\rm d}\chi\,V(\chi)- \int_{x'}^L {\rm d}\chi\,V(\chi)
  \Bigg)
  \bar\Delta(x,x'),\nonumber
  \end{multline}
  which we will neglect so that $\delta \rho_{+-}(x)\approx u(x)$. 

For long wires, $k_F L\gg 1$, we neglect terms
  oscillating with $4k_F$ and get

  \begin{equation}\label{<uu>}
\big<u(x)u(\bar{x})\big>\approx 0,\qquad  \big<u(x)u^*(\bar{x})\big>=L\left(\frac{V_0}{\hbar v_F}\right)^2
  \left(1-\frac{2|x-\bar{x}|}{L}\right).
  \end{equation}
  The prefactor $L(V_0/\hbar v_F)^2=L/\ell$ can be related to the reflection coefficient of the wire. To see
  this we consider the Dyson equation for the retarded Green function to second order in the disorder

  \begin{multline}
  {\mathscr G}^r(x,x')\simeq {\mathscr G}_0^r(x,x') +\int_0^L {\rm d}\chi\,{\mathscr G}_0^r(x,\chi)V(\chi)
  {\mathscr G}_0^r(\chi,x')\\
  +\iint_0^L {\rm d}\chi{\rm d}\chi'\,{\mathscr G}_0^r(x,\chi)V(\chi){\mathscr G}_0^r(\chi,\chi')
  V(\chi'){\mathscr G}_0^r(\chi',x'),
  \end{multline}
  and from the Fisher--Lee relation\index{Fisher--Lee relation} \cite{fisher1981} we notice that 
\begin{equation}
{\cal R}=1-{\cal T}=
  1- (\hbar v_F)^2\big|{\mathscr G}^r(L,0)\big|^2.
\end{equation}
 It then follows that to second
  order in the disorder \index{length scales!, mean free path}
\begin{equation}
\big<{\cal R}\big> \simeq L\left(\frac{V_0}{\hbar v_F}\right)^2 =\frac{L}{\ell}.
\end{equation}

\begin{figure}
  \begin{center}
 \epsfig{file=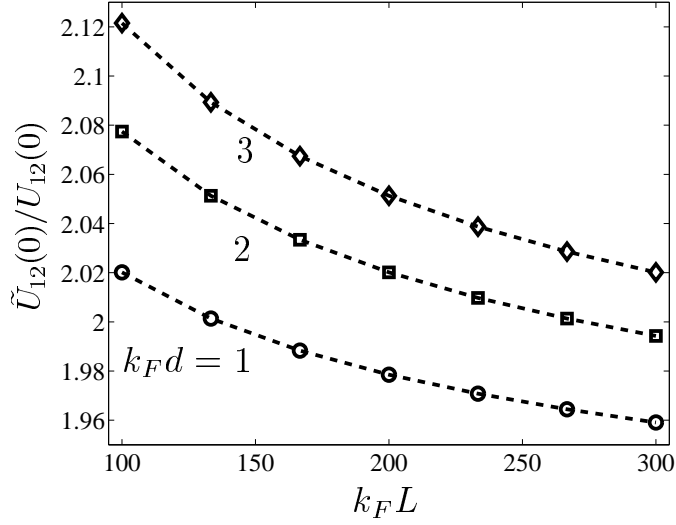, height=0.79\columnwidth,clip,angle=-90}
  \end{center}
  \caption[Comparison of  $\widetilde{U}_{12}(0)$ to $U_{12}(0)$ for a long-range interaction]{Comparison of  $\widetilde{U}_{12}(0)$ to $U_{12}(0)$ for a long-range interaction. The plot shows a numerical evaluation (indicated by the symbols) of the ratio $\widetilde{U}_{12}(0)/U_{12}(0)$ as a function of $k_FL$ for three different separations $k_Fd=1,2,3$. The dashed lines are guides to the eye.}
  \label{factor2}
  \end{figure}

This means that
\begin{multline}
  \big< \delta\Delta(x,x')\delta\Delta(\bar{x},\bar{x}')\big>
\approx \big<{\cal R}\big>\big\{f(x'-\bar{x}',x-\bar{x})+f(x-\bar{x},x'-\bar{x}')\\-f(x'-\bar{x},x-\bar{x}')- f(x-\bar{x}',x'-\bar{x}) \big\}/2,
  \end{multline}
where we have introduced the function

\begin{equation}
f(z,z')=(1-2|z|/L)\cos2k_F(z').
\end{equation}
For the fluctuations we thus get $4^2=16$ terms, but only 4 of these terms with the form
$\cos2k_F(x'-\bar{x}')\cos2k_F(y'-\bar{y}')$ are finite for long wires, $k_FL\gg 1$. Eq.~(\ref{fluc1}) now becomes

\begin{multline}
  \big<(\delta G_{21})^2\big>_{\rm uc} \simeq {\cal C}^2  \big<{\cal R}_1\big>
  \big<{\cal R}_2\big> \int  U_{12}(x,y)U_{12}(x',y')  U_{12}(\bar{x},\bar{y})U_{12}(\bar{x}',\bar{y}')\\
\times f(x-\bar{x},x'-\bar{x}')f(y-\bar{y},y'-\bar{y}'),
  \end{multline}
and introducing 
\begin{multline}
\widetilde{U}_{12}^2(0)\equiv 
\iiiint_{0}^{L}
 {\rm d}x_{1}{\rm d}x_{2}{\rm d}x_{1}'{\rm d}x_{2}'\\
\times U_{12}(x_{1},x_{2}) U_{12}(x_{1}',x_{2}')
  \Big(1-\tfrac{2|x_{1}-x_{1}'|}{L}\Big)\Big(1-\tfrac{2|x_{2}-x_{2}'|}{L}\Big)
  \end{multline}
we get
\begin{equation}\label{deltaG21C}
  \big<(\delta G_{21})^2\big>_{\rm uc} \simeq   {\cal C}^2  \frac{1}{2}\big<{\cal R}_1\big> \big<{\cal R}_2\big> U_{12}^2(2k_F)
  \widetilde{U}_{12}^2(0).
  \end{equation}
This is in close agreement with the form suggested in Eq.~(\ref{estimatewires}). To estimate the relative magnitude we divide Eq.~(\ref{deltaG21C}) by Eq.~(\ref{mean2kF}) and get

\begin{equation}\label{analytic}
\frac{\big<(\delta G_{21})^2\big>_{\rm uc}^{1/2}}{\big<G_{21}\big>_{\rm uc}}
\simeq \frac{\big[2\big<{\cal R}_1\big> \big<{\cal R}_2\big> U_{12}^2(2k_F)\widetilde{U}_{12}^2(0)\big]^{1/2}}{U_{12}^2(2k_F)}.
  \end{equation}
From Eqs.~(\ref{mean2kF},\,\ref{analytic}) it follows that the
  relative magnitude of the fluctuations is of the order 
\begin{equation}
\frac{\big<(\delta G_{21})^2\big>_{\rm uc}^{1/2}}{\big<G_{21}\big>_{\rm uc}}\sim \big<{\cal
    R}\big>\frac{U_{12}(0)}{U_{12}(2k_F)},
\end{equation}
since $\widetilde{U}_{12}(0)\sim U_{12}(0)$ within a numerical factor of order $2$, see Fig.~\ref{factor2}. This is the key result on one-dimensional wires reported the first time in Ref.~[E] and later studied in more detail in Refs.~[F,\,G,\,H].

 Even though $\big<{\cal
    R}\big>\ll 1$ the effect of long-range\index{coupling!, long-range}
  interaction may compensate this since $U_{12}(0)/U_{12}(2k_F)\gg 1$, see Fig.~\ref{fig:Uq}.  This results in relative
  fluctuations of order unity, {\it i.e.} fluctuations comparable to
  the mean value.  As we shall see in the next section,
  Eq.~(\ref{analytic}) fully accounts for our numerical results.

\subsection{Mutually correlated versus mutually un-correlated disorder}

We will now relax the assumption of mutually un-correlated disorder\index{disorder!, mutually un-correlated},
see Eq. (\ref{DeltaDelta_uc}). In fact we will study the other extreme
limit where the disorder potentials of the two wires are mutually
fully correlated\index{disorder!, mutually correlated} ($\rm c$), {\it i.e.}

\begin{equation}\label{disorder_c}
V_1(x)=V_2(x),\qquad\big<V_i(x)\big>=0,\qquad \big<V_i(x)V_j(x')\big>_{\rm c}=V_0^2\delta(x-x').
\end{equation}

  \begin{figure}
  \begin{center}
 \epsfig{file=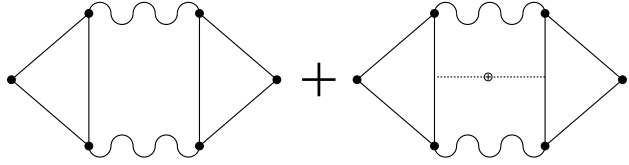, height=0.69\columnwidth,clip,angle=-90}
  \end{center}
  \caption[Diagrams for the mean drag with correlated disorder]{Diagrams for the mean drag with correlated disorder, $\big<G_{21}\big>_{\rm c}$. The first term is the ballistic limit and the second term shows an example of the lowest order diagram in the diagrammatic
  expansion of the mean value with a $ V_0^2 \propto 1/k_F\ell$ dependence. }
  \label{fig:mean_correlated}
  \end{figure}

The calculation of the corrections to the mean value in the case of fully correlated disorder is not that different from the calculation of the fluctuations in the case of un-correlated disorder. Fig.~\ref{fig:mean_correlated} shows an example of a lowest order diagram which contributes in the case of correlated disorder. Again, we consider back scattering in the disorder channel so that in the second term the impurity line carries momentum $2k_F$. This allows one of the interaction lines to carrying also $2k_F$ while the other one carries zero momentum, $q\sim 0$. For the mean value we thus get the estimate

\begin{equation}
\big<G_{21}\big>_{\rm c}\sim  U_{12}^2(2k_F)+ \big<{\cal
    R}\big>\,U_{12}(0) U_{12}(2k_F),
\end{equation}
and from Eq.~(\ref{mean2kF}) it thus follows that
\begin{equation}\label{enhancement_mean}
\big<G_{21}\big>_{\rm c}\sim \Big[1+\big<{\cal
    R}\big>\, U_{12}(0)\big/ U_{12}(2k_F)\Big]\times \big<G_{21}\big>_{\rm uc}.
\end{equation}
Since $\big<{\cal
    R}\big> \,U_{12}(0)/ U_{12}(2k_F)$ can be of order unity for long-range coupling\index{coupling!, long-range} it means that correlation of the disorder may enhance the mean value by up to a factor of order two compared to un-correlated disorder. Similarly to the case of fluctuations this effect is due to a combination of back scattering in the disorder channel and forward scattering in the Coulomb channel. The case of correlated disorder was recently studied for bi-layer systems where it was also found to enhance the mean value \cite{gornyi1999}. 

For the fluctuations the calculation can be carried out using simple diagrammatic arguments.
We again consider the diagrams in Fig.~\ref{fig:fluctuations_diagrams}.
 For correlated disorder both
  of the diagrams (c) and (d) contribute equally whereas for
  un-correlated disorder only the diagram (c) is relevant. More
  generally, for each topologically different diagram contributing in
  the case of un-correlated disorder there are two similar diagrams
  contributing equally in case of correlated disorder. Of course there
  are also other possible diagrams in case of correlated disorder but to
  lowest order in $1/k_F\ell$ this means that

  \begin{equation}
  \frac{\big<(\delta G_{21})^2\big>_{\rm c}}{\big<(\delta G_{21})^2\big>_{\rm uc}}
  \simeq \frac{\big<a\big>+\overbrace{\big<b\big>}^{=\left<a\right>}+\big<c\big>+\overbrace{\big<d\big>}^{=\left<c\right>}}{\big<a\big>+\big<c\big>}=2,
  \end{equation}
  where $\big<a\big>$, $\big<b\big>$, $\big<c\big>$, and $\big<d\big>$ refer symbolically to the
  diagrams in Fig.~\ref{fig:fluctuations_diagrams} averaged over disorder. This means that the
  fluctuations in the case of correlated disorder will be enhanced by a
  factor of $\sqrt{2}$ compared to the case of un-correlated disorder,
 \begin{equation}\label{sqrt2}
\big<(\delta G_{21})^2\big>_{\rm c}^{1/2}\simeq \sqrt{2}\times \big<(\delta G_{21})^2\big>_{\rm uc}^{1/2}.
  \end{equation}
Though these arguments may at this point seem simplistic, the result turns out to be correct as we shall see later when comparing to numerics.

\section[Quasi-ballistic regime --- numerics]{Quasi-ballistic regime --- numerics}

  The study of the statistical properties of disordered systems is an
  interesting example where Eq.~(\ref{G21lattice}) can be applied to a
  large ensemble of different disorder configurations. For the disorder we need a way to represent a short-range correlated disorder\index{disorder!, short-range correlated} potential, see Eqs.~(\ref{disorder_uc},\,\ref{disorder_c}). Here, we use the most often used model; the Anderson model \cite{anderson1958} which the first time was introduced in the study of one-dimensional localization.

\subsection{Anderson model}
In the Anderson model\index{disorder!, Anderson model} with
  diagonal disorder \cite{anderson1958,kramer1993} the site potentials $V_{ni}$ at site $n$ of a
  given wire $i$ are statistically independent with each site energy taken
  from a uniform distribution of width $W$ centered around zero, {\it i.e.}\index{disorder!, uniform distribution}

\begin{equation}\label{P(V)}
p(V_{ni})=\Theta(W/2-V_{ni})/W.
\end{equation}
This means that
\begin{equation}
\big<V_{ni}\big>=0,\qquad \big<V_{ni}V_{n'j} \big>_{\rm c}=\delta_{nn'}\frac{W^2}{12} ,\qquad \big<V_{ni}V_{n'j} \big>_{\rm uc}=\delta_{ij}\big<V_{ni}V_{n'j} \big>_{\rm c}.
\end{equation} 
Again, the disorder strength $W$ can be related to the back-scattering mean free path. For a tight-binding chain with $N$ sites and no disorder we have
eigenstates $\psi_n^k=N^{-1/2} \exp(i k n a)$. We calculate the rate
  for back-scattering from Fermi's golden rule\index{Fermi's golden rule}, suppressing the wire index $i$,

  \begin{equation}
  \frac{1}{\tau(k)}=\sum_{k'} \frac{2\pi}{\hbar}
  \left| \big< \psi^{k'} \big| V\big| \psi^k\big> \right|^2
  \delta(\varepsilon-\varepsilon') \frac{1- kk'/|k||k'|}{2},
  \end{equation}
  where the disorder potential $V$ is treated as a
  perturbation. The factor following the delta function restricts the $k'$ sum to those relevant for back-scattering, {\it i.e.} $k'$ opposite to $k$. For a chain with large $N$ we take the sum into an
  integral and since $ \delta(\varepsilon-\varepsilon')=\left|\hbar
    v_k\right|^{-1}[\delta(k-k')+\delta(k+k')] $ we get
  \begin{equation}
  \frac{1}{\tau(k)}=\frac{a}{N\hbar^2 \left| v_k\right|}
  \sum_{nm} V_n\,V_m .
  \end{equation}
The corresponding mean free path\index{length scales!, mean free path} is given by $\ell(k)=v_k
  /\big<{\tau^{-1}(k)}\big>$ and for the dispersion in Eq.~(\ref{cosine_band})  we get 

  \begin{equation}\label{ell_F}
  \ell=12a(4\gamma\varepsilon_F - \varepsilon_F^2)/W^2,
  \end{equation}
at the Fermi level. This result agrees with
  Ref.~\cite{kramer1993} except for the trivial constant shift of the energy by $2\gamma$.

  \subsection{Numerical results}

  We will first consider two special cases: {\it i)} both wires being
  disordered, but with $V_1$ and $V_2$ mutually un-correlated (as in
  Ref.~[E]) and {\it ii)} both wires being disordered
  and fully correlated, {\it i.e.}  $V_1=V_2$ (as first suggested in
  Ref.~\cite{gornyi1999}). Finally, {\it iii)} we consider the more general case where the two wires have partly a correlated (common) disorder potential $V_{\rm c}$ and partly an un-correlated $V_{\rm uc}$ one, \index{disorder!, mutually correlated}\index{disorder!, mutually un-correlated}\index{disorder!, partly correlated}

\begin{equation}
V_1=V_{\rm c} + V_{{\rm uc},1},\qquad V_2=V_{\rm c}+V_{{\rm uc},2},\qquad W^2 = W_{\rm c}^2+W_{\rm uc}^2,
\end{equation} 
but with the impurity strength adjusted so that the mean free path is the same as in the two special cases (the last equation).

We consider
  quarter-filled bands ($\varepsilon_F=\gamma$) and wires with $N=100$
  lattice points so that $k_FL =(\pi/3)\times 100$. The separation is
  $k_F d= 1$ and we assume long-range Coulomb
  interaction,\index{coupling!, long-range}

  \begin{equation}\label{U(r)}
  \big\{U_{12}\big\}_{nn'}=\frac{e^2}{4\pi \epsilon_0\epsilon_r\sqrt{
  (n-n')^2 a^2 +d^2}}.
  \end{equation}

\begin{figure}
  \begin{center}
\begin{minipage}[c]{0.65\textwidth}
 \epsfig{file=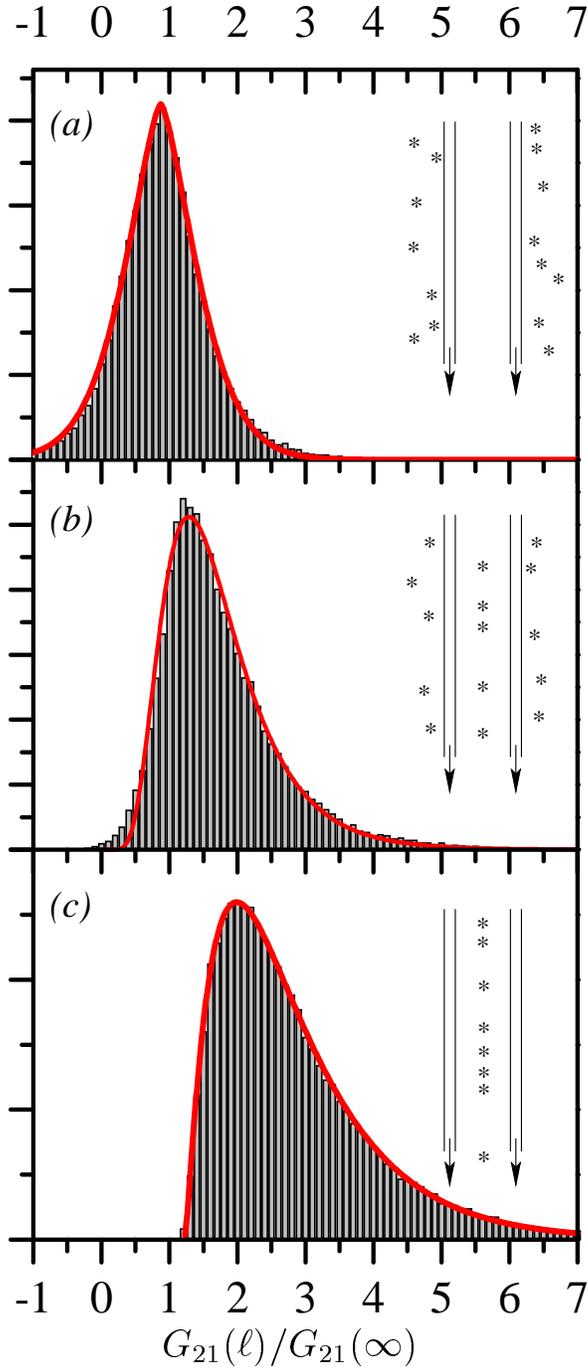, width=\textwidth,clip}
\end{minipage}\hfill
\begin{minipage}[c]{0.31\textwidth}
  \caption[Drag conductance histograms]{Histograms for $G_{21}(\ell)$ normalized by the ballistic result $G_{21}(\infty)$. The histograms are based on $>10^4$ disorder configurations and for all three histograms the mean free path is $\ell=36L$.
    Panel (a) is for the situation of mutually un-correlated disorder, panel (b) for partly correlated disorder ($W_{\rm c}=W_{\rm uc}$), and panel (c) is for mutually fully correlated disorder.\index{disorder!, mutually correlated}\index{disorder!, mutually un-correlated}\index{disorder!, partly correlated}}
  \label{fig:histograms}
\end{minipage}
\end{center}
  \end{figure}

  For the case $\ell = 36L$ (this corresponds to $W=\varepsilon_F/10$)
  the disorder has as expected \cite{abrikosov1981} almost no effect on the
  Landauer conductance, {\it i.e.} $\big<G_{ii}\big>\sim 2e^2/h$ with
  vanishing fluctuations. However, for the transconductance the
  situation is very different. Panel (a) of Fig.~\ref{fig:histograms}
  shows a typical histogram of $G_{21}(\ell)/G_{21}(\infty)$ for $\ell =36 L$ in the case of un-correlated
  disorder.\index{disorder!, mutually un-correlated} Depending on the disorder configuration $G_{21}(\ell)$ can
  be either higher or lower than in the ballistic regime. We
  emphasize that the histogram peaks close to the ballistic value and
  not at zero drag, {\it i.e.} the mean drag is finite and positive in agreement with Eq.~(\ref{mean2kF}). The
  variance is of the same order as the mean value so that sign reversal
  for some disorder realizations is possible. The later is represented
  by the negative tail in the histogram. The sign of the drag is thus
  arbitrary in the sense that both positive and negative drag can be
  observed.

  \begin{figure}
  \begin{center}
\epsfig{file=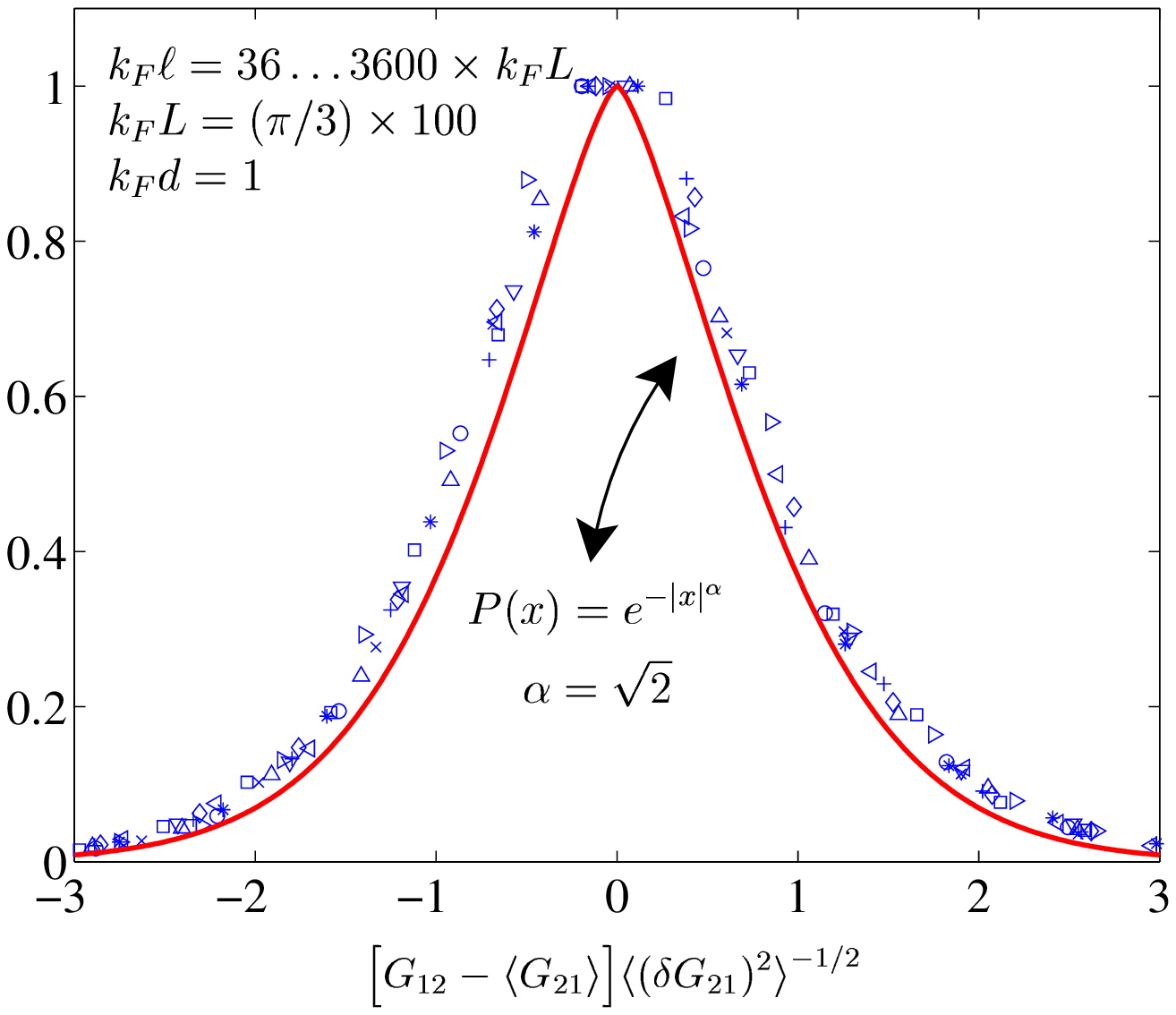, height=0.99\columnwidth,clip,angle=-90}
  \end{center}
  \caption[Rescaling of drag conductance, $k_F\ell$ (uc)]{Rescaling of the distribution of the drag conductance $G_{21}$ for mutually un-correlated disorder. Distributions for $k_FL\sim 100$ and $k_Fd=1$ for different values of the mean free path $k_F\ell$.}
  \label{fig:rescaling_ell}
  \end{figure}

\begin{figure}
  \begin{center}
\epsfig{file=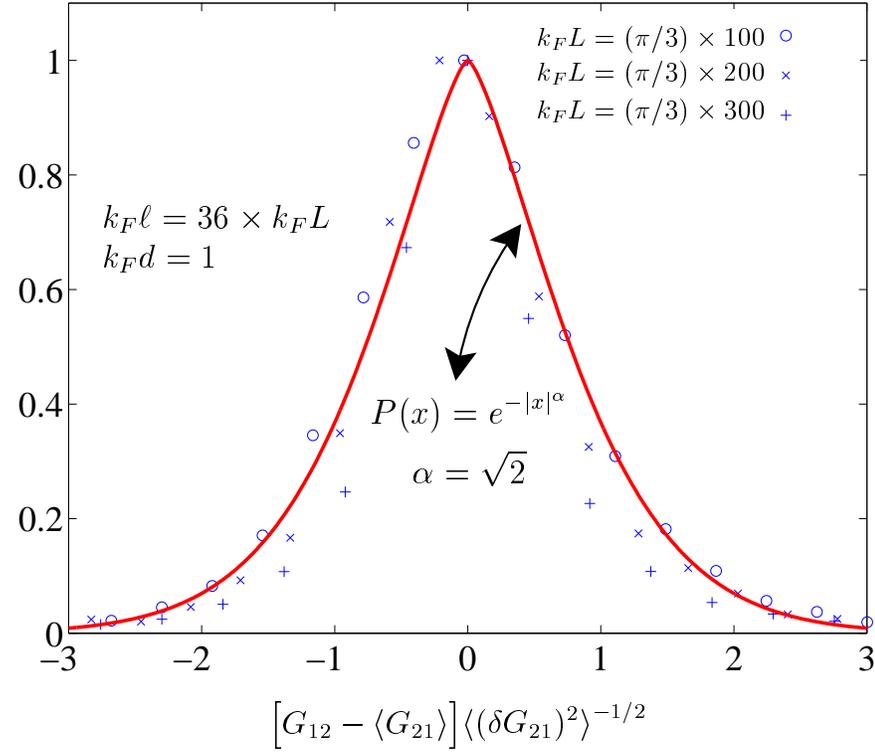, height=0.99\columnwidth,clip,angle=-90}
  \end{center}
  \caption[Rescaling of drag conductance distributions, $k_FL$ (uc)]{Rescaling of the distribution of the drag conductance $G_{21}$ for mutually un-correlated disorder. Distributions for different values of $k_FL$ for $k_Fd=1$ and a fixed ratio of $k_F\ell/k_FL$.}
  \label{fig:rescaling_L}
  \end{figure}

\begin{figure}
  \begin{center}
\epsfig{file=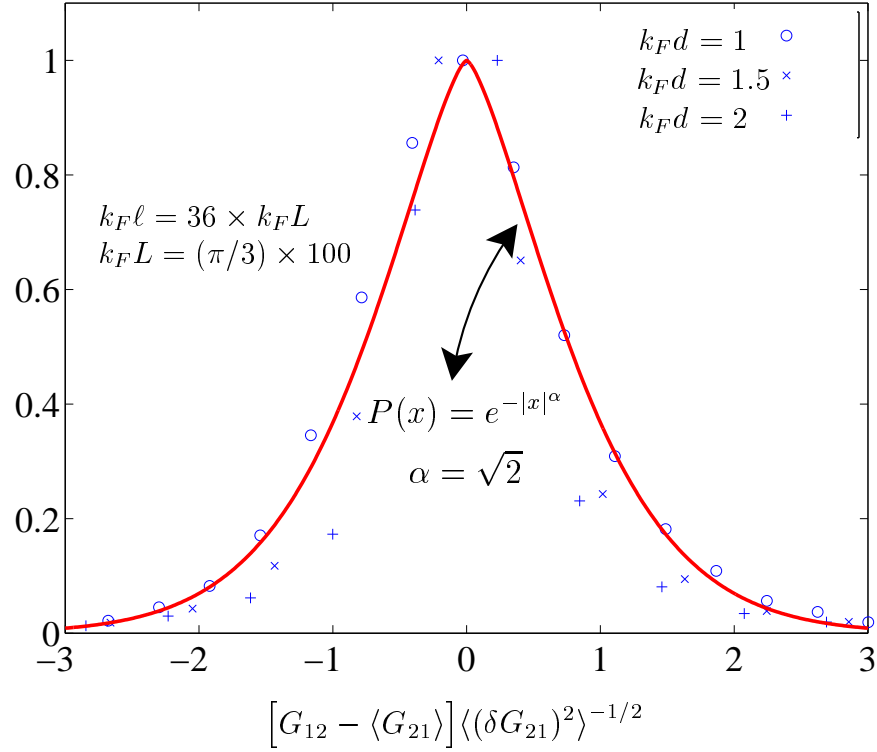, height=0.99\columnwidth,clip,angle=-90}
  \end{center}
  \caption[Rescaling of drag conductance distributions, $k_Fd$ (uc)]{Rescaling of the distribution of the drag conductance $G_{21}$ for mutually un-correlated disorder. Distributions for different values of $k_Fd$ for $k_F\ell = 36\times k_FL$ and $k_FL\sim 100$.}
  \label{fig:rescaling_d}
  \end{figure}

For the distribution of $G_{21}$ in the case of un-correlated disorder we find that \index{disorder!, mutually un-correlated}\index{conductance!, drag $G_{21}$}

\begin{equation}\label{Puc}
P_{\rm uc}(x)\propto
  \exp\Bigg[-\Bigg|\frac{x-x_0}{x_1}\Bigg|^\alpha\Bigg],\qquad \alpha\simeq 1.4,
\end{equation}
 fits surprisingly well to the data. In fact we have performed these fits to
  histograms for $k_F\ell$ ranging from $10^3$ to $10^5$ with $k_FL$ in
  the range from $100$ to $300$. By rescaling of $G_{21}$,

\begin{equation}
G_{21}\longrightarrow \frac{G_{21}-\big<G_{21}\big>}{\big<(\delta G_{21})^2\big>^{1/2}},
\end{equation} 
it is possible to let
  all histograms fall onto the same curve. There are three parameters which may be varied: $\ell$, $L$, and $d$. For simplicity we illustrate this by fixing two of the parameters and varying the third one. Figs.~\ref{fig:rescaling_ell},~\ref{fig:rescaling_L},~\ref{fig:rescaling_d} show the rescaling for fixed $L$ and $d$, fixed $\ell$ and $d$, and fixed $\ell$ and $L$, respectively. 

Note that the distribution is
  non-universal in the sense that it depends on the range of the interaction $U_{12}$. It should also be stressed that the particular form in Eq.~(\ref{Puc}) is not unique and that other similar functions might be fitted equally well \cite{cheianov}.

  \begin{figure}
  \begin{center}
\epsfig{file=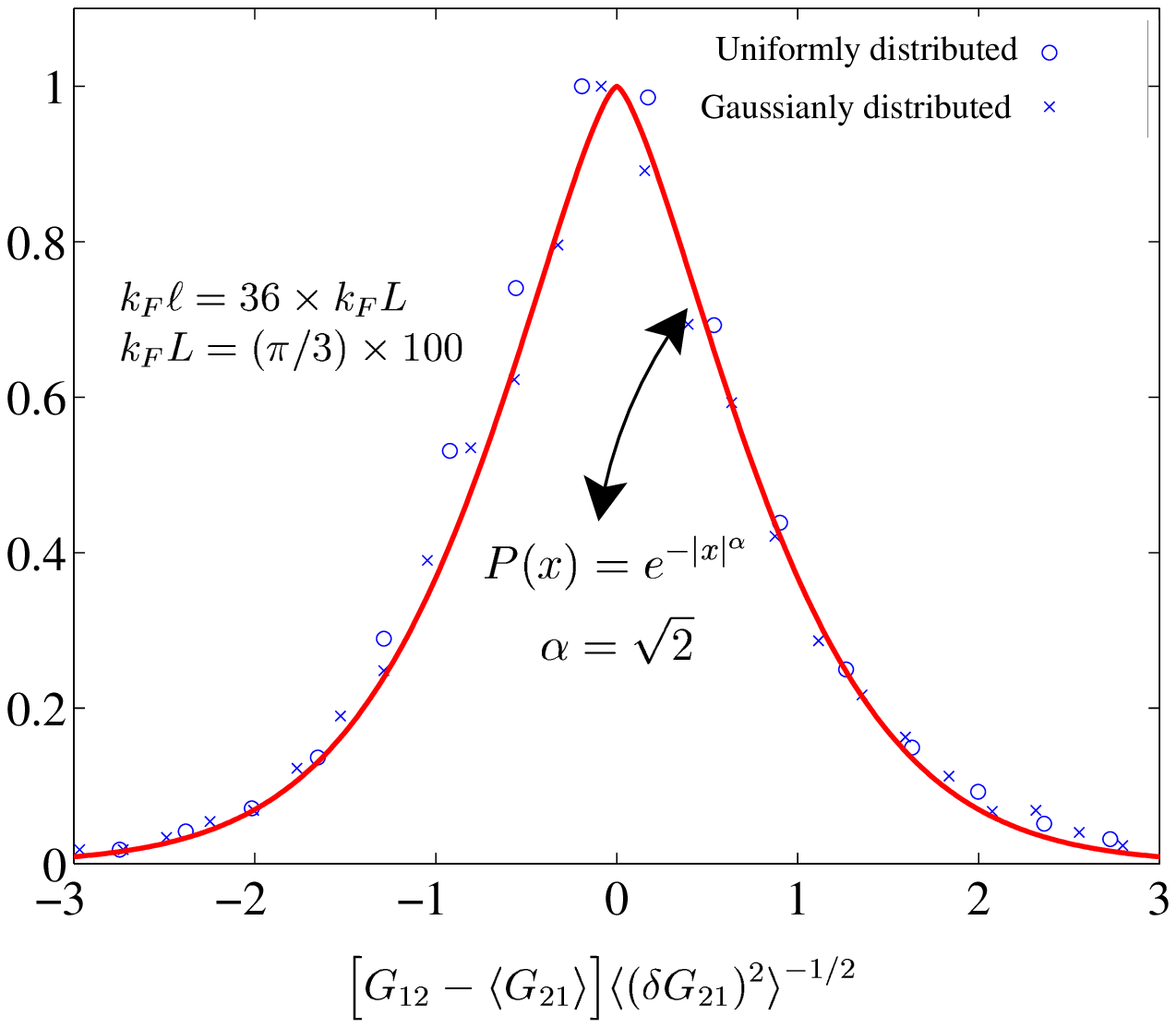, height=0.99\columnwidth,clip,angle=-90}
  \end{center}
  \caption[Comparison of two models for the disorder]{Rescaling of the distribution of the drag conductance $G_{21}$ for uniformly and Gaussianly distributed on-site energies.\index{disorder!, uniform distribution}\index{disorder!, Gaussian distribution}}
  \label{fig:rescaling_stat}
  \end{figure}

In Fig. \ref{fig:rescaling_stat} we compare the results of the Anderson model for the case of a uniform distribution, Eq.~(\ref{P(V)}), to a Gaussian distribution, but with the same mean value and variance. As seen the distribution of the drag conductance $G_{21}$ does not depend on the particular choice of disorder model.\index{disorder!, uniform distribution}\index{disorder!, Gaussian distribution}

  \begin{figure}
  \begin{center}
\epsfig{file=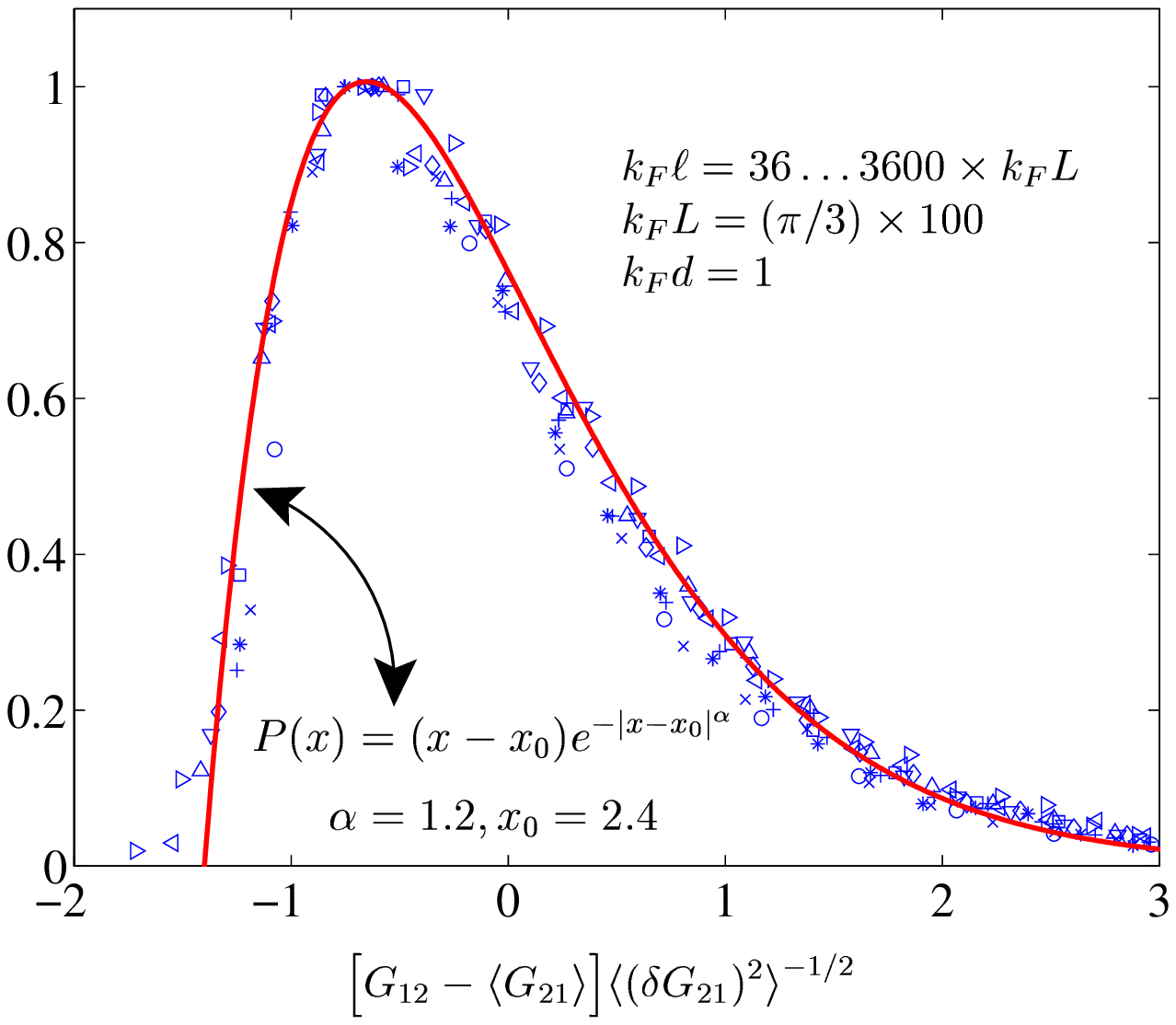, height=0.99\columnwidth,clip,angle=-90}
  \end{center}
  \caption[Rescaling of drag conductance distributions, $k_F\ell$ (c)]{Rescaling of the distribution of the drag conductance $G_{21}$ for mutually correlated disorder. Distributions for $k_Fd=1$ and $k_FL\sim 100$ for different values of $k_F\ell$. }
  \label{fig:rescaling_c}
  \end{figure}

  For the same system parameters but now with fully correlated disorder\index{disorder!, mutually correlated}
  we get a very different distribution as seen in panel (c) of
  Fig.~\ref{fig:histograms}. Again we find that rescaling is possible, see Fig.~\ref{fig:rescaling_c}.

  \begin{figure}
  \begin{center}
\epsfig{file=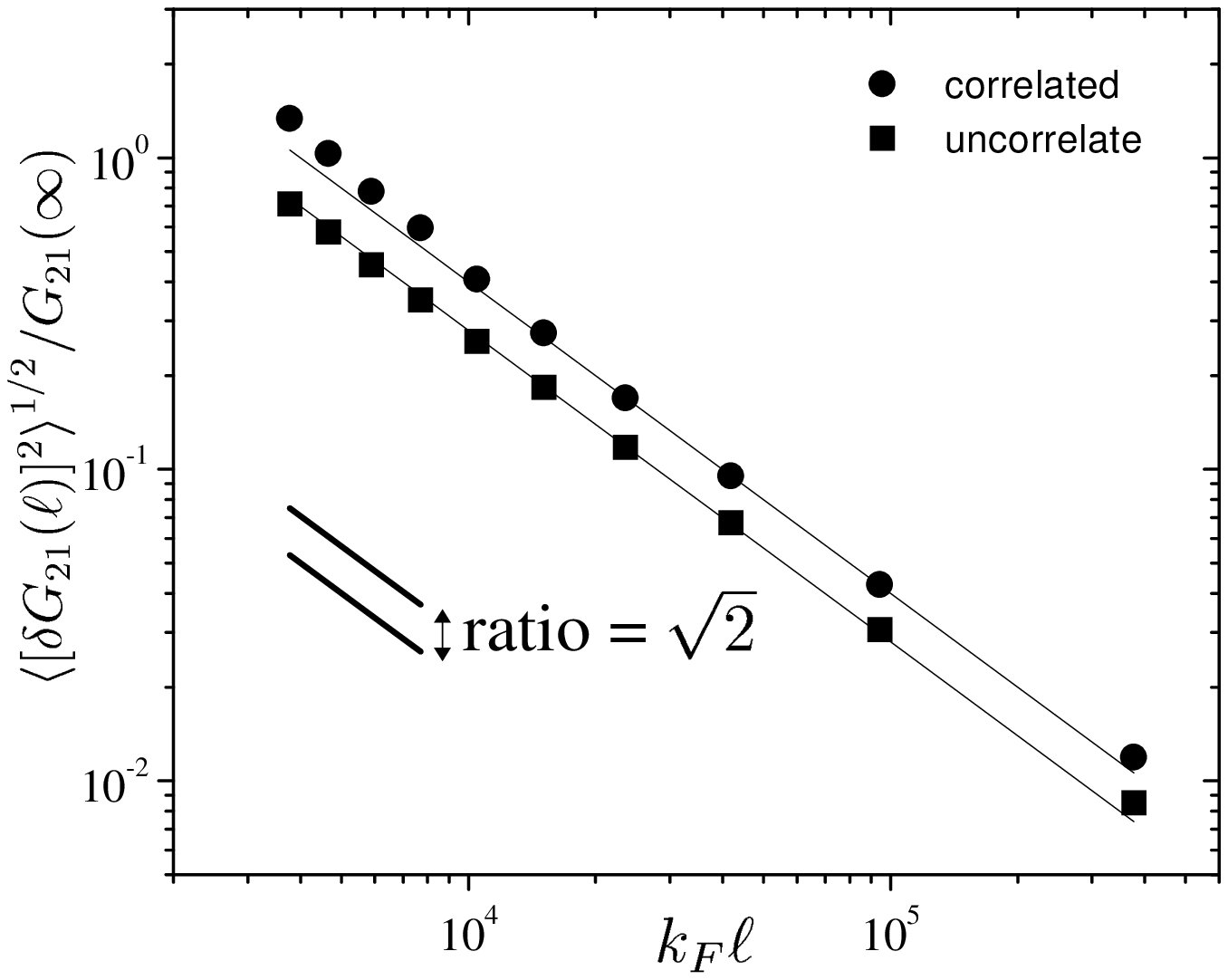, width=0.99\columnwidth,clip}
  \end{center}
  \caption[Disorder dependence of drag conductance fluctuations]{Plot of fluctuations $\big<[\delta G_{21}(\ell)]^2\big>^{1/2}$ as a
  function of the mean free path $k_F\ell$. In both cases the expected $1/k_F\ell$
  behavior is born out by the numerical calculations.
  The full lines are Eq.~(\ref{analytic}) with no free parameters. The expected
  enhancement, Eq.~(\ref{sqrt2}), for correlated disorder by a factor of $\sqrt{2}$
  compared to un-correlated disorder is also confirmed by the numerical calculations.}
  \label{fig:fluctuations}
  \end{figure}

In this situation the mean transconductance
  is enhanced by approximately a factor of two compared to the un-correlated case. This confirms, at least
  qualitatively, the prediction of Eq.~(\ref{enhancement_mean}). The fluctuations are also enhanced.  However, since the mean
  transconductance is enhanced by almost a factor of two compared to the
  ballistic limit this also means that there is no disorder
  configurations giving rise to negative drag.

  In Fig.~\ref{fig:fluctuations} we show the dependence of the
  fluctuations on the mean free path $\ell$ which has the expected
  $1/k_F\ell$ dependence. We also notice that
  correlated disorder gives rise to slightly larger fluctuations compared
  to un-correlated disorder. In fact they are exactly enhanced by a
  factor of $\sqrt{2}$ as predicted, see Eq.~(\ref{sqrt2}).

For partly correlated disorder we as expected get fluctuations which are enhanced by a numerical factor between unity and $\sqrt{2}$. For the histograms we observe how the, say, histogram in the case of un-correlated disorder evolves into that of correlated disorder when increasing the degree of correlation, see panel (b) of Fig.~\ref{fig:histograms}. It should however be stressed that for partly correlated disorder the histogram is not a simple superposition of the histograms in the two extreme limits.

To summarize we find that our perturbative results for the fluctuations are in excellent agreement with our numerical studies. Furthermore, the numerical studies give insight into, not only the statistical moments of $G_{21}$, but in fact the full distribution $P(G_{21})$ which we presently have no analytical predictions for.

%% file: dots.tex
\chapter{Quantum dots}\label{chap:dots}

A quantum dot is a sub-micron cavity with electronic properties governed by the phase-coherence of quantum mechanics. It can {\it e.g.} be one of Nature's own small metallic nano-particles or it can be man-made by confining the electronic degrees of freedom in a semiconductor. Quantum dots exhibit many exciting properties both in their charge transport and in their optical response. Quantum dots with the classical motion being chaotic is also a field of tremendous interest often referred to as quantum chaos.

Here, we consider two Coulomb coupled semiconductor quantum dots such as suggested in Fig.~\ref{SAMPLES_dots}. Our aim is to calculate the statistical properties of the drag conductance with the aid of random matrix theory.\index{quantum!, dots} 

\section{Statistical averages}

Reviews of the statistical theory of disordered and chaotic systems are given in Refs.~\cite{beenakker1997,mirlin2000,alhassid2000}. For the application of random matrix theory to the current matrix we will mainly follow the notation used in the review of Beenakker \cite{beenakker1997}.

\subsection{Mutually un-correlated disorder}

We consider an ensemble of mesoscopic chaotic systems, as suggested in Fig.~\ref{SAMPLES_dots}. We assume that the region where the subsystems couple by Coulomb interaction have mutually un-correlated disorder\index{disorder!, mutually un-correlated} so that, suppressing the integration variables,
\begin{subequations}
\begin{eqnarray}
\big<G_{21}\big>&\propto& \int U_{12}U_{12} \big<\Delta_1\big> \big<\Delta_2\big>,\\
\big<G_{21}^2\big>&\propto& \int U_{12}U_{12} U_{12}U_{12} \big<\Delta_1\Delta_1\big> \big<\Delta_2\Delta_2\big>.
\end{eqnarray}
\end{subequations}
We consider the low temperature limit and start from Eq.~(\ref{Deltai_matrix_T^2}) --- at some later point we turn to the formulation in Eq.~(\ref{Deltai_matrix_T^2_S}). The task is to calculate $\big<\Delta_i\big>$ and $\big<\Delta_i\Delta_i\big>$ which we do using random matrix theory.

\subsection{Correlation of current and wave functions?}

To lowest order in $1/k_F\ell$ the average of two wave functions is, see {\it e.g.} Ref.~\cite{aleiner1995},
\begin{equation}
\langle\phi_{\gamma}^{\ast}(x)\phi_{\delta}(y)\rangle=\delta_{\gamma\delta}\langle\phi_{\gamma}^{\ast}(x)\phi_{\gamma}(y)\rangle
\approx\delta_{\gamma\delta} \frac{\delta\varepsilon}{2\pi}\,\Big<2\pi\sum_\gamma \phi_{\gamma}^{\ast}(x)\phi_{\gamma}(y) \delta({\mathscr E}_\gamma)\Big>, 
\end{equation}
where $\delta\varepsilon$ is the mean level spacing. Introducing the spectral function
\begin{equation}
{\mathscr A}({\boldsymbol r},{\boldsymbol r}')=2\pi\sum_\alpha \phi_\alpha^*({\boldsymbol r})\phi_\alpha({\boldsymbol r}')\delta({\mathscr E}_{\alpha}),
\end{equation}
we get
\begin{equation}\label{aleinerresult}
\langle\phi_{\gamma}^{\ast}(x)\phi_{\delta}(y)\rangle \approx
\delta_{\gamma\delta}\frac{\delta\varepsilon}{2\pi}
\big<{\mathscr A}(x-y)\big>, 
\end{equation}
where the average spectral function $\langle {\mathscr A}(x,y)\rangle =\langle {\mathscr A}(x-y)\rangle$ at the Fermi level is given by, suppressing the energy subscript,\index{spectral function}\index{length scales!, mean free path}
\begin{equation}\label{<A>}
\langle {\mathscr A}(r)\rangle\simeq
(m/2\hbar^2)\exp(-r/2\ell)J_0(k_{F} r).
\end{equation}
Next, consider the average
\begin{equation}
\big< I_{\alpha\beta}\phi_{\gamma}^{\ast}(x)\phi_{\delta}
(y)\big> \propto   (\partial_{x_{1}}-\partial_{x_{2}})
\big<\phi_{\alpha}^{\ast}(x_{1})\phi_{\beta}(x_{2})\phi_{\gamma}^*
(x)\phi_{\delta}(y)\big> \Big|_{x_{1}=x_{2}},
\end{equation}
where to lowest order in $1/k_{F}\ell$
\begin{multline}\label{4av}
\big< \phi_{\alpha}^*(x_{1})\phi_{\beta}(x_{2})\phi_{\gamma}^*(x)\phi_{\delta}(y)\big>   \simeq \big<\phi_{\alpha}^*(x_{1})\phi_{\beta}(x_{2})\big>\big<\phi_{\gamma}^*(x)\phi_{\delta
}(y)\big>\\  +\big<\phi_{\alpha}^*(x_{1})\phi_{\delta}(y)\big>\big<\phi_{\beta}(x_{2})\phi_{\gamma}^*(x)\big>.
\end{multline}
Due to current conservation the points $x_{1}=x_{2}$ can be chosen freely. Taking them to be outside the chaotic region, the decay of the spectral function makes the second term in Eq.~(\ref{4av}) vanish if $x$ or $y$ are inside. The first term amounts to performing the average over $I_{\alpha\beta}$ and $\phi_{\gamma
}^{\ast}(x)\phi_{\delta}(y)$ separately. Similar arguments hold for
higher order averages so that to lowest order in $1/k_F\ell$

\begin{equation}\label{<Delta>}
\frac{\big< \Delta_i(\omega,{\boldsymbol r},{\boldsymbol
r}') \big>}{\hbar\omega\,(2\pi)^2\hbar } \simeq \Imag
\sum_{\alpha\beta\gamma}\big<
I_{\alpha\gamma}^i\big> 
 \big<\rho_{\alpha\beta}^i({\boldsymbol r})\rho_{\beta\gamma}^i({\boldsymbol r}')
\delta({\mathscr E}_\alpha)\delta({\mathscr E}_\beta)\delta({\mathscr E}_\gamma)\big>,
\end{equation}
and since $I$ and $\rho$ are Hermitian

\begin{multline}\label{<DeltaDelta>}
\frac{\big< \Delta_i(\omega,{\boldsymbol r},{\boldsymbol
r}') \Delta_i(\tilde\omega,{\boldsymbol s},{\boldsymbol
s}') \big>}{\hbar\omega\,\hbar\tilde\omega\,(2\pi)^4\hbar^2} \simeq \frac{1}{(2i)^2}
\sum_{\alpha\beta\gamma}\sum_{\tilde\alpha \tilde\beta\tilde\gamma}\big<\delta({\mathscr E}_\gamma)\delta({\mathscr E}_{\tilde\gamma})\big> \\
\times\Big\{\big<
I_{\alpha\gamma}^iI_{\tilde\alpha\tilde\gamma}^i\big>\big< \rho_{\alpha\beta}^i({\boldsymbol r})\rho_{\beta\gamma}^i({\boldsymbol r}') \rho_{\tilde\alpha\tilde\beta}^i({\boldsymbol s})\rho_{\tilde\beta\tilde\gamma}^i({\boldsymbol s}')\delta({\mathscr E}_\alpha)\delta({\mathscr E}_\beta)\delta({\mathscr E}_{\tilde\alpha})\delta({\mathscr E}_{\tilde\beta}) \big>\\
-\big<I_{\alpha\gamma}^iI_{\tilde\gamma\tilde\alpha}^i\big>\big< \rho_{\alpha\beta}^i({\boldsymbol r})\rho_{\beta\gamma}^i({\boldsymbol r}')
\rho_{\tilde\beta\tilde\alpha}^i({\boldsymbol s})\rho_{\tilde\gamma\tilde\beta}^i({\boldsymbol s}')\delta({\mathscr E}_\alpha)\delta({\mathscr E}_\beta)\delta({\mathscr E}_{\tilde\alpha})\delta({\mathscr E}_{\tilde\beta}) \big>\\
-\big<I_{\gamma\alpha}^iI_{\tilde\alpha\tilde\gamma}^i\big>\big<\rho_{\beta\alpha}^i({\boldsymbol r})\rho_{\gamma\beta}^i({\boldsymbol r}')
 \rho_{\tilde\alpha\tilde\beta}^i({\boldsymbol s})\rho_{\tilde\beta\tilde\gamma}^i({\boldsymbol s}')\delta({\mathscr E}_\alpha)\delta({\mathscr E}_\beta)\delta({\mathscr E}_{\tilde\alpha})\delta({\mathscr E}_{\tilde\beta}) \big>\\
 +\big<I_{\gamma\alpha}^iI_{\tilde\gamma\tilde\alpha}^i\big>\big<\rho_{\beta\alpha}^i({\boldsymbol r})\rho_{\gamma\beta}^i({\boldsymbol r}')
\rho_{\tilde\beta\tilde\alpha}^i({\boldsymbol s})\rho_{\tilde\gamma\tilde\beta}^i({\boldsymbol s}')\delta({\mathscr E}_\alpha)\delta({\mathscr E}_\beta)\delta({\mathscr E}_{\tilde\alpha})\delta({\mathscr E}_{\tilde\beta}) \big>\Big\}.
\end{multline}
For the averages we have used that eigenvalues and eigenfunctions are un-correlated so that there is quite some freedom in where to place the delta functions. We note that in these expressions the triangle functions are only given to lowest order in $\omega$ so that the right hand sides do not depend on $\omega$. 

\section{Random matrix theory}

In order to calculate $\big<
I_{\alpha\gamma}\big>$ and $\big<
I_{\alpha\gamma}I_{\tilde\alpha\tilde\gamma}\big>$ that enter Eqs.~(\ref{<Delta>},\,\ref{<DeltaDelta>}) we apply standard random matrix theory\index{random matrix theory} for chaotic quantum dots \cite{beenakker1997}. By chaotic we mean that the classical motion in the dot can be regarded as chaotic on larger time scales than the ergodic time; the time needed for a classical trajectory to explore the entire phase-space. We assume that the dot is coupled to the leads via ballistic point contacts with $N$ modes. Then the distribution of the $2N\times 2N$ scattering matrix $S$, see Eq.~(\ref{Smatrix}), is given by the circular ensemble\index{circular ensemble}; the scattering matrix $S$ is uniformly distributed in the unitary group and is only subject to the constraint of a possible time-reversal symmetry.\index{scattering!, matrix}\index{matrix!, scattering}

\subsection{Circular ensemble}

Now we consider a unitary $M\times M$ matrix $U$ ($UU^\dagger=1$) which is distributed uniformly over the unitary ensemble with no further constraints; for $M=1$ this is to say that $U=\exp i\theta$ is uniformly distributed over the unit circle. Here, we will only need a few results which can be found in Refs.~\cite{beenakker1997,mello1990}. The most important one is that if we consider a polynomial function

\begin{equation}
f(U)=\prod_{i=n}^p U_{\alpha_n a_n}\prod_{m=1}^q U_{\beta_m b_m}^*,
\end{equation}
then the average $\big<f(U)\big>$ is zero unless $p=q$ and the sets of ``left'' indices coincide, $\{\alpha_n\}=\{\beta_n\}$, and similarly for the ``right'' indices, $\{a_n\}=\{b_n\}$. The first constraint means that

\begin{equation}
\big<U\big>=0,\qquad \big<U^2\big>=0.
\end{equation}
The other constraints are most easily illustrated for $p=q=1$ where

\begin{equation}\label{p=q=1}
\big<U_{\alpha a}U_{\beta b}^*\big>=\frac{1}{M}\delta_{\alpha\beta}\delta_{ab}.
\end{equation}
For $p=q=2$ the result is slightly more complicated

\begin{multline}\label{p=q=2}
\big<U_{\alpha_1 a_1}U_{\alpha_2 a_2}U_{\beta_1 b_1}^*U_{\beta_2 b_2}^*\big>=\\\frac{1}{M^2-1}\big[\delta_{\alpha_1\beta_1}\delta_{a_1b_1}\delta_{\alpha_2\beta_2}\delta_{a_2b_2}+\delta_{\alpha_1\beta_2}\delta_{a_1b_2}\delta_{\alpha_2\beta_1}\delta_{a_2b_1}\big]\\
- \frac{1}{M(M^2-1)}\big[\delta_{\alpha_1\beta_1}\delta_{a_1b_2}\delta_{\alpha_2\beta_2}\delta_{a_2b_1}+\delta_{\alpha_1\beta_2}\delta_{a_1b_1}\delta_{\alpha_2\beta_1}\delta_{a_2b_2}\big].
\end{multline}
General results for $p=q=3$ and $p=q=4$ have been given by Mello \cite{mello1990}. We will need the expression for $p=q=4$ but since it contains $(4!)^2=576$ terms its application is not a simple task. Instead we consider the Gaussian approximation which is the leading-order term in the average \cite{beenakker1997}

\begin{equation}\label{gaussian}
\big<f(U)\big>= M^{-p}\delta_{pq}\sum_{\cal P}\prod_{j=1}^p \delta_{\alpha_j\beta_{{\cal P}(j)}} \delta_{a_j b_{{\cal P}(j)}}+{\cal O}(M^{-p-1}), 
\end{equation} 
where $\cal P$ is all the $p!$ permutations of the numbers $1,2,\ldots p$.

\subsection{Current matrix elements}

The task is to calculate $\big<I_{\alpha\gamma}\big>$ and $\big<I_{\alpha\gamma}I_{\tilde\alpha\tilde\gamma}\big>$ but since we, as in Eq.~(\ref{aleinerresult}), may sum over the indices and divide by the number of modes we might as well consider\index{scattering!, matrix}\index{matrix!, scattering}\index{matrix!, current}

\begin{equation}
\sum_{\alpha\gamma}\big<I_{\alpha\gamma}\big>,\qquad \sum_{\alpha\gamma\tilde\alpha\tilde\gamma}\big<I_{\alpha\gamma}I_{\tilde\alpha\tilde\gamma}\big>.
\end{equation}
Starting from the current matrix
\begin{equation}
I=\frac{\hbar}{2m}\big[\tau^3-S^\dagger\tau^3 S \big],\qquad \tau^3=\begin{pmatrix}1&0\\0&-1\end{pmatrix},
\end{equation}
the first average becomes
\begin{equation}\label{SumI}
\sum_{\alpha\gamma}\big<I_{\alpha\gamma}\big> =-\frac{\hbar}{2m}\sum_{\alpha\gamma} \big<\{S^\dagger\tau^3 S\}_{\alpha\gamma}\big>,
\end{equation}
 and similarly it can be shown that
\begin{equation}\label{SumII}
\sum_{\alpha\gamma\tilde\alpha\tilde\gamma}\big<I_{\alpha\gamma}\,I_{\tilde\alpha\tilde\gamma}\big>=\left(\frac{\hbar}{2m}\right)^2\sum_{\alpha\gamma\tilde\alpha\tilde\gamma} \big<\{S^\dagger\tau^3 S\}_{\alpha\gamma} \{S^\dagger\tau^3 S\}_{\tilde\alpha\tilde\gamma}\big>.
\end{equation}

We will first consider the case of time-reversal symmetry (usually denoted by the symmetry index $\beta=1$ \cite{beenakker1997}) and then afterwards discuss the effect of breaking time-reversal symmetry by a weak magnetic field ($\beta=2$). Finally, we summarize the results in the asymptotic limit $N\gg 1$.

\subsection{Time-reversal symmetry}

In the presence of time reversal symmetry $S=S^T$ which motivates the introduction of a unitary $2N\times 2N$ matrix $U$ by $S=UU^T$. From Eq.~(\ref{p=q=2}) we get\index{scattering!, matrix}\index{matrix!, scattering}

\begin{multline}
\big<\{S^\dagger \tau^3 S\}_{\alpha\gamma}\big>_{\beta=1}=\sum_\nu \tau_{\nu\nu}^3\big<[(UU^T)^\dagger]_{\alpha \nu} [UU^T]_{\nu\gamma}\big>\\
=\sum_{\nu xy} \tau_{\nu\nu}^3\big< U_{\nu y}U_{\gamma y}U_{\nu x}^*U_{\alpha x}^* \big>
=\sum_{\nu x} \tau_{\nu\nu}^3\big< U_{\nu x}U_{\gamma x}U_{\nu x}^*U_{\alpha x}^* \big>\\
=\underbrace{\sum_x}_{=M}\sum_{\nu} \tau_{\nu\nu}^3\Big[
\frac{1}{M^2-1}\delta_{\nu\alpha}\delta_{\gamma\nu}
- \frac{1}{M(M^2-1)}\delta_{\nu\alpha}\delta_{\gamma\nu}\big]
\Big]
=\frac{1}{M+1} \tau_{\alpha\gamma}^3,
\end{multline}
where $M=2N$. This means that Eq.~(\ref{SumI}) becomes

\begin{equation}\label{SumIbeta1}
\big<I_{\alpha\gamma}\big>_{\beta=1}\propto \tau_{\alpha\gamma}^3,\qquad\sum_{\alpha\gamma}\big<I_{\alpha\gamma}\big>_{\beta=1} =0.
\end{equation}
Similarly,
\begin{multline}
\big<S^\dagger_{\alpha \nu} S_{\nu\gamma} S^\dagger_{\tilde\alpha\tilde\nu} S_{\tilde\nu\tilde\gamma}\big>=\big<[(UU^T)^\dagger]_{\alpha \nu} [UU^T]_{\nu\gamma} [(UU^T)^\dagger]_{\tilde\alpha\tilde\nu} [UU^T]_{\tilde\nu\tilde\gamma}\big>\\
=\sum_{xyzu}\big<  U_{\nu y} U_{\gamma y}  U_{\tilde\nu u}U_{\tilde\gamma u}U_{\nu x}^*U_{\alpha x}^* U_{\tilde\nu z}^* U_{\tilde\alpha z}^*\big>
\end{multline}
The average is zero unless for the indices 
\begin{equation}
\{yyuu\}=\{xxzz\},\qquad \{ \nu \gamma \tilde\nu\tilde\gamma\}=\{\nu\alpha\tilde\nu\tilde\alpha\},
\end{equation}
which in principle allows for simplifications. Here we will just consider the Gaussian approximation, Eq.~(\ref{gaussian}). The calculation is in principle simple although long. However, with the aid of {\it Mathematica} Eq.~(\ref{SumII}) becomes

\begin{equation}\label{SumIIbeta1}
\sum_{\alpha\gamma\tilde\alpha\tilde\gamma}\big<I_{\alpha\gamma}\,I_{\tilde\alpha\tilde\gamma}\big>_{\beta=1}
= \left(\frac{\hbar}{2m}\right)^2 \frac{M(M^3+3M^2+6M)}{M^3}.
\end{equation}
The same calculation based on the full expression of Mello \cite{mello1990} is of course possible as soon as the $(4!)^2=576$ terms have been fed into the computer. The $4!=24$ terms in the Gaussian approximation are on the other hand easily generated by the computer itself, see Eq.~(\ref{gaussian}). 

\subsection{Breaking of time-reversal symmetry}

In the absence of time reversal symmetry $S$ is unitary with no further constraints so that Eq.~(\ref{p=q=1}) gives

\begin{equation}
\big<S^\dagger \tau^3 S\big>_{\alpha\beta}=
\sum_{\nu} \tau_{\nu\nu}^3 \big< U_{\nu\beta}U_{\nu\alpha}^*\big>=\frac{1}{M}
\delta_{\alpha\beta}\Tr\tau^3= 0,
\end{equation}
and thus Eq.~(\ref{SumI}) gives

\begin{equation}\label{SumIbeta2}
\big<I_{\alpha\gamma}\big>_{\beta=2}\propto \tau_{\alpha\gamma}^3,\qquad \sum_{\alpha\gamma}\big<I_{\alpha\gamma}\big>_{\beta=2} =0.
\end{equation}
In the same way
\begin{multline}
\big<S^\dagger_{\alpha \nu} S_{\nu\gamma} S^\dagger_{\tilde\alpha\tilde\nu} S_{\tilde\nu\tilde\gamma}\big>=\big< U_{\nu\gamma} U_{\tilde\nu\tilde\gamma}U_{\nu \alpha}^* U_{\tilde\nu \tilde \alpha}^*\big>
=\frac{1}{M^2-1}\big[\delta_{\gamma\alpha}\delta_{\tilde\gamma\tilde\alpha}+\delta_{\nu\tilde\nu}\delta_{\gamma\tilde\alpha}\delta_{\tilde\gamma\alpha}\big]
\\- \frac{1}{M(M^2-1)}\big[\delta_{\gamma\tilde\alpha}\delta_{\tilde\gamma \alpha}+\delta_{\nu\tilde\nu}\delta_{\gamma \alpha}\delta_{\tilde\gamma \tilde\alpha}\big],
\end{multline}
which means that

\begin{equation}\label{StSStS}
\big<\big[S^\dagger\tau^3 S\big]_{\alpha\gamma} \big[S^\dagger\tau^3 S\big]_{\tilde\alpha\tilde\gamma}\big>= M \Big[\frac{1}{M^2-1}\delta_{\gamma\tilde\alpha}\delta_{\tilde\gamma\alpha}
- \frac{1}{M(M^2-1)}\delta_{\gamma \alpha}\delta_{\tilde\gamma \tilde\alpha}\Big].
\end{equation}
From Eq.~(\ref{SumII}) it now follows that

\begin{equation}\label{SumIIbeta2}
\sum_{\alpha\gamma\tilde\alpha\tilde\gamma}\big<I_{\alpha\gamma}\,I_{\tilde\alpha\tilde\gamma}\big>_{\beta=2}
= \left(\frac{\hbar}{2m}\right)^2 \frac{M^2}{M+1}.
\end{equation}
Using the Gaussian approximation, Eq.~(\ref{gaussian}), instead we get the result $(\hbar/2m)^2 \,M$.

\subsection{Asymptotic limit}

From Eqs. (\ref{SumIbeta1},\,\ref{SumIbeta2}) we have\index{matrix!, current}
\begin{equation}\label{proptotau3}
\big<I_{\alpha\gamma}\big>_{\beta=1,2}=\frac{\hbar}{2m}\tau_{\alpha\gamma}^3,\qquad\sum_{\alpha\gamma}\big<I_{\alpha\gamma}\big>_{\beta=1,2}=0,
\end{equation}
independently of $M=2N$ and $\beta$. In fact, the last equation holds not only in the asymptotic limit $M\gg 1$ but also for any arbitrary number of modes. From Eqs. (\ref{SumIIbeta1},\,\ref{SumIIbeta2}) we get

\begin{equation}
\sum_{\alpha\gamma\tilde\alpha\tilde\gamma}\big<I_{\alpha\gamma}\,I_{\tilde\alpha\tilde\gamma}\big>_{\beta=1,2}
\simeq \left(\frac{\hbar}{2m}\right)^2 M,\qquad M\gg 1,
\end{equation}
which also does not depend on $\beta$. From Eq.~(\ref{proptotau3}) it now also follows that
\begin{equation}
\big< I_{\alpha\gamma} I_{\tilde\alpha\tilde\gamma}\big>_{\beta=1,2}= {\rm const.}\times \tau_{\alpha\gamma}^3\tau_{\tilde\alpha\tilde\gamma}^3 + \left(\tfrac{\hbar}{2m}\right)^2 \big< \{S^\dagger\tau^3 S\}_{\alpha\gamma}\{S^\dagger\tau^3
S\}_{\tilde\alpha\tilde\gamma}\big>_{\beta=1,2}
\end{equation}
which in the asymptotic limit reduces to
\begin{equation}\label{IIgaussian}
\big< I_{\alpha\gamma} I_{\tilde\alpha\tilde\gamma}\big>_{\beta=1,2}
\simeq {\rm const.}\times \tau_{\alpha\gamma}^3\tau_{\tilde\alpha\tilde\gamma}^3+  \left(\frac{\hbar}{2m}\right)^2 M^{-1}\delta_{\alpha\tilde\gamma}\delta_{\tilde\alpha\gamma} ,\qquad M\gg 1.
\end{equation}
This is our main result of random matrix theory and the result does not depend on $\beta$. For the fluctuations that we are going to calculate this means that there will be no effect of breaking time-reversal symmetry by a weak magnetic field in contrast to the case of universal conductance fluctuations \cite{altshuler1985,lee1985} where there is a factor-of-two reduction of the fluctuations when breaking time-reversal symmetry \cite{beenakker1997}. We notice the misprint in Ref.~[E] where the factor $M^{-1}$ in Eq.~(\ref{IIgaussian}) has a wrong power which unfortunately propagates throughout the paper.

\section{Statistical properties of the triangle function}

\subsection{Zero mean value}

Since $\big< I_{\alpha\gamma}\big> \propto
\tau^3_{\alpha\gamma}= \tau^3_{\gamma\gamma}\delta_{\alpha\gamma}$ we can in the second average in $\langle
\Delta \rangle $ replace $\gamma$ by $\alpha$, see Eq.~(\ref{<Delta>}). The second average does not depend on the direction of the current and thus we get $\langle \Delta \rangle =0 $ which in turn gives
\begin{equation} 
 \langle G_{21} \rangle =0.
\end{equation}
This is the first out of two key results for quantum dots. Experimentally, ensemble averaging could be performed similarly to the case of universal conductance fluctuations by magneto-conductance experiments \cite{marcus1992}, by varying the shape of the dots \cite{chan1995}, or by changing the Fermi level \cite{keller1996}. The prediction of zero average drag can be considered as a test of the degree of ergodicity. Our second key result concerns the finiteness of the fluctuations which we study below.

\subsection{Finite fluctuations}

Let us now calculate the fluctuations by using Eq.~(\ref{IIgaussian}) in Eq.~(\ref{<DeltaDelta>}). The first term of Eq.~(\ref{IIgaussian}) does not contribute for the same reasons as for the mean value. For the second term we perform the $\tilde\gamma$ and $\tilde\alpha$ sums according to Eq.~(\ref{sum_scatteringstates}) and since the sum and the energy integral in Eq.~(\ref{sum_scatteringstates}) can be performed independently due to eigenstates and eigenvalue being un-correlated we get\index{scattering!, sum}

\begin{equation}
\sum_{\alpha\beta\gamma}\sum_{\tilde\alpha \tilde\beta\tilde\gamma}\big<\delta({\mathscr E}_\gamma)\delta({\mathscr E}_{\tilde\gamma})\big> 
 \big<
I_{\alpha\gamma}^iI_{\tilde\alpha\tilde\gamma}^i\big>\ldots\longrightarrow
\frac{1}{(2\pi)^2}\sum_{\tilde\alpha \tilde\beta\tilde\gamma}  \sum_{\tilde\alpha \tilde\gamma}\big<
I_{\alpha\gamma}^iI_{\tilde\alpha\tilde\gamma}^i\big>\ldots.
\end{equation}
With the Kronecker deltas $
\delta_{\alpha\tilde\gamma}\delta_{\tilde\alpha\gamma}$ in the second term of Eq.~(\ref{IIgaussian}) we can now perform the $\tilde\gamma$ and $\tilde\alpha$ sums which gives

\begin{eqnarray}
&&\frac{\big< \Delta_i(\omega,{\boldsymbol r},{\boldsymbol
r}') \Delta_i(\tilde\omega,{\boldsymbol s},{\boldsymbol
s}') \big>}{\hbar\omega\,\hbar\tilde\omega\,(2\pi)^4\hbar^2}\simeq \frac{1}{4(2\pi)^2\hbar^2(2i)^2(2N)(2\pi)^4}\nonumber\\
&&\qquad\qquad\times\Big\{\big< {\mathscr A}_i({\boldsymbol r},{\boldsymbol s}') {\mathscr A}_i({\boldsymbol r}',{\boldsymbol r}) {\mathscr A}_i({\boldsymbol s},{\boldsymbol r}') {\mathscr A}_i({\boldsymbol s}',{\boldsymbol s})  \big>\nonumber\\
&&\qquad\qquad-\big< {\mathscr A}_i({\boldsymbol r},{\boldsymbol s}) {\mathscr A}_i({\boldsymbol r}',{\boldsymbol r}) {\mathscr A}_i({\boldsymbol s},{\boldsymbol s}') {\mathscr A}_i({\boldsymbol s}',{\boldsymbol r}')  \big>\nonumber\\
&&\qquad\qquad-\big<{\mathscr A}_i({\boldsymbol r},{\boldsymbol r}') {\mathscr A}_i({\boldsymbol r}',{\boldsymbol s}') {\mathscr A}_i({\boldsymbol s},{\boldsymbol r}) {\mathscr A}_i({\boldsymbol s}',{\boldsymbol s}) \big>\nonumber\\
&&\qquad\qquad+\big<{\mathscr A}_i({\boldsymbol r},{\boldsymbol r}') {\mathscr A}_i({\boldsymbol r}',{\boldsymbol s}) {\mathscr A}_i({\boldsymbol s},{\boldsymbol s}') {\mathscr A}_i({\boldsymbol s}',{\boldsymbol r}) \big>\Big\}.
\end{eqnarray}
Here, we have introduced the spectral function at the Fermi level \index{spectral function}. 

\section{Fluctuations of drag conductance}

To lowest order in $1/k_F\ell$ we replace each spectral function by its average and use $\big<{\mathscr A}({\boldsymbol r},{\boldsymbol r}')\big>=\big<{\mathscr A}({\boldsymbol r}',{\boldsymbol r})\big>$, so that

\begin{equation}
\frac{\big< \Delta_i(\omega,{\boldsymbol r},{\boldsymbol
r}') \Delta_i(\tilde\omega,{\boldsymbol s},{\boldsymbol
s}') \big>}{\pi^2\,\hbar\omega\,\hbar\tilde\omega} \simeq \frac{{\cal F}_i({\boldsymbol r},{\boldsymbol
r}',{\boldsymbol s},{\boldsymbol
s}')}{2(2N)(2\pi)^4},
\end{equation}
where

\begin{multline}
{\cal F}_i({\boldsymbol r},{\boldsymbol
r}',{\boldsymbol s},{\boldsymbol
s}')=\big< {\mathscr A}_i({\boldsymbol r},{\boldsymbol r}')\big>\big< {\mathscr A}_i({\boldsymbol s},{\boldsymbol s}')\big>\\
\times\Big[\big< {\mathscr A}_i({\boldsymbol r},{\boldsymbol s})\big>\big< {\mathscr A}_i({\boldsymbol r}',{\boldsymbol s}')  \big>-\big< {\mathscr A}_i({\boldsymbol r},{\boldsymbol s}')\big>\big< {\mathscr A}_i({\boldsymbol r}',{\boldsymbol s}) \big>\Big].
\end{multline}
Since
\begin{equation}
\big<(\delta G_{21})^2\big>=\big<G_{21}^2\big>-\overbrace{\big<G_{21}\big>^2}^{=0}=\big<G_{21}^2\big>,
\end{equation}
we perform the $\omega$-integration in Eq.~(\ref{G21dc}) and get 

\begin{multline}
\big<(\delta G_{21})^2\big>^{1/2}\simeq \frac{e^2}{h}\frac{ (kT)^2}{3\, 2^4\, N}\Big[\int U_{12}({\boldsymbol
r}_1,{\boldsymbol r}_2) U_{12}({\boldsymbol r}_1',{\boldsymbol r}_2')U_{12}({\boldsymbol
s}_1,{\boldsymbol s}_2)U_{12}({\boldsymbol s}_1',{\boldsymbol s}_2')\\
\times{\cal F}_1({\boldsymbol r}_1,{\boldsymbol r}_1',{\boldsymbol s}_1,{\boldsymbol s}_1') {\cal F}_2({\boldsymbol
s}_2,{\boldsymbol s}_2',{\boldsymbol r}_2,{\boldsymbol r}_2')\Big]^{1/2}.
\end{multline}
To obtain an estimate of the fluctuations we note that $\mathscr A$ in Eq.~(\ref{<A>}) is largest for $k_Fr<1$ and on long length scales 
\begin{equation}
\big<{\mathscr A}({\boldsymbol r})\big>\approx \pi/(4\varepsilon_F \,k_F\ell)\,\delta({\boldsymbol r}),\qquad k_F\ell\gg 1.
\end{equation}
 Using that approximation for all spectral functions is too crude, but

\begin{multline}
{\cal F}_i({\boldsymbol r},{\boldsymbol
r}',{\boldsymbol s},{\boldsymbol
s}')\approx\left(\frac{\pi}{4\varepsilon_F \,k_F\ell}\right)^2\big< {\mathscr A}({\boldsymbol r},{\boldsymbol r}')\big>^2\\
\times\big[\delta({\boldsymbol r}-{\boldsymbol s})\delta({\boldsymbol r}'-{\boldsymbol s}')-\delta({\boldsymbol r}-{\boldsymbol s}')\delta({\boldsymbol r}'-{\boldsymbol s})\big],
\end{multline}
still gives a finite answer

\begin{multline}
\big<(\delta G_{21})^2\big>^{1/2}\simeq \frac{e^2}{h}\frac{ \pi^2}{3\,2^8\, N}\frac{1}{(k_F\ell)^2}\left(\frac{kT}{\varepsilon_F}\right)^2\Big[\int \big< {\mathscr A}({\boldsymbol r}_1,{\boldsymbol r}_1')\big>^2\big< {\mathscr A}({\boldsymbol s}_2,{\boldsymbol s}_2')\big>^2\\
\times\big[ U_{12}({\boldsymbol
r}_1,{\boldsymbol s}_2) U_{12}({\boldsymbol r}_1',{\boldsymbol s}_2')- U_{12}({\boldsymbol
r}_1',{\boldsymbol s}_2) U_{12}({\boldsymbol r}_1,{\boldsymbol s}_2')\big]^2
\Big]^{1/2}.
\end{multline}

Due to the peaked behavior of the spectral functions the mixed terms give a vanishing contribution and introducing the new variables 
\begin{equation}
{\boldsymbol u}={\boldsymbol r}_1-{\boldsymbol r}_1',\qquad {\boldsymbol v}={\boldsymbol s}_2-{\boldsymbol s}_2',\qquad {\boldsymbol y}={\boldsymbol r}_1'-{\boldsymbol s}_2',
\end{equation}
we get

\begin{multline}
\big<(\delta G_{21})^2\big>^{1/2}\simeq \frac{e^2}{h}\frac{ \pi^2}{3\,2^8\,N}\frac{1}{(k_F\ell)^2}\left(\frac{kT}{\varepsilon_F}\right)^2\\
\times\Big[2{\cal A}\int \big< {\mathscr A}({\boldsymbol u},0)\big>^2\big< {\mathscr A}({\boldsymbol v},0)\big>^2 U_{12}^2({\boldsymbol u}+{\boldsymbol
y},{\boldsymbol v}) U_{12}^2({\boldsymbol y},0)\Big]^{1/2},
\end{multline}
where ${\cal A}$ is the interaction area. Assuming that the screening length $r_s\ll \ell$,\index{length scales!, mean free path}\index{length scales!, screening length} the range of $\big<{\mathscr A}\big>$ is longer than the range of $U_{12}$ and we further approximate\index{coupling!, screened}\index{length scales!, screening length}

  \begin{figure}
  \begin{center}
\begin{minipage}[c]{0.65\textwidth}
 \epsfig{file=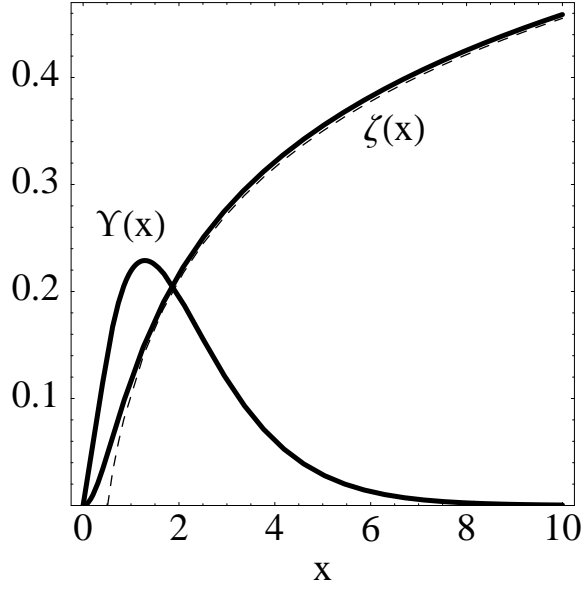, width=\textwidth,clip}
\end{minipage}\hfill
\begin{minipage}[c]{0.31\textwidth}
\caption[Plot of $\Upsilon(x)$ and $\zeta(x)$]{Plot of  $\Upsilon(x)$ and $\zeta(x)$ as a function of $x$. The dashed line shows the logarithmic approximation to the latter.}
\label{fig:rmtfunctions}
\end{minipage}
\end{center}
  \end{figure}


\begin{multline}
\big<(\delta G_{21})^2\big>^{1/2}\simeq \frac{e^2}{h}\frac{ \pi^2}{3\,2^8\,N}\frac{1}{(k_F\ell)^2}\left(\frac{kT}{\varepsilon_F}\right)^2\sqrt{2{\cal A}}\\
\times\Big[\int  {\rm d}{\boldsymbol v}  \big< {\mathscr A}({\boldsymbol v},0)\big>^4\int {\rm d}{\boldsymbol u}\, U_{12}^2({\boldsymbol u},0)\int {\rm d}{\boldsymbol y}\, U_{12}^2({\boldsymbol y},0)\Big]^{1/2}.
\end{multline}
 The integral over the spectral function gives
\begin{equation}
\int {\rm d}{\boldsymbol r} \big<{\mathscr A}({\boldsymbol r},0)\big>^4=2\pi\int_0^\infty {\rm d}r\,r \big<{\mathscr A}(r)\big>^4 =\frac{\pi}{2^7}\frac{\zeta(k_F\ell)k_F^{6}}{\varepsilon_F^4},
\end{equation}
where
\begin{equation}
\zeta(x)=\int_0^\infty dy\,ye^{-2y/x}J_0^4(y)\approx \frac{3}{2\pi^2}\ln 2x,\qquad x\gg 1.
\end{equation}
The approximate result can be shown with the aid of L'Hospital's rule. Similarly, for a screened interaction \index{coupling!, screened}\index{length scales!, screening length}
\begin{equation}
\bar{U}_{12}(r)=U_{12}(r)\exp(-r/r_s),\qquad U_{12}(r)=\frac{e^2}{4\pi\epsilon_0\epsilon_r r},
\end{equation}
we get
\begin{multline}
\int {\rm d}{\boldsymbol r} U_{12}^2({\boldsymbol r},0)=2\pi\int_0^\infty {\rm d}r\,r \bar{U}_{12}^2(\sqrt{r^2+d^2})\\=2\pi  U_{12}^2(d)d^2\Gamma_0(2d/r_s)=\frac{\pi}{2}  U_{12}^2(d) \Upsilon(2d/r_s) r_s^2,
\end{multline}
where 
\begin{equation}
\Gamma_\alpha(x)=\int_x^\infty dt\, t^{\alpha-1}e^{-t},
\end{equation}
is the incomplete Gamma function and
\begin{equation}
\Upsilon(x)=x^2\Gamma_0(x).
\end{equation}
The two introduced functions $\zeta(x)$ and $\Upsilon(x)$ are shown in Fig.~\ref{fig:rmtfunctions}. Collecting things we now obtain the estimate


\begin{equation}
\big<(\delta G_{21})^2\big>^{1/2}\simeq \varkappa \frac{e^2}{h}\left(\frac{kT}{\varepsilon_F}\cdot\frac{ U_{12}(d)}{\varepsilon_F}\right)^2\,\frac{\Upsilon(\tfrac{2d}{r_s})\, r_s^2\,k_F\sqrt{{\cal A}}\,\ln 2k_F\ell}{\ell^2\,N},
\end{equation}
where the numerical prefactor is given by
\begin{equation}
\varkappa =2^{-11}\sqrt{\pi^5/6}\simeq 3.5\times 10^{-3}.
\end{equation}
This result is reported in Ref.~[F] which generalizes the result first given in Ref.~[E] for $d\sim r_s$.

  For typical numbers (such as those in Table \ref{TABLE_samples} for the quantum dots shown in Fig.~\ref{SAMPLES_dots}) we estimate the drag fluctuations to be of the order of $0.1$ Ohm. As we have already noted the zero mean value and the magnitude of the finite fluctuations are not changed by breaking of time reversal symmetry, {\it i.e.} $\beta=1\longrightarrow 2$. The prediction of zero mean value and finite fluctations can be considered as a test of the degree of ergodicity.

%% file: summary.tex
\chapter{Summary}\label{chap:summary}

In the following we give a brief summary on the mesoscopic drag formalism, the numerical method, and the applications to quantum wires and quantum dots. Finally, concluding remarks are given.
 
\section{Mesoscopic formalism}
Starting from Kubo formalism we have calculated the DC drag conductance $G_{21}$ which relates the frictionally mediated response $I_2$ in subsystem $2$ to a driving field $V_1$ in subsystem $1$. The calculation is based on a second order perturbation expansion in the inter-subsystem Coulomb interaction $U_{12}$. The drag conductance,

\begin{multline}\label{summary_G21}
G_{21} =\frac{e^2}{h}
\iiiint{\rm d}{\boldsymbol r}_1'
{\rm d}{\boldsymbol r}_2'
{\rm d}{\boldsymbol r}_1''
{\rm d}{\boldsymbol r}_2''\,U_{12}({\boldsymbol r}_1',{\boldsymbol r}_2') U_{12}({\boldsymbol
 r}_1'',{\boldsymbol r}_2'')\\
\times{\mathscr P}\int_{-\infty}^\infty \hbar\,{\rm
    d}\omega\,\frac{{\Delta}_1(\omega,{\boldsymbol r}_1',{\boldsymbol r}_1'') {\Delta}_2(-\omega,{\boldsymbol r}_2',{\boldsymbol r}_2'')}{2kT\sinh^2(\hbar\omega/2kT)},
\end{multline}
is expressed in terms of two ``triangle'' functions
$\Delta_i=-\big<\hat{I}_i\hat{\rho}_i\hat\rho_i\big>$. Whereas often $U_{12}({\boldsymbol r}_1,{\boldsymbol r}_2)=U_{12}({\boldsymbol r}_1-{\boldsymbol r}_2)$ we for disordered systems with broken translation symmetry have ${\Delta}_i(\omega,{\boldsymbol r}_i,{\boldsymbol r}_i')\neq{\Delta}_i(\omega,{\boldsymbol r}_i-{\boldsymbol r}_i')$. This for instance means that the outcome of Eq.~(\ref{summary_G21}) can have both signs depending on the given disorder potential. 

Our calculations, which are restricted to Fermi liquids, demonstrate how we with the aid of Wick's theorem can formulate the $\Delta$'s in terms of the single-particle properties of the two subsystems.
The single-particle properties can be expressed by {\it i)} the wave functions in terms of the matrix elements of the current $I$ and particle-density $\rho$, {\it ii)} by the spectral function ${\mathscr A}$, {\it iii)} or by scattering states where the current matrix $I$ is expressed in terms of the scattering matrix $S$. Furthermore, we have derived results for the low-temperature limit and shown that $G_{21}\propto T^2$ as is also the case for the rate for ordinary particle-particle scattering. This formalism provides the framework for Refs.~[E,\,F,\,G,\,H].

\section{Numerical method}

Using the finite differences approximation we have mapped Eq.~(\ref{summary_G21}) onto a lattice and also formulated how the spectral function can be obtained by use of a simple numerical method. The resulting formula for $G_{21}$ can be written as a trace over matrices and in the low temperature regime we get

\begin{equation}\label{summary_trace}
G_{21} = \frac{e^2}{h}(kT)^2 \frac{\gamma^2}{3} \Tr\big\{ U_{21}\tilde{\Delta}_1 U_{12} \tilde{\Delta}_2\big\},
\end{equation}
where $\tilde{\Delta}$ is expressed in terms of the spectral function ${\mathscr A}_{\varepsilon_F}$ at the Fermi level. Here, $\gamma$ is the hopping element of the tight-binding-like Hamiltonian (band-width is $4\gamma$). Eq.~(\ref{summary_trace}) is very similar to existing trace formulas for the Landauer conductance $G_{ii}$. The implementation is straight forward and the method provides a very efficient tool for investigating drag in different geometries and/or for different disorder configurations. In this work the method is applied to ensembles of disordered wires.

\section{Quantum wires}
For one-dimensional quantum wires we have shown how in the ballistic regime analytical progress can be made. For long wires, $k_FL\gg 1$ we find that 
\begin{equation}
G_{21}\propto  U_{12}^2(2k_F),\qquad \ell \longrightarrow \infty,
\end{equation} 
where $U_{12}(q)$ is the Fourier transformed interaction and $\ell$ the mean free path. For quasi-ballistic wires the corrections to the mean-value are small so that 

\begin{equation}
\big<G_{21}\big>\propto  U_{12}^2(2k_F)+{\cal O}(1/k_F\ell).
\end{equation} 
For the fluctuations we apply perturbation theory and to leading order in the disorder strength we find that 
\begin{equation}
\big<(\delta G_{21})^2\big>\propto \big<{\cal R}\big>^2U_{12}^2(2k_F)U_{12}^2(0),
\end{equation}
where $\big<{\cal R}\big>\simeq L/\ell\ll 1$ is the mean reflection probability of the wires. For long-range interaction the relative magnitude 

\begin{equation}
\frac{\big<(\delta G_{21})^2\big>^{1/2}}{\big<G_{21}\big>}\sim \big<{\cal R}\big> \frac{U_{12}(0)}{U_{12}(2k_F)},
\end{equation} 
can be of the order of unity because $U_{12}(0)\gg U_{12}(2k_F)$. The large fluctuations can be understood from a combination of back-scattering in the disorder channel and forward-scattering in the Coulomb channel --- breaking the translational invariance by disorder allows for transfered momenta different from $2k_F$ and this gives rise to large drag fluctuations [E].

This we have also studied numerically and we found excellent agreement with the perturbative results [E]. In addition to the statistical properties the numerical studies have also given insight about the full distribution of $G_{21}$ for different natures of the disorder [F,\,G,\,H]. If the disorder potentials of the two wires are mutually fully correlated we diagrammatically find that mean value and fluctuations are enhanced compared to the un-correlated case, for the latter by a factor $\sqrt{2}$. These predictions are also confirmed numerically [F,\,G,\,H].

\section{Quantum dots}
For coupled chaotic quantum dots we have used random matrix theory to calculate the statistical properties. For the mean value we find that there is no drag on average, $\big<G_{21}\big>=0$. The physical explanation is that in the chaotic regime the induced current has no preferred direction and electrons will with equal probabilities escape the dot through either of its two leads. This implies finite fluctuations which is also what random matrix theory predicts. The estimate that we get for the fluctuations is of the form [E,\,F]

\begin{equation}
\big<(\delta G_{21})^2\big>^{1/2}\propto \frac{ r_s^2\,k_F\sqrt{{\cal A}}}{\ell^2\,N},
\end{equation}
where the screening length $r_s$ is of the order of the separation $d$, $\cal A$ is the interaction area, and $N$ the number of modes. Using realistic numbers we estimate that the drag fluctuations can be of the order of $0.1$ Ohm.

\section{Concluding remarks}

Rojo ends his 1998 review on electron-drag \cite{rojo1999} by the sentence: ``{\it The magnitude of the challenge seems to equal that of the progress made so far in the field}''. Since then the study of the mesoscopic regime has been initiated \cite[E]{narozhny2000,narozhny2001} and though there has been substantial theoretical progress I belive the major challenge is on the experimental side. However, the results reported in this thesis and attached publications [E,\,F,\,G,\,H] and by others \cite{narozhny2000,narozhny2001} illustrate the potential reward of overcoming the experimental challenges: Although the mesoscopic
fluctuations of the Coulomb drag conductance are non-universal they are an extreme
example of mesoscopic fluctuations --- they can be of the order of,
or even exceed, the mean value! Probably other mesoscopic phenomena are to be discovered and I hope that this work may serve as an inspiration for experimentalists to conduct Coulomb drag measurements in the mesoscopic regime.